\documentclass[11pt]{article}
\usepackage{graphicx,calc,epsfig} 
\usepackage{a4wide}
\usepackage{amsmath,amssymb,amsfonts} 
\usepackage[bf]{caption2}
\def\D{\displaystyle}

\def\ket#1{| #1 \rangle}

\def\bb1{\textup{\small{1}} \kern-3.6pt \textup{1}\ }

\DeclareFontFamily{U}{rsfs}{}         
\DeclareFontShape{U}{rsfs}{m}{n}{<5> rsfs5 <6><7> rsfs7          %
  <8><9><10><10.95><12><14.4><17.28><20.74><24.88> rsfs10}{}     %
\DeclareMathAlphabet{\mathfs}{U}{rsfs}{m}{n}                     %
\begin{document}
  \thispagestyle{empty}
  \vspace*{-1cm}
  \hspace*{\fill}\parbox{3.2cm}{{\small MPI--PhT--2000/39}
          \newline {\small LMU--TPW--00/22}}
  \begin{center}
    \vspace*{2 truecm}
    \Large\bf A generalized Hamiltonian Constraint Operator \\
        in Loop Quantum Gravity and its simplest \\
        Euclidean Matrix Elements\\[1.2 truecm]
    \normalsize \rm
    \renewcommand{\thefootnote}{\fnsymbol{footnote}}
     Marcus Gaul ${}^{2,3,4,}$\footnote[2]{\ttfamily{mred@mppmu.mpg.de}}
     and Carlo Rovelli ${}^{1,2,}$\footnote[3]
                {\ttfamily{carlo@rovelli.org}}\\[0.8 truecm]
    \renewcommand{\thefootnote}{\arabic{footnote}}
    {\small\it
      ${}^1$Department of Physics and Astronomy,
        University of Pittsburgh, Pittsburgh,\\
        PA 15260, USA\\[0.1 truecm]
      ${}^2$Centre de Physique Th\'eorique, CNRS Luminy, 
        F-13288 Marseille, France\\[0.1 truecm]
      ${}^3$Max--Planck--Institut f\"ur Physik, F\"ohringer Ring 6,
        D-80805 M\"unchen, Germany\\[0.1 truecm]
      ${}^4$Sektion Physik, Ludwig--Maximilians--Universit\"at,
        Theresienstr.\ 37,\\ D-80333 M\"unchen, Germany}\\
    \vspace{7mm}
    {\small (March 8, 2001)}\\
    \vspace{6mm}
    \begin{abstract}
      We study a generalized version of the Hamiltonian constraint
      operator in nonperturbative loop quantum gravity.  The
      generalization is based on admitting arbitrary irreducible
      $SU(2)$ representations in the regularization of the operator,
      in contrast to the original definition where only the
      fundamental representation is taken.  This leads to a
      quantization ambiguity and to a family of operators with the
      same classical limit.  We calculate the action of the Euclidean
      part of the generalized Hamiltonian constraint on trivalent
      states, using the graphical notation of Temperley-Lieb
      recoupling theory.  We discuss the relation between this
      generalization of the Hamiltonian constraint and crossing
      symmetry.
    \end{abstract}
  \vspace{18mm}
  \end{center}
  Short title: Generalized Hamiltonian in Loop Quantum Gravity
\newpage


\section{Introduction}

Loop quantum gravity is a canonical approach to the quantization of
general relativity which has undergone lively progress in the last
decade, yielding a well-defined framework for the nonperturbative
formulation of background-independent quantum field theory (for a
review, see \cite{Rovelli97a}).  One of the elusive issues in this
approach is the identification of the physically correct Hamiltonian
constraint operator (HCO), encoding the dynamics of classical general
relativity.

The correct form of the HCO has long been searched
\cite{loops,Early-hamiltonians,Late-hamiltonians}.  A mathematically
well-defined and anomaly-free HCO was found by Thiemann in 1996
\cite{Thiemann96a,Thiemann96b}, developing ideas and techniques
introduced in \cite{HamiltonianRS} and \cite{HamiltonianR}.  Although
certain doubts concerning the correctness of the classical limit of
this operator have been raised \cite{Gambinietal97,%
LewandowskiMarolf97,Smolin96}, ongoing work on coherent states and
semi-classical quantum gravity 
\cite{Thiemann00a,ThiemannWinkler00a,ThiemannWinkler00b} should clarify 
the issue, and the Thiemann operator remains a very appealing candidate
for the HCO of the physical theory.  On the other hand, the
construction of the Thiemann operator involves some arbitrary choices,
or quantization ambiguities, and the possibility of exploring
alternatives is open.  In this work, we analyze one of the
quantization ambiguities entering the definition of the HCO, and a
corresponding variant of Thiemann's HCO. In particular, we study a
family of operators $\hat{\mathcal{H}}^m$ labelled by irreducible
$SU(2)$ representations\footnote{We label $SU(2)$ irreps with the
``color" $m=2j$, which is twice the spin, thus $m\in \mathbb{N}$.} $m$, 
all having the same classical limit, namely the classical Hamiltonian
constraint of general relativity.  Thiemann's HCO corresponds to the
fundamental representation $m=1$.  In a nutshell, the HCO requires a
gauge invariant point-splitting-like regularization, which is obtained
by using the trace of the holonomy of the gravitational connection. 
It turns out that by choosing the representation-$m$ trace, we obtain
a distinct operator with the same classical limit: the
$\hat{\mathcal{H}}^m$ version of the HCO.

The HCO operator introduced in References \cite{HamiltonianRS} and
\cite{HamiltonianR} adds a link of color 1 to the nodes of the spin
network states.  Thiemann's operator $\hat{\mathcal{H}}^1$, modeled on
the former, acts in the same way.  On the other hand, the operators
$\hat{\mathcal{H}}^m$ for arbitrary $m$, which we study here, act on
the spin network states by adding a link of color $m$.  The
possibility of this variant of the HCO has been suggested also by
Roberto De Pietri and Laurent Freidel \cite{DPF}.

This extension is motivated by the spacetime covariant formulation of
the theory.  A (Euclidean) ``path integral''-like sum-over-histories
approach can be formally derived from the canonical theory
\cite{ReisenbergerRovelli97}.  Histories are represented by spin
foams, that is, branched colored 2-dimensional surfaces (2-complexes)
\cite{Baez98,Baez99}.  The resulting model has close connections with
topological field theories, as well as their non-topological
extensions like the Barrett-Crane \cite{BarrettCrane98} model, and to
simplicial models of quantum gravity.  The key ingredient of a spin
foam model is its vertex amplitude.  In the spin foam model arising
from loop quantum gravity, the vertex amplitude is given by the matrix
elements of the Hamiltonian constraint.  The problem of finding the
correct HCO is thus translated into the problem of finding the correct
vertex amplitude.  In the covariant framework, however, we have the
advantage of manifest 4d general covariance.  In particular, if we
consider the Euclidean sector, the vertex amplitude should be
rotationally invariant \cite{ReisenbergerRovelli97}, a requirement
denoted \emph{crossing symmetry}.  It is easy to see, using
counter-examples, that Thiemann's HCO does \emph{not} yield a crossing
symmetric vertex amplitude.  The reason is that crossing symmetry
rotates the link added by the constraint into links of the state acted
upon, but the first one has always color 1, while the latter have arbitrary
colors.  In searching for a crossing symmetric variant of Thiemann's
HCO, one is thus naturally led to consider HCO's
$\hat{\mathcal{H}}^m$ that add links of arbitrary color $m$.  In this
paper, we show in detail that quantization ambiguities do indeed allow
us to define such operators, and we study the action of these
operators on trivalent states.  On the other hand, we shall not
address here the problem of the existence of a crossing symmetric
linear combination of such operators $\hat{\mathcal{H}}=\sum_m c_{m}
\hat{\mathcal{H}}^m$, which would actually define a crossing symmetric
quantization of the Hamiltonian constraint of general relativity.

The paper is organized as follows.  In the following section we
construct the generalized HCO's $\hat{\mathcal{H}}^m$, closely
following Thiemann's construction \cite{Thiemann96a,Thiemann96b}.  We
show that these form a family of anomaly-free, classically equivalent
operators.  We also show that no analogous ambiguity emerges for
similar generalizations of simpler operators like area or volume.  In
other words, the ambiguity is a feature of the complications of the
dynamical operator, not a generic ambiguity in the formalism. 

We then restrict our attention to the Euclidean part of the
$\hat{\mathcal{H}}^m$, and we compute their action on generic
(gauge-invariant) trivalent vertices.  To this aim, we use the
powerful graphical computational techniques of the tangle-theoretic
Temperley-Lieb recoupling theory.  This work generalizes the results of 
Reference \cite{Borissovetal97} on Thiemann's HCO to arbitrary $m$.
We obtain a final form that is appropriate for further considerations
concerning crossing symmetry. 

In appendix \ref{app:ThiemannsHCO}, the particular case which
corresponds to the original definition of the Hamiltonian constraint
operator is addressed.  Its matrix elements have already been computed
in \cite{Borissovetal97}, giving the opportunity for a consistency
check.  Restrictions of the general expression to $\mathcal{H}^1$
confirm our result.  Appendix \ref{app:recoupl_theory} outlines the
basic facts and the most commonly used identities of recoupling theory
that are needed throughout the text.

For the general framework of loop quantum gravity and for details on
the computational tools we use here, see
\cite{GaulRovelli99,Rovelli97a,DePietriRovelli96,DePietri97,Borissovetal97}.


\section{The generalized Hamiltonian}
\label{sec:Hamiltonian}


\subsection{Classical Theory}
We begin by reviewing the construction of Thiemann's HCO. The starting
point is the classical Lorentzian Hamiltonian constraint $\mathcal{C}$
of density weight one.  Using real Ashtekar-Barbero variables
\cite{Barbero95}, this can be written as
\begin{equation}
  \label{BarberoHamiltonian}
  \mathcal{C} = \frac{1}{\sqrt{\det(q)}} \, 
        \left( \epsilon^{ij}{ }_k E^a_i E^b_j F_{ab}^k - 
        4\, K_a^i K_b^j E^a_{[i} E^b_{j]} \right)  ~. 
\end{equation}
Here $a,b$ are tensorial indices on the compact spatial manifold
$\Sigma$, and $i,j,k$ are $su(2)$ indices.  Square brackets denote
antisymmetrization.  The inverse densitized triad (an $su(2)$ valued
vector density) has components defined by $E^a_i := \det(e^j_b)
e^a_i$, where $e^i_a$ is the triad on $\Sigma$, and the real $SU(2)$
Ashtekar-Barbero connection\footnote{That is, the Immirzi parameter in
$A^i_a:= \Gamma^i_a + \beta K_a^i$ is set to $\beta=1$.} is $A^i_a :=
\Gamma^i_a + K_a^i$.  On $\Sigma$ we have the induced metric $q_{ab}$,
whose inverse satisfies $\det(q)\, q^{ab} = E^a_i E^b_j\,
\delta^{ij}$, as well as the extrinsic curvature $K_{ab}$.  Using the
triad, one obtains $K_a^i = K_{ab} E^{bi}/\sqrt{\det(q)}$ by
transforming one spatial index into an internal one.  Furthermore,
$\Gamma^i_a$ is the spin connection compatible with the triad.  The
variables $(A^i_a,E^a_i)$ form a canonically conjugate pair, whose
fundamental Poisson brackets are $\{A^i_a(x),E^b_j(y)\} = G\,
\delta^a_b \delta^i_j \delta(x,y)$, $G$ being $16 \pi G_N c^{-3}$ with
Newtons constant $G_N$.  Finally, $F_{ab}^k$ are the components of
the curvature of the connection, given by $F_{ab}^k = 2 \partial_{[a}
A^k_{b]} + \epsilon_{ij}{ }^k A_a^i A_b^j$.

We express Lie algebra valued quantities in 
terms of a basis of (anti-hermitian) $SU(2)$ generators $\tau_i$, 
satisfying $[\tau_i,\tau_j]=\epsilon_{ij}{}{}^k \tau_k$. That is, 
we write $A_a = A^i_a \tau_i$, $E^a = E^{ai} \tau_i$ and so on. 
The Hamiltonian constraint (\ref{BarberoHamiltonian}) can then be 
written as 
\begin{eqnarray}
  \label{Ham}
  \mathcal{C}\!\!\!&=&\!\!\! - \:  \frac{2}{\sqrt{\det(q)}} \,
    \mbox{Tr} \left((F_{ab} - 2\, [K_a, K_b])[E^{a},E^{b}]\right).
    \label{Ham2}
\end{eqnarray} 
As realized by Thiemann \cite{Thiemann96a}, a more convenient starting
point for the quantization is given by the polynomial expression for
the densitized Hamiltonian constraint 
\begin{eqnarray}
  \mathcal{C} &=& - 2 \, \left[ \, \frac{2}{G}\, \epsilon^{abc}\,\mbox{Tr} 
     \big(F_{ab} \{ A_c,V \}\big) - \frac{8}{G^3}\, 
       \epsilon^{abc} \, \mbox{Tr} \big(\{ A_a,K \} \{ A_b,K \} 
       \{ A_c,V \}\big) \right]   
        \label{general_Ham} \\
    &=:& \mathcal{H} - \mathcal{T} 
        \label{HamE_T} ~,
\end{eqnarray}
where $V$ is the volume of $\Sigma$,
\begin{equation}
  \label{class_volume}
  V(\Sigma) = \int_{\Sigma} \mbox{d}^3x \sqrt{|\det(q)|}
        = \int_{\Sigma} \mbox{d}^3x \sqrt{\frac{1}{3!} | \epsilon_{abc}
        \epsilon^{ijk} E^a_i E^b_j E^c_k |} ~.
\end{equation}
$K$ is the integrated trace of the densitized extrinsic curvature 
of $\Sigma$,
\begin{equation}
  \label{integrated_K}
  K = \int_{\Sigma} \mbox{d}^3x \sqrt{\det(q)}\, K_{ab} q^{ab} 
        = \int_{\Sigma} \mbox{d}^3x \, K^i_a E^a_i ~,
\end{equation}
and we denote the Euclidean Hamiltonian constraint with $\mathcal{H}$, and
the `Lorentzian term' with $\mathcal{T}$ .  Using the relations
\begin{equation}
  \label{keyID1}
  \frac{[E^a,E^b]^i}{\sqrt{\det(q)}} = \frac{2}{G} \, 
        \epsilon^{abc} \{A_c^i,V\} 
\end{equation}
and 
\begin{equation}
  \label{keyID2}
  K^i_a = \frac{1}{G} \{A_a^i,K\} ~.
\end{equation}
it is straightforward to see that (\ref{HamE_T}) is equal to
(\ref{BarberoHamiltonian}).  Furthermore, one takes advantage of the
fact that the integrated extrinsic curvature (\ref{integrated_K}) is
the time derivative of the volume, i.e. can be written as the Poisson
bracket of volume and (Euclidean) Hamiltonian constraint at lapse
equal to one.  Using this in (\ref{general_Ham}) we see that the
Hamiltonian constraint can be entirely expressed in terms of the
volume and the connection.  
In this paper we restrict ourselves to the
study of the Euclidean constraint.  Its smeared form is
\begin{eqnarray}
  \mathcal{H}[N] &=&   
       \int_\Sigma d^3x \, N(x)\, \mathcal{H}(x)\\
     &=&  -\frac{4}{G}\, \int_\Sigma d^3x \, N(x) 
      \, \epsilon^{abc}\, \mbox{Tr} \big(F_{ab} 
       \{ A_c, V \}\big) ~,
        \label{H_E_N}   
\end{eqnarray} 
where $N(x)$ is the lapse function.  

\begin{figure}[b]
  \centerline{\mbox{\epsfig{file=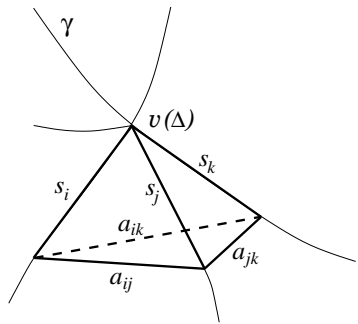,width=5cm}}}
  \begin{center}
    \parbox{9cm}{\caption[]{\label{tetrahedron} \small
        An elementary tetrahedron $\Delta \in T$ constructed
        by adapting it to a graph $\gamma$ which
        underlies a cylindrical function.}}
  \end{center}
\end{figure}

The regularization of $\mathcal{H}[N]$ is obtained by approximating
$F_{ab}$ and $A_c$, which do not have direct quantum analogues, with
holonomies of the connection around ``small" loops, which do.  We fix
an arbitrary triangulation $T$ of the manifold $\Sigma$ into
elementary tetrahedra with analytic edges.  Consider a tetrahedron
$\Delta$, and a vertex $v$ of this tetrahedron.  Denote the three
edges that meet at $v$ as $s_i$, $i=1,2,3$.  Denote $a_{ij}$ the edge
connecting the two end-points of $s_{i}$ and $s_{j}$ which are not on
$v$.  That is, $s_{i}$, $s_{j}$ and $a_{ij}$ form a triangle.  We
denote this triangle as $\alpha_{ij} := s_{i} \circ a_{ij} \circ
s_j^{-1}$.  When we want to stress that a vertex, a segment or a loop
belongs to the tetrahedron $\Delta$, we write $v(\Delta)$ ,
$s_i(\Delta)$, $\alpha_{ij}(\Delta)$ and so on.  Figure
\ref{tetrahedron} illustrates the construction.
We decompose the smeared Euclidean constraint (\ref{H_E_N}) into a sum
of one term per each tetrahedron of the triangulation
\begin{eqnarray}
  \label{Ham_T1}
  \mathcal{H}[N] = \sum_{\Delta \in T}
        \frac{-4}{G}\, \int_\Delta d^3x \, N(x) 
          \, \epsilon^{abc}\, \mbox{Tr} \big(F_{ab} 
            \{ A_c, V \}\big) ~.
          \label{Ham_T2}
\end{eqnarray}
Finally, we consider the holonomy $h_e :=
\mathcal{P}\exp(-\int_e A) = \mathcal{P}\exp(-\int_e A^i_a
\tau_i dx^a)$ of the connection along edges $e$, and we define the 
(classical) {\em regularized\/} Euclidean Hamiltonian constraint as 
\begin{equation}
   \mathcal{H}_{T}[N]
     :=  \sum_{\Delta \in T} \mathcal{H}_{\Delta} [N] ~,
        \label{H_delta:class_equiv}
 \end{equation} 
where 
\begin{equation}
   \label{H_delta}
   \mathcal{H}_{\Delta}[N]
     := - \frac{2}{3 G} \, 
        N(v(\Delta)) \, \epsilon^{ijk} \,
        \mbox{Tr}\Big[h_{\alpha_{ij}(\Delta)} 
        h_{s_{k}(\Delta)} \big\{ 
        h^{-1}_{s_{k}(\Delta)},V\big\}\Big] ~. 
\end{equation} 
A straightforward calculation, using the expansions 
\begin{equation} 
h_{s_{k}} = 1+A_{a}(v) s_{k}^a +O(s^2), 
\label{exp1}
\end{equation} 
and
\begin{equation} 
h_{\alpha_{ij}}= 1+ \frac{1}{2} F_{ab}(v) s_{i}^a s_{j}^b 
  + O(|s_{i}\times s_{j}|), 
\label{exp2}
\end{equation} 
shows that for a fixed value of the connection and triad, the
expression (\ref{H_delta:class_equiv}) converges to the Hamiltonian
constraint (\ref{Ham_T2}) if the triangulation is sufficiently fine. 
That is, the lattice spacing of the triangulation $T$ acts as a 
regularization parameter.  We write
\begin{equation}
    \mathcal{H}_{T}[N] \;\;
     \stackrel{T\to \infty}{\longrightarrow} \;\;
     \mathcal{H}[N] 
\end{equation}
to indicate that, for a fixed value of the fields, the r.h.s. can be
made arbitrary close to the l.h.s, by taking a sufficiently fine 
triangulation. Here $T\to \infty$ denotes the continuum limit of
finer and finer triangulations of $\Sigma$ in the sense of more 
and more tetrahedra.

Since the volume and the holonomy have corresponding quantum
operators, expression (\ref{H_delta:class_equiv}) can 
immediately be transformed into a quantum operator, yielding the
regularized HCO. Remarkably, this operator converges to a well-defined
limit when we take finer and finer triangulations.

Let us now introduce our alternative regularization.  Notice that the
trace in Equation (\ref{Ham_T2}) is over the $su(2)$
algebra; on the other hand, the trace used in the regularization
(\ref{H_delta}) is over the $SU(2)$ group.   
However, there are many traces over  $SU(2)$. Given an irreducible 
representation of spin $j$ and color $m=2j$, we can write
\begin{equation} 
\mbox{Tr}_{m}[U]=\mbox{Tr}[R^{(m)}(U)] ~,
\end{equation}
where $R^{(m)}$ is the matrix representing $U$ in the representation
$m$. What happens if we replace the trace $\mbox{Tr}$ with the trace 
$\mbox{Tr}_{m}$ in the regularization of the constraint? 
Let us define
\begin{equation}
   \mathcal{H}^m_{\Delta}[N]
     := \frac{2}{3 G C(m)} \, 
        N(v(\Delta)) \, \epsilon^{ijk} \, 
        \mbox{Tr}_{m}\Big[h_{\alpha_{ij}(\Delta)} 
        h_{s_{k}(\Delta)} \big\{ 
        h^{-1}_{s_{k}(\Delta)},V\big\}\Big] ~,
\end{equation} 
where $C(m)$ is a constant that we will fix in a moment. Equivalently, 
\begin{equation}
   \label{Hm_delta:classical}
   \mathcal{H}^m_{\Delta}[N]
     := \frac{2}{3 G C(m)} \, 
        N(v(\Delta)) \, \epsilon^{ijk} \,
        \mbox{Tr}\Big[h^{(m)}_{\alpha_{ij}(\Delta)} 
        h^{(m)}_{s_{k}(\Delta)} \big\{ 
        h^{(m)-1}_{s_{k}(\Delta)},V\big\}\Big] ~,
\end{equation} 
where $h^{(m)}=R^{(m)}(h)$.  Clearly, $\mathcal{H}^m_{\Delta}[N]$ is
in general distinct from $\mathcal{H}_{\Delta}[N]$.  However, it is
straightforward to verify that (for a suitable value of $C(m)$) it
converges to the same value, namely to the classical Hamiltonian
constraint, for a sufficiently fine triangulation. Indeed, using
again the expansions of the holonomy,
\begin{eqnarray}
  h_{s_k}^{(m)} \!\!&\simeq&\!\! \bb1^{\!\!(m)} + A_a^j 
   \tau_j^{(m)}
        s_k^a(\Delta)~, \\
  h_{\alpha_{ij}}^{(m)} \!\!&\simeq&\!\! \bb1^{\!\!(m)} + \frac{1}{2} 
        \, F^{k}_{ab}  \tau_k^{(m)} s_i^a(\Delta) s_j^b(\Delta) ~,
\end{eqnarray}
where $ \tau_j^{(m)}$ are the generators of the irreducible 
representation of color $m$, and using 
\begin{equation}
  \mbox{Tr} \left(\tau_i^{(m)} \tau_j^{(m)}\right) =
       - \,\frac{1}{12}\, m(m+1)(m+2)\, \delta_{ij} ~,
\end{equation}
we see immediately that if we pose 
\begin{equation} 
  C(m) = \frac{1}{12}\, m(m+1)(m+2) ~,
\end{equation} 
then
\begin{equation}
   \label{Hm_delta:class_equiv}
   \mathcal{H}^m_{T}[N]
     :=  \sum_{\Delta \in T} \mathcal{H}^m_{\Delta} [N] 
 \end{equation} 
converges to the Hamiltonian constraint for a sufficiently fine
triangulation, precisely as $\mathcal{H}_{\Delta}[N]$.  Thus,
$\mathcal{H}^{m}_{\Delta}[N]$ is a different quantity than
$\mathcal{H}_{\Delta}[N]$, but the difference between the two goes
to zero as the triangulation is refined.  That is,
\begin{equation}
    \mathcal{H}^m_{T}[N] \;\;
     \stackrel{T\to \infty}{\longrightarrow} \;\;
     \mathcal{H}[N] ~.
\end{equation}


\subsection{Quantum Theory}

The quantization of the regularized Hamiltonian is performed by
replacing the classical variables volume\footnote{ As clarified in
\cite{Lewandowski96}, there are two versions of the volume operator,
$\hat{V}_{RS}$ introduced in \cite{RovelliSmolin95} and $\hat{V}_{AL}$
introduced in \cite{AshtekarLewand98}.  Their difference stems from a
different regularization procedure.  In the case of a generic,
non-planar trivalent vertex, which we consider in Section
\ref{sec:action}, the two are equivalent.} and holonomy by the
corresponding quantum operators via $V \to \hat{V} \equiv
\hat{V}_{AL}$ and $h^{(m)}_e \to \hat{h}^{(m)}$.  Moreover, Poisson
brackets $\{\cdot\,,\cdot \}$ turn into commutators
$[\cdot\,,\cdot]/i\hbar$.

Classically, the regularized Euclidean constraints
$\mathcal{H}^m_{T}[N]$ (\ref{Hm_delta:class_equiv}) are different
objects for distinct colors $m$.  In the limit of arbitrary fine 
triangulation,
they are the same.  Quantizing these quantities, we obtain a family of
regularized operators $\hat{\mathcal{H}}^m_{T}[N]$.  The limit in
which the regularization is finer and finer is well-defined in the
quantum theory (by restricting to the diffeomorphism invariant states)
\cite{HamiltonianRS,HamiltonianR,Thiemann96a,Thiemann96b}.
However, the key point is that this limit turns out to give different 
operators for different $m$'s.

Following \cite{Thiemann96a,Borissovetal97}, the HCO operator is
defined by adapting the triangulation $T$ to the graph $\gamma$ of the
basis state $\psi_{\gamma}$ on which the operator is going to act, see
Fig.  \ref{tetrahedron}.  That is, there is a procedure for fixing a
triangulation $T^\gamma$ for each graph $\gamma$, and if we define 
\begin{equation}
  \label{Ham_operatorDef}
  \hat{\mathcal{H}}^m[N] \, \psi_\gamma := 
     \hat{\mathcal{H}}^m_{T^\gamma} [N] \, \psi_\gamma = 
     \sum_{\Delta \in T^\gamma} 
     \hat{\mathcal{H}}^m_{\Delta} [N] \, \psi_\gamma ~,
\end{equation}
one shows that a finer triangulation would yield the same
operator.  We refer to Thiemann's papers for the discussion of these
technicalities, which play no special role here.  By replacing the
classical quantities with quantum operators, and the Poisson brackets
with commutators, we obtain the operator associated to a single
tetrahedron:
\begin{eqnarray}
   \label{Ham_operator}
     \hat{\mathcal{H}}^m_{\Delta}[N] &=& - \frac{2 i}{3 l_0^2 C(m)} \, 
        N(v(\Delta)) \, \epsilon^{ijk} \,
        \mbox{Tr} \left(\hat{h}^{(m)}[\alpha_{ij}] \,
        \hat{h}^{(m)}[s_{k}]\, \Big[ 
        \hat{h}^{(m)}[s^{-1}_{k}] \, ,\hat{V} \Big] \right) \\  
        &=:& N_v \hat{\mathcal{H}}^m_{\Delta} ~.
        \label{NxH}
\end{eqnarray}
Here we have introduced $N_v := N(v(\Delta))$ and $l_0^2 = \hbar G = 16
\pi G_N \hbar c^{-3} = 16 \pi l^2_{Planck}$. 

Let us comment on factor ordering.  It is obvious that the factor
ordering of the operators in (\ref{Ham_operator}) or (\ref{NxH}) is
not the only possible one that has (\ref{Hm_delta:classical}) as its
classical limit.  For the $m=1$ case it has been shown in
\cite{Borissovetal97} that two possible orderings of the operators
exist.  Since the argument that is given in this article is in fact 
independent of any color $m$, the same is valid here.  Hence the two natural 
orderings for $\hat{\mathcal{H}}^m_{\Delta}$ are
\begin{eqnarray}
  \label{ordering1a}
  \lefteqn{\hat{\mathcal{H}}^m_{\Delta (1)} = - \frac{2 i}{3 l_0^2 C(m)} \, 
    \epsilon^{ijk} \, \mbox{Tr}\left(
    \frac{\hat{h}^{(m)}[\alpha_{ij}] - \hat{h}^{(m)}[\alpha_{ji}]}{2}\,
    \hat{h}^{(m)}[s_{k}]\, \Big[ \hat{h}^{(m)}[s^{-1}_{k}] 
    \, ,\hat{V} \Big] \right)} \hspace{0.7cm}\\
  && \longrightarrow ~
  \frac{2 i}{3 l_0^2 C(m)} \, \epsilon^{ijk} \,
  \mbox{Tr}\left(
    \frac{\hat{h}^{(m)}[\alpha_{ij}] - \hat{h}^{(m)}[\alpha_{ji}]}{2}\,
        \hat{h}^{(m)}[s_{k}]\, \hat{V} \,
        \hat{h}^{(m)}[s^{-1}_{k}] \right)
  \label{ordering1b}
\end{eqnarray}
and
\begin{eqnarray}
  \label{ordering2a}
  \lefteqn{\hat{\mathcal{H}}^m_{\Delta (2)} = - \frac{2 i}{3 l_0^2 C(m)} \, 
    \epsilon^{ijk} 
    \, \mbox{Tr} \left( \hat{h}^{(m)}[s_{k}]\, \Big[ 
    \hat{h}^{(m)}[s^{-1}_{k}] \, ,\hat{V} \Big] \,
    \frac{\hat{h}^{(m)}[\alpha_{ij}] - \hat{h}^{(m)}[\alpha_{ji}]}{2}
    \right)} \hspace{1cm}\\
  && \longrightarrow ~
    \frac{2 i}{3 l_0^2 C(m)} \, \epsilon^{ijk} \,
    \mbox{Tr} \left( \hat{h}^{(m)}[s_{k}]\, \hat{V} \,
    \hat{h}^{(m)}[s^{-1}_{k}] \,
    \frac{\hat{h}^{(m)}[\alpha_{ij}] - \hat{h}^{(m)}[\alpha_{ji}]}{2}
    \right) ~.
  \label{ordering2b}
\end{eqnarray}
Equations (\ref{ordering1a}) and (\ref{ordering2a}) are direct
consequences of the ordering choices. But it turns out that only
(\ref{ordering1b}) and (\ref{ordering2b}) have non-vanishing
actions on cylindrical functions, independent of the valence of
the underlying graph.

We proceed with the construction of the adapted triangulation
$T^\gamma$ that was left out above.  The action of the HCO on spin
network states is such that only vertices of the triangulation
corresponding to nodes of the graph $\gamma$ contribute.  Because of
this, the triangulation can be split into parts around vertices and
another part for the rest of $\Sigma$.  Consider an $n$-valent vertex
$v$.  Assign to each triple $(e_i, e_j, e_k)$ of edges adjacent to the
vertex an adapted tetrahedron $\Delta$ of the triangulation, such that
its basepoint $v(\Delta)$ coincides with $v$.  The segments $s_i$,
$s_j$ and $s_k$ introduced above, are chosen as parts of the edges
incident to $v$.  The loop $\alpha_{ij}$ (a ``triangle''), which is
build from $s_i$ and $s_j$, forms the base of the tetrahedron, that is
finally spanned by the three segments $s_i$, $s_j$ and $s_k$. 
Thiemann has given a procedure to construct seven more tetrahedra
around each vertex, based on the above one.  This is performed in such
a way that the vertex is always completely enclosed by the eight
tetrahedra, independent of the fineness of the triangulation.  The
vertex triangulation is concluded by repeating this procedure for each
set of the $E(v) = \binom{n}{3}$ unordered triples of edges adjacent
to $v$.  The rest of the 3-manifold $\Sigma$ is triangulated
arbitrarily.

Further simple manipulations of (\ref{Ham_operatorDef}) finally allow 
to write the generalized HCO as
\begin{equation}
  \hat{\mathcal{H}}^m_{\gamma}[N] \, \psi_\gamma = 
  \sum_{v \in \mathcal{V}(\gamma)} 8 N_v
  \sum_{v(\Delta) = v} \hat{\mathcal{H}}^m_{\Delta} 
  \,\frac{p_{\Delta}}{E(v)} \, \psi_\gamma ~,
\end{equation}
where $\mathcal{V}(\gamma)$ is the set of vertices of $\gamma$.
Moreover, $p_{\Delta}$ is one, whenever 
$\Delta$ is a tetrahedron having three edges coinciding with three
edges of the spin network state, that meet at the vertex $v$.
In the other cases $p_{\Delta}$ equals zero.

As first realized in \cite{HamiltonianRS}, the continuum limit of the
HCO turns out to be trivial in the quantum theory.  In the
diffeo\-mor\-phism-invariant context, which is the regime in which the
operator is indeed well-defined, the regulator dependence drops out
trivially without ever taking the limit explicitly.  More concretely,
two operators $\mathcal{H}$ and $\mathcal{H}'$ that are related by a
refinement of an adapted triangulation (for a fixed graph) differ from
each other only by the size of the loops $\alpha_{ij}$.  More
precisely, their actions on fixed spin network states based on
$\gamma$ differ only by a diffeomorphism which `moves' the segment
$a_{ij}$ of $\alpha_{ij}$.  If $\phi$ is a diffeomorphism invariant
state, we have therefore $\langle \phi \mathcal{H}\psi \rangle = 
\langle \phi \mathcal{H}' \psi \rangle$, and therefore the (dual) 
action of $\mathcal{H}$ and
$\mathcal{H}'$ on $\phi$ is the same.  Hence the restriction of
the HCO's on the (dual) diffeomorphism invariant states is independent
from the refinement of the triangulation.

As in Thiemann's definition, it is easy to see that for each fixed $m$ 
the operators \linebreak
$\{ \hat{\mathcal{H}}^m [N] \mid m \in \mathbb{N}_+
\}_{\gamma}$ are anomaly free.  The proof that is given in
\cite{Thiemann96a} is independent of any representation in the sense
used here, hence it can be adopted for the generalized case as well. 
That is
\begin{equation}
  \Big[ \hat{\mathcal{H}}^m [N], 
        \hat{\mathcal{H}}^m [M] \Big] \,\psi_{\gamma}
        = 0 ~,
\end{equation}
for any two lapse functions $N, M$ and cylindrical functions (or spin
network states) $\psi_\gamma$ when evaluated on a
diffeo\-mor\-phism-invariant state.  Commutators of constraints in
different representations $m\neq m'$ will be considered elsewhere.

To put it in a nutshell, we end up with a finite, well-defined,
consistent and diffeomorphism-covariant family of anomaly-free
Euclidean constraint operators giving rise to a new quantization 
ambiguity with respect to the $SU(2)$ color $m$. 

The quantization of the kinetic term $\mathcal{T}^m$ in
(\ref{general_Ham}) will not be studied in this article.  It can
straightforwardly be carried out in the same fashion as above,
resulting eventually in the generalized version of the full Lorentzian
HCO.


\subsection{A brief note on the quantization ambiguity}
\label{subsec:QuantAmbig}

Before concluding this section, we comment on the meaning of the
quantization ambiguity we have found.  Holonomies appear generically
in the regularization of quantum operators in loop quantum gravity. 
Thus one may wonder whether all operators are plagued by the same
quantization ambiguity as the Hamiltonian constraint.  This fact would
shed some doubts on the results on the spectrum of area and volume
\cite{RovelliSmolin95}, which are central results in loop quantum
gravity.  Here we show that this is not the case.  That is, the
quantization ambiguity associated to the choice of the representation
in which to take the holonomy is a consequence of the complication of
the Hamiltonian constraint, and not a generic feature in loop quantum 
gravity.  In particular, area and volume operators do not change if we 
quantize them using holonomies in an arbitrary representation $m$. 

Detailed quantizations of the volume operator involving various
technicalities, can be found for the loop as well as the connection
representation in \cite{AshtekarLewand98,DePietriRovelli96}. 
Generalizing to arbitrary colors does not introduce new complications. 
Let us take for example the volume operator defined in
\cite{RovelliSmolin95}.  The classical volume (\ref{class_volume}) of
a 3-dimensional spatial region $\mathcal{R} \subset \Sigma$ is given
by
\begin{equation}
  \label{class_vol}
  V_\mathcal{R} = \int_{\mathcal{R}} d^3 x 
        \, \sqrt{\det q(x) }\: = \int_{\mathcal{R}} d^3 x \, 
        \sqrt{|\det E(x) |} ~.
\end{equation}
This expression is regularized via a point-splitting procedure as
follows.  $\mathcal{R}$ is partitioned into small, $\epsilon$-sized
cubic cells $I_\epsilon$.  Consider $\det E(x_I)$ for an arbitrary
$x_I \in I_\epsilon$.  Point-split by placing the triads at three
distinct points $\sigma$, $\tau$ and $\rho$ of a small, closed loop
$\alpha$, which lies entirely inside a cube $I_\epsilon$ and satisfies
$\alpha \cap \partial I_\epsilon = \{\sigma,\tau,\rho\}$ on the
boundary $\partial I_\epsilon$. Consider the loop variable 
\begin{equation}
\label{Tabc}
  T^{abc}[\alpha](\sigma,\tau,\rho) =
  - \mbox{Tr} \left(E^a(\sigma)\, h_\alpha(\sigma,\tau) \,
        E^b(\tau)\, h_\alpha(\tau,\rho) \, 
        E^c(\rho)\, h_\alpha(\rho,\sigma) \, \right) ~,
\end{equation}
The trace is defined over the fundamental color $m=1$ matrix 
representation of the involved holonomies and triads, i.e. in terms
of the usual trace of matrix products.
The relation of (\ref{Tabc}) to the volume can easily be seen in the
limit of vanishing regulator. We expand $h = 1 + \mathcal{O}(\epsilon)$,
and for smooth $E^{ai}$, $E^{a}(\sigma) = E^{ai}(x_I) \, \tau_i 
+ \mathcal{O}(\epsilon)$.
Then we get
\begin{eqnarray}
  T^{abc}[\alpha](\sigma,\tau,\rho) = 
  \frac{1}{4} \,\epsilon^{abc} \det E(x_I) + \mathcal{O}(\epsilon) ~,
\end{eqnarray}  
where $\mbox{Tr} (\tau_i \tau_j \tau_k) = - \frac{1}{4} \, \epsilon_{ijk}$ 
is used. We define now
\begin{equation}
  \label{regularized_volume}
  V_{\epsilon} = \sum_{I_\epsilon} \sqrt{|V^{2}_{I_\epsilon}|} ~,
\end{equation} 
and
\begin{equation}
  V^{2}_{I_\epsilon} = \frac{1}{12 \, \epsilon^6}
  \int_{\partial I_\epsilon} d\sigma^2 \int_{\partial I_\epsilon} d\tau^2 
  \int_{\partial I_\epsilon} d\rho^2 \,
  n_a(\sigma) n_b(\tau) n_c(\rho) \,T^{abc}[\alpha](\sigma,\tau,\rho) ~.
\end{equation}
Here the $n_k$ are normal one-forms on the boundary of the cube $I_\epsilon$.
In the limit of vanishing regulator $\epsilon$, we obtain 
\begin{equation}
  V^{2}_{I_\epsilon} \; \longrightarrow \; \det E(x_I) ~,
\end{equation}
hence (\ref{regularized_volume}) tends to the exact volume 
$V_{\mathcal{R}}$.
The corresponding regularized quantum operator $\hat V_{\epsilon}$,
which thus avoids the ill-defined local operator
$\sqrt{|\det{\hat{E}(x)}|}$, 
is simply obtained by replacing $T^{abc}$ with the
operator $\hat T^{abc}$ in the expressions above.  It can be shown
that it tends to a well-defined operator in the limit:
\begin{equation}
\lim_{\epsilon \to 0} \hat V_{\epsilon}= \hat V. 
\end{equation} 

The alternative regularization is obtained by replacing (\ref{Tabc})
with a color-$m$ loop variable $\overset{{\mbox{\it\tiny m}}}{T}{}^{abc}$, 
using holonomies $h^{(m)}$ and $su(2)$ generators in the 
corresponding matrix representation (which in turn defines the generalized 
trace),
\begin{eqnarray}
   \lefteqn{\overset{{\mbox{\it\tiny m}}}{T}{}^{abc}[\alpha](\sigma,\tau,\rho)}
   \hspace{5mm} \nonumber \\
   & & = - \mbox{Tr} \left(E^{ai}(\sigma) \tau_i^{(m)}\, 
     h^{(m)}_\alpha(\sigma,\tau) 
     \, E^{bj}(\tau) \tau_j^{(m)}\, h^{(m)}_\alpha(\tau,\rho) \, 
     E^{ck}(\rho) \tau_k^{(m)} h^{(m)}_\alpha(\rho,\sigma) \right) \, .
\end{eqnarray}
Expanding triads and holonomies, we obtain for a small regulator,
\begin{eqnarray}
  \overset{{\mbox{\it\tiny m}}}{T}{}^{abc}[\alpha](\sigma,\tau,\rho) =
    \frac{C(m)}{2} \,\epsilon^{abc} \det E(x_I) + \mathcal{O}(\epsilon) ~.
\end{eqnarray}
The required trace of a product of three generalized generators is
\begin{equation}
  \label{triple_tau_trace}
  \mbox{Tr} \left(\tau_i^{(m)} \tau_j^{(m)} \tau_k^{(m)} \right) =
       - \,\frac{C(m)}{2}\, \epsilon_{ijk} =
       - \,\frac{1}{24}\, m(m+1)(m+2)\, \epsilon_{ijk} ~.
\end{equation}
It is easy to see that 
\begin{equation}
  V^m_{\epsilon} = \sum_{I_\epsilon} \sqrt{|V^{2}_{(m) I_\epsilon}|} ~,
\end{equation} 
where 
\begin{equation}
  V^{2}_{(m) I_\epsilon} = \frac{1}{24\, C(m) \, \epsilon^6}
  \int_{\partial I_\epsilon} d\sigma^2 \int_{\partial I_\epsilon} d\tau^2 
  \int_{\partial I_\epsilon} d\rho^2 \,
  n_a(\sigma) n_b(\tau) n_c(\rho) \,
  \overset{{\mbox{\it\tiny m}}}{T}{}^{abc}[\alpha](\sigma,\tau,\rho) ~,
\end{equation}
tends to the exact volume $V_{\mathcal{R}}$ in the
limit of vanishing regulator $\epsilon$ as well. Replacing the 
corresponding quantum operators, we obtain the regularized operator 
$\hat V^m_{\epsilon}$. 

Now the key point is that, unlike to what happens with the HCO, it can 
be easily shown that 
\begin{equation}
  \label{vol_op_limit}
  \lim_{\epsilon \to 0} \hat V^m_{\epsilon} = \hat V ~.
\end{equation} 
In order to make the limiting procedure well-defined, an appropriate 
operator topology needs to be introduced, 
see for example \cite{DePietriRovelli96}.
Note that the r.h.s. of (\ref{vol_op_limit}) is independent of $m$.  
This is because the
operator adds links of color $m$, but these links are shrunk to zero
in the limit, and the recoupling algebra turns out to give precisely
the same factor as in the classical case.  Hence there is no quantization
ambiguity with respect to $SU(2)$ representations for the volume
operator. In a similar fashion, one can verify that the same is also
true for the area operator.


\section{The action of $\mathcal{H}^m$ on trivalent vertices}
\label{sec:action}

In the following we calculate explicitly the action of generalized
Euclidean Hamiltonian constraint operator on trivalent vertices.


\subsection{Remarks on the Computational Tools}

Calculations are performed in the spin network basis.  Graphical
techniques, namely Penrose' graphical binor calculus (roughly speaking
a diagrammatic way of performing $SU(2)$ tensor calculations), can be
introduced in the connection representation, for example to represent
spin network states or operators.  This method, in turn, is equivalent
to the graphical description in the loop representation
\cite{DePietri97}, which satisfies the basic axioms of the
tangle-theoretic formulation of Temperley-Lieb recoupling theory
\cite{KauffmanLins94}.  Calculations can thus be carried out nicely by
applying powerful graphical computational techniques on planar graphs
in a well-defined way.  We have listed the most relevant identities
for this article in appendix \ref{app:recoupl_theory}.  These methods
provide the basic computational tools for the following.

Consider a spin network state over an oriented colored graph $\gamma$. 
In the planar binor representation its planar projection is drawn over
a ribbon graph (or net).  This extended, i.e. thickened drawing of the
graph is itself an oriented two-dimensional surface with 
non-trivial topology.  Each (ribbon) edge represents an irreducible
$SU(2)$ tensor labelled by the color of the representation in which
the tensor lives.  Furthermore, to each, in general, $n$-valent vertex
is associated an intertwining tensor that is graphically represented
in terms of a virtual trivalent expansion.  The internal edges of this
expansion are denoted as `virtual' since they do not have any real
finite extension on the spatial hypersurface $\Sigma$, but rather
reflect the index pattern of combinations of Clebsch-Gordan
coefficients.  This virtual region is drawn as a `blowing up' of the
vertex to a dashed circle surrounding it.  Mathematically, the
expansion is justified by the Wigner-Eckart theorem which states that
an invariant tensorial intertwiner that represents the coupling of $n$
representations of a compact group, can be given in terms of
Clebsch-Gordan coefficients.  In terms of the diagrammatic language,
this is represented as a trivalent decomposition.  Hence any arbitrary
spin network state can be expanded in terms of (partially virtual)
trivalent spin network states, which thus form a basis.

The main advantage of the planar binor representation stems from some
fundamental relations that the graphically represented spin network
states satisfy.  Roughly speaking, they obey the formal identities
that define the Temperley-Lieb recoupling theory described in
\cite{KauffmanLins94}.  This allows the application of powerful
formulae along the edges of ribbon nets and within virtual vertices. 
These concepts will become more transparent in the next subsections,
where explicit calculations are being performed.  Note that
orientations of individual edges are not relevant in the planar binor
representation.  Nevertheless, the orientation of those colored edges
which are connected to a 3-vertex has to be fixed with respect to
each other, for example by assigning a cyclic order to
them\footnote{The reason for this is the twist property
(\ref{twist}).}.

Returning to the main subject, a further remark is advisable.  Note
that both, the generalized as well as Thiemann's original HCO, are
$SU(2)$ gauge-invariant, but are defined in terms of
non-gauge-invariant operators.  Hence non-gauge-invariant states
appear in the course of the calculations.  Although we haven't
mentioned these states yet, there are no inconsistencies. 
Gauge-invariant spin network states are straightforwardly generalized
to non-gauge-invariant \emph{extended spin network states}, see
\cite{Borissovetal97,AshtekarLewand97}.  Non-gauge-invariance simply
implies the existence of a free tensor index, which in the graphical
language is represented by an open (virtual) edge at a vertex, that is
not connected to any external line.


\subsection{Evaluating the action of $\hat{\mathcal{H}}^m$ 
        on a single 3-vertex}
We consider the Hamiltonian constraint operator 
$\hat{\mathcal{H}}^m_{T}[N]$ in (\ref{Ham_operatorDef}). 
It acts independently on single vertices, hence it suffices to 
continue with one of its basic building blocks, namely the 
local operators $\hat{\mathcal{H}}_{\Delta}^{m}$ 
in (\ref{ordering1b}) or (\ref{ordering2b}), depending on the
chosen operator ordering. They act on 
single vertices $v$ of the graph $\gamma$ underlying the state which 
is acted upon. We consider here the first ordering choice, i.e.
\begin{equation}
  \label{H_m_Delta} 
  \hat{\mathcal{H}}^m_{\Delta (1)} \, \ket{v} \equiv
  \hat{\mathcal{H}}^m_{\Delta} \, \ket{v} = \frac{2 i}{3 l_0^2 C(m)} \, 
  \epsilon^{ijk} \, \mbox{Tr} \left(
  \frac{\hat{h}^{(m)}[\alpha_{ij}] - \hat{h}^{(m)}[\alpha_{ji}]}{2} \,
  \hat{h}^{(m)}[s_{k}]\, \hat{V} \,  
  \hat{h}^{(m)}[s^{-1}_{k}] \right) \ket{v} ~,
\end{equation}
Recall that $\alpha_{ji} = \alpha^{-1}_{ij}$ and 
$\hat{h}[s_k^{-1}] = \hat{h}^{-1}[s_k]$ for any segment $s_k$. 
We will be investigating its action 
on a single trivalent vertex for an arbitrary but fixed color $m$.

\begin{figure}[t]
  \begin{equation}
  \Bigg|\!\! \begin{array}{c}\mbox{\epsfig{file=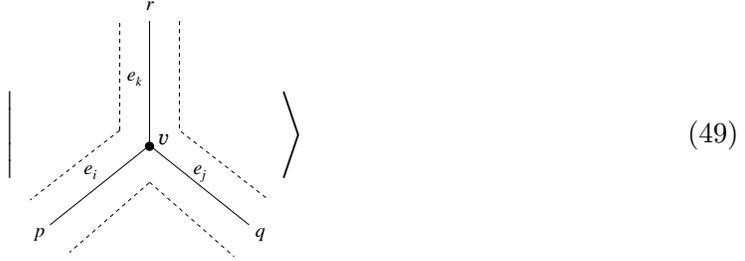,width=3.5cm}}
    \end{array} \!\!\!\Bigg\rangle \end{equation}
  \begin{center} \small
    \parbox{9cm}{\caption[]{\label{3vertexstate} \small
        The graphical representation of the part of the 
        spin network state containing the vertex
        $\ket{v(p,q,r)}$. Only the region around the vertex, 
        i.e. its adjacent edges, are shown.}}
  \end{center}
\end{figure}

Note that the form of (\ref{H_m_Delta}) is almost the same as in 
\cite{Borissovetal97}. The difference in the generalized expression 
is just the additional index $m$. The trivalent vertex is denoted by 
$\ket{v(p,q,r)} \equiv \ket{v}$, whereas $p,q$ and $r$ are the
colors of the adjacent edges $e_i,\, e_i,\, e_k$, see Fig. 
\ref{3vertexstate}.

We proceed by applying the operators emerging in (\ref{H_m_Delta})
successively, performing the summation over $i,j,k$ in the end.
The operator $\hat{h}^{(m)}[s^{-1}_{k}]$, which is itself not
gauge-invariant, corresponds to the holonomy along a segment 
$s_k$ with reversed orientation\footnote{We mentioned earlier 
that the orientation of edges is irrelevant. However, this is 
only true in the gauge-invariant case. 
Here we have to pay attention, since two open endings 
can only be combined when their orientations match.}. 
It attaches an open color-$m$ loop segment to the 
edge $e_k$, creating a new vertex on it, and 
altering the color between the two vertices, i.e.
\begin{eqnarray}
  \label{holonomy_action1} 
  \hat{h}^{(m)}[s^{-1}_{k}]~ \Bigg| \!\!
    \begin{array}{c}\mbox{\epsfig{file=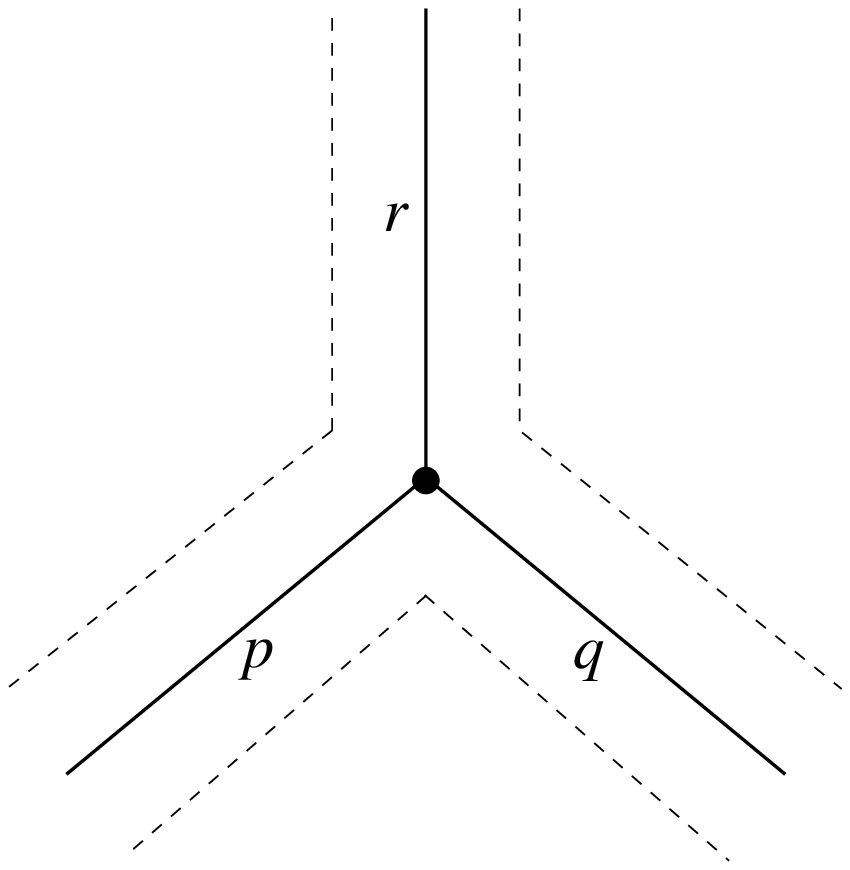,width=3.5cm}}
    \end{array} \!\!\Bigg\rangle 
      ~= &\Bigg| \! 
        \begin{array}{c}\mbox{\epsfig{file=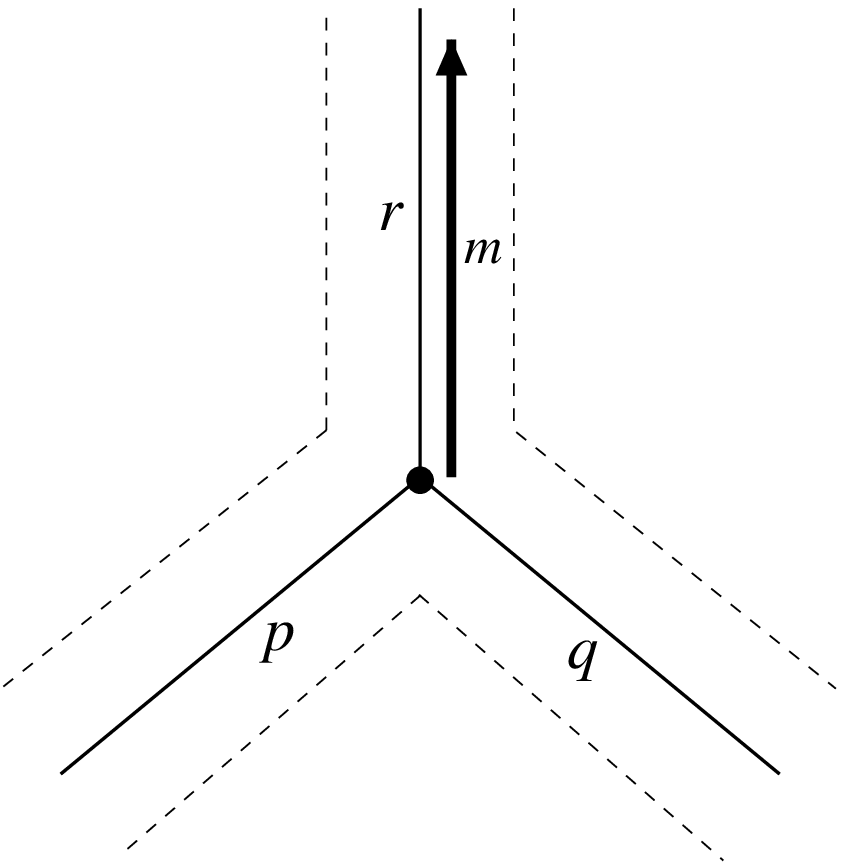,width=3.5cm}}
        \end{array} \!\!\Bigg\rangle& \\
      \label{holonomy_action2}  
      = \;\;\sum_c \,
        \frac{\begin{array}{c}\setlength{\unitlength}{.5 pt}
          \begin{picture}(40,35)
            \put(16,19){$\scriptstyle c$} 
            \put(17,15){\oval(34,30)[l]}        
            \put(23,15){\oval(34,30)[r]} 
            \put(17,-6){\line(0,1){12}}\put(23,-6){\line(0,1){12}}      
            \put(17,-6){\line(1,0){6}}\put(17,6){\line(1,0){6}}
            \put(17,30){\line(1,0){6}}
          \end{picture}\end{array}} 
        {\begin{array}{c}\setlength{\unitlength}{.5 pt}
          \begin{picture}(40,42)
            \put(17,33){$\scriptstyle r$}
            \put(13,17){$\scriptstyle m$} 
            \put(16, 3){$\scriptstyle c$} 
            \put(20,15){\oval(40,30)} \put( 0,15){\line(1,0){40}} 
            \put( 0,15){\circle*{3}}  \put(40,15){\circle*{3}}
          \end{picture}
        \end{array}} &\Bigg| \!\!
    \begin{array}{c}\mbox{\epsfig{file=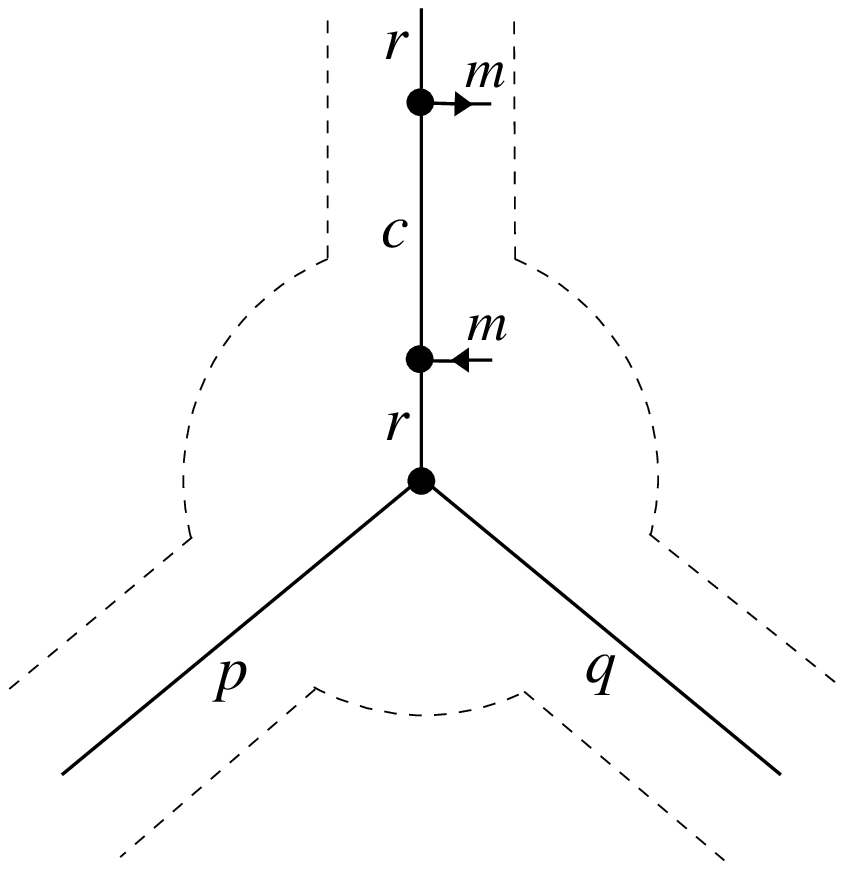,width=3.5cm}} 
    \end{array} \!\!\Bigg\rangle& ~.
\end{eqnarray} 
This follows since the segment $s_k$, whose one end lies in the 
original vertex, is entirely contained in $e_k$. The 
corresponding color-$r$ holonomy in the spin network state 
is tensorized with the color-$m$ parallel propagator that is assigned
to $s_k$. The resulting state is in general not irreducible. Decomposing 
it along the edge is graphically performed by using the edge addition 
formula (\ref{edge_addition}), as shown in (\ref{holonomy_action2}).
Consequently, a free index (i.e. an unconnected edge) in the color-$m$ 
representation is now located at the vertex, making it 
non-gauge-invariant, while 
another vertex is created on the edge $e_k$. The admissibility 
or Clebsch-Gordan conditions determine the color range of the
segment between the two vertices to $c=|r-m|,\,|r-m|+2, \ldots, (r+m)$.
Furthermore, the decomposition also fixes the intertwiner
by means of the virtual edge of color $r$.


\subsubsection{The action of $\hat{V}$}
\label{subsubsec:volumeaction}
In the next step, the volume operator acts on the non-gauge-invariant
state $\hat{h}^{(m)}[s^{-1}_{k}]\, \ket{v}$. Its
action has been calculated in 
\cite{DePietriRovelli96} and applied to Thiemann's Hamiltonian constraint 
operator in \cite{Borissovetal97}. We summarize for completeness the
basic facts about the calculation of its matrix elements,
and slightly extend and apply them to the generalized case.

The kinematical Hilbert space $\mathsf{H}$ has a basis of spin 
network states\footnote{In the diffeomorphism invariant context, which is 
the proper realm of the Hamiltonian constraint operator, the Hilbert
space has a natural basis labelled by so-called \emph{s-knots}.
These are equivalence classes of spin networks under diffeomorphisms.}.
Consider the finite-dimensional subspace $\mathfrak{h} \subset \mathsf{H}$
of states which are based 
on a fixed graph and a fixed coloring of the real edges, but 
arbitrary `virtual' edges, i.e. intertwiners. 
Since $\hat{V}$ modifies neither the 
graph nor the edge colorings, but only the intertwiners,
$\mathfrak{h}$ is an invariant subspace of $\mathsf{H}$ under 
the action of the volume operator. The volume operator has
to be diagonalized in this $D$-di\-men\-si\-onal space. How can $D$ be
determined? Let $d_i$ be the number of 
compatible colorings of a virtual trivalent decomposition of the 
vertex $i$. In other words, $d_i$ is the dimension of the 
intertwiner space (which is of 
course invariant under the specific virtual recoupling scheme, 
or basis chosen for the decomposition).
For each vertex, its valence and the coloring of the external edges 
determine $d_i$\footnote{Consider two simple examples for 
a 4-valent vertex. It we choose the colors of the adjacent edges 
to be $(2,3,4,5)$, the number of admissible colorings of 
the virtual edge is equal to $d=3$. For the assignment 
$(5,5,5,5)$ the subspace has dimension $d=6$. In both cases, $d$ 
is independent of the recoupling scheme.}.
Hence the (finite) dimension $D$ of the invariant 
subspace $\mathfrak{h}$ is given by the product of the dimensions of 
the intertwiner spaces of the vertices,
$D \equiv \mbox{dim}\, \mathfrak{h} = \prod_i d_i$. 

In the generalization of the HCO we adopt the 
same (unmodified!) volume operator that is used in 
the original $m=1$ construction --- the Ashtekar-Lewandowski one. 
Summarized in terms of its action on a cylindrical function 
$\psi_\gamma$, it is given by
\begin{equation}
  \label{volume}
  \hat{V} \, \psi_\gamma = \sum_{v \in \mathcal{V}(\gamma)}
        \hat{V}_v \, \psi_\gamma ~,
\end{equation}
where 
\begin{equation}
  \label{V_v}
  \hat{V}_v = l^3_0\, \sqrt{\left| \frac{i}{16\cdot3!} 
        \sum_{e_{I}\cap e_{J} \cap e_{K}=v} 
        \epsilon(e_{I},e_{J},e_{K})\, \hat{W}_{[IJK]} 
        \right|} ~.
\end{equation}
The first sum extends over the set $\mathcal{V}(\gamma)$ of vertices
of the underlying graph, while the sum in (\ref{V_v}) extends
over all triples $(e_I, e_J, e_K)$ of edges adjacent to
a vertex. The orientation factor 
$\epsilon(e_{I},e_{J},e_{K})$ is
$+1$ if the tangents $(\dot{e}_{I},\dot{e}_{J},\dot{e}_{K})$ 
at the vertex are positively oriented, $-1$ for negative 
orientation, and $0$ in the case of degenerate, i.e. linearly 
dependent or planar edges.
Besides, edges meeting in an $n$-vertex are assumed to be 
outgoing. 

\begin{figure}[t]
  \begin{displaymath}
    \hat{W}_{[abc]}\;
    \begin{array}{c}
        \epsfig{file=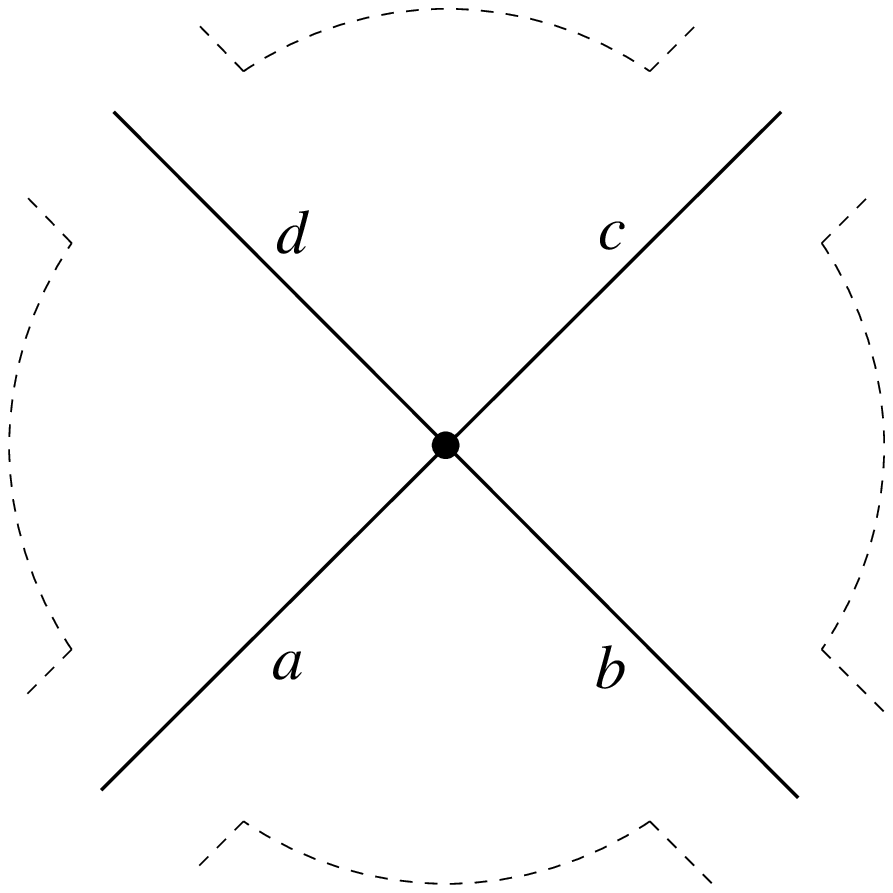,width=3.5cm}
    \end{array}
    =\, a b c 
    \begin{array}{c}
        \mbox{\epsfig{file=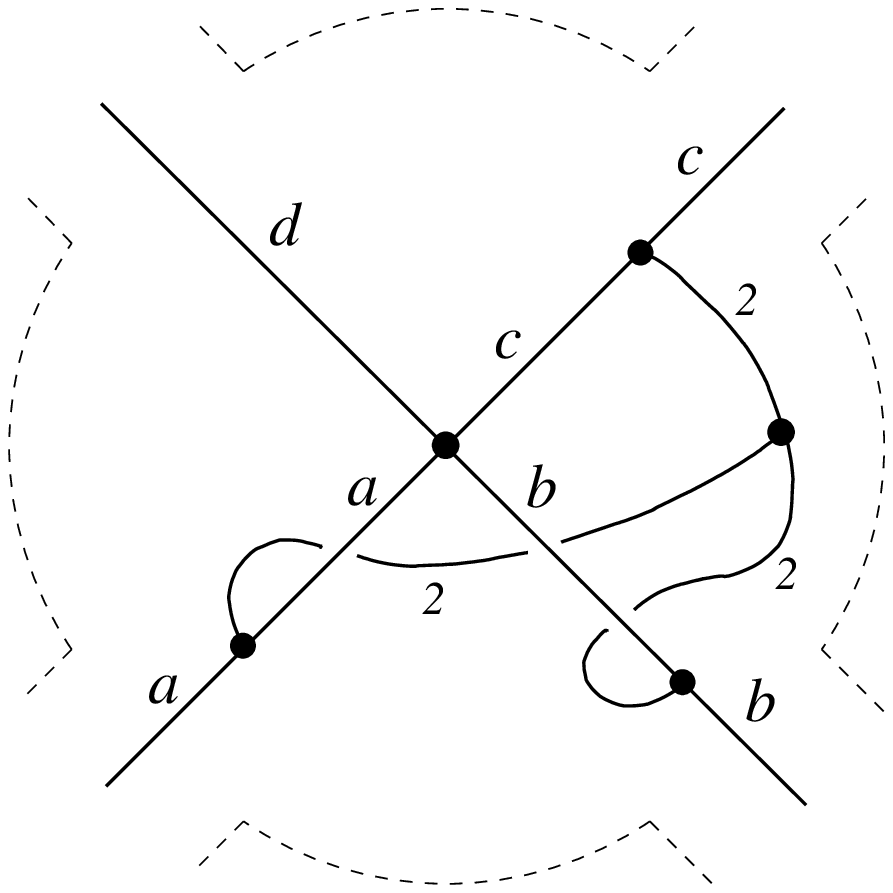,width=3.5cm}}
    \end{array} 
    \sim \, a b c \sum_i
    \begin{array}{c}
        \mbox{\epsfig{file=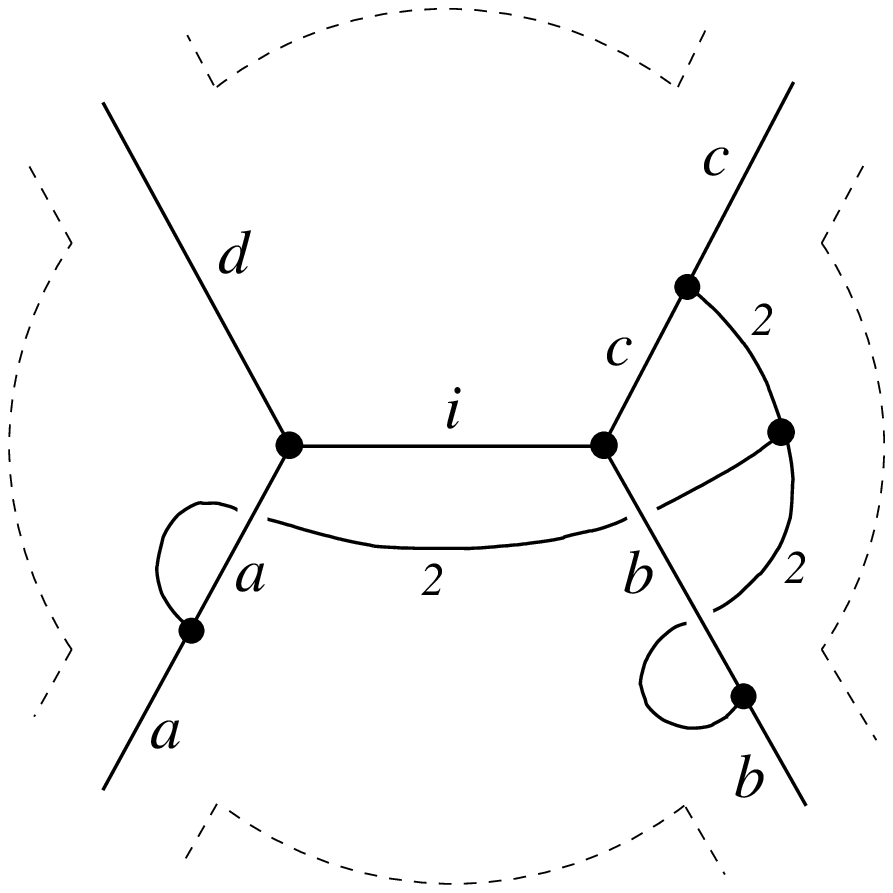,width=3.5cm}}
    \end{array} 
  \end{displaymath}
  \begin{center}
    \parbox{13cm}{\caption[]{\label{grasping} The vertex 
        operator $\hat{W}_{[abc]}$ `grasps' the indicated edges
        of a 4-valent vertex (the edges are labelled by
        their colors). To avoid sign confusion, the three graspings
        have to be performed `on the same side' of the involved edges.
        The last term illustrates a virtual trivalent decomposition 
        of the intertwiner.}}
  \end{center}
\end{figure}

The essential part of the operator (\ref{V_v}) is given by
$\hat{W}_{[IJK]}$ (the `square' of the volume)
that acts on the finite dimensional intertwiner space of 
an $n$-valent vertex $v_n$. Its action is described in terms 
of the `grasping' of any three distinct edges $e_I,\,e_J$ and 
$e_K$ adjacent to $v_n$. For the explicit definition of the 
grasping operation in the loop as well as the connection 
representation, see
\cite{AshtekarLewand98,RovelliSmolin95}. 
Graphically, the triple grasping has been
constructed in the planar binor representation \cite{DePietri97}, 
and is performed as follows, see Fig. \ref{grasping}. 
Three color-$2$ edges that intersect in a vertex $v'$, 
are being attached to three distinct adjacent edges $(e_I, e_J, e_K)$
of $v_n$, creating a single new vertex on each of these edges. Since
for every such triple of edges, $\hat{W}_{[IJK]}$ in (\ref{V_v})
affects only the intertwiner associated to $v_n$, its graphical
action is performed in the virtual dashed circles that represent 
the colored vertices. 
Restricting the action to \emph{real} edges only, 
the volume operator is equally well-defined on non-gauge-invariant 
vertices\footnote{This might of course also be seen
in the connection representation, where the volume operator is
defined in terms of left-invariant vector fields, hence over the
whole space of generalized connections.}.

Consider our specific case of a non-gauge-invariant 3-vertex. 
It turns out that gauge-invariant 4-valent vertices are of 
particular interest for this case.
They are used to mimic the volume's action on non-gauge-invariant 
$3$-valent vertices as follows. 
The volume operator `grasps' triples of real edges 
$(e_I, e_J, e_K)$ adjacent to a vertex. For the 
non-gauge-invariant 3-vertex this gives a single term.
Non-degenerate, gauge-invariant $n$-valent vertices contribute
one term for each triple. Thus the grasping of 
a $4$-vertex which is constructed from a non-gauge-invariant 
3-vertex by assuming that the virtual fourth 
edge be a real one, results in four terms.
It is immediately realized that the term, in which the originally 
virtual edge remains ungrasped is
identical to the single term that arises from the  
non-gauge-invariant 3-valent vertex. 
This is insofar important, since all computations 
concerning the volume, have as yet been performed in the 
gauge-invariant context only, e.g. in \cite{DePietriRovelli96}. 
But with the above identification these results can equally 
well be used for non-gauge-invariant cases.

Regarding the computations of the matrix elements 
for the generalized HCO, this applies as follows. The virtual
part of the vertex in (\ref{holonomy_action2}), i.e. the interior of the 
dashed circle, is a non-gauge-invariant 3-vertex on which the volume 
operator acts according to (\ref{H_m_Delta}). 
We obtain
\begin{eqnarray}
  \hat{V} \left( \hat{h}^{(m)}[s^{-1}_{k}] \, \ket{v} \right)
        &=& \sum_c \,  
        \frac{\begin{array}{c}\setlength{\unitlength}{.5 pt}
          \begin{picture}(40,35)
            \put(16,19){$\scriptstyle c$} 
            \put(17,15){\oval(34,30)[l]}        
            \put(23,15){\oval(34,30)[r]} 
            \put(17,-6){\line(0,1){12}}\put(23,-6){\line(0,1){12}}      
            \put(17,-6){\line(1,0){6}}\put(17,6){\line(1,0){6}}
            \put(17,30){\line(1,0){6}}
          \end{picture}\end{array}}
        {\begin{array}{c}\setlength{\unitlength}{.5 pt}
          \begin{picture}(40,42)
            \put(17,32.5){$\scriptstyle r$}
            \put(14,17){$\scriptstyle m$} 
            \put(16, 3){$\scriptstyle c$} 
            \put(20,15){\oval(40,30)} \put( 0,15){\line(1,0){40}} 
            \put( 0,15){\circle*{3}}  \put(40,15){\circle*{3}}
          \end{picture}
        \end{array}}
        \: \hat{V}_v \;\;\Bigg|\!\!
        \begin{array}{c}\setlength{\unitlength}{1.3 pt}
        \begin{picture}(50,40)
          \put(1,0){${\scriptstyle q}$}\put(1,34){${\scriptstyle p}$}
          \put(36,14){${\scriptstyle m}$}\put(46,34){${\scriptstyle c}$}
          \put(20,18){\line(-1,-1){15}}\put(20,18){\line(-1,1){15}}
          \put(30,18){\line(1,1){15}}\put(30,18){\line(1,-1){8}}
          \put(20,18){\line(1,0){10}}\put(24,20){${\scriptstyle r}$}
          \put(20,18){\circle*{3}}\put(30,18){\circle*{3}}
          \bezier{20}(5,18)( 5,33)(25,33)
          \bezier{20}(45,18)(45,33)(25,33)
          \bezier{20}(5,18)(5,3)(25,3)
          \bezier{20}(45,18)(45,3)(25,3)
       \end{picture}
    \end{array} \Bigg\rangle \nonumber \\
        &=& \sum_c \, \frac{l_0^3}{4} \,  
        \frac{\begin{array}{c}\setlength{\unitlength}{.5 pt}
          \begin{picture}(40,35)
            \put(16,19){$\scriptstyle c$} 
            \put(17,15){\oval(34,30)[l]}        
            \put(23,15){\oval(34,30)[r]} 
            \put(17,-6){\line(0,1){12}}\put(23,-6){\line(0,1){12}}      
            \put(17,-6){\line(1,0){6}}\put(17,6){\line(1,0){6}}
            \put(17,30){\line(1,0){6}}
          \end{picture}\end{array}}
        {\begin{array}{c}\setlength{\unitlength}{.5 pt}
          \begin{picture}(40,42)
            \put(17,32.5){$\scriptstyle r$}
            \put(14,17){$\scriptstyle m$} 
            \put(16, 3){$\scriptstyle c$} 
            \put(20,15){\oval(40,30)} \put( 0,15){\line(1,0){40}} 
            \put( 0,15){\circle*{3}}  \put(40,15){\circle*{3}}
          \end{picture}
        \end{array}} 
           \, \sqrt{\left| i \hat{W}_{[pqc]} \right|}
           \;\;\Bigg|\!\!
     \begin{array}{c}\setlength{\unitlength}{1.3 pt}
        \begin{picture}(50,40)
          \put(1,0){${\scriptstyle q}$}\put(1,34){${\scriptstyle p}$}
          \put(36,14){${\scriptstyle m}$}\put(46,34){${\scriptstyle c}$}
          \put(20,18){\line(-1,-1){15}}\put(20,18){\line(-1,1){15}}
          \put(30,18){\line(1,1){15}}\put(30,18){\line(1,-1){8}}
          \put(20,18){\line(1,0){10}}\put(24,20){${\scriptstyle r}$}
          \put(20,18){\circle*{3}}\put(30,18){\circle*{3}}
          \bezier{20}(5,18)( 5,33)(25,33)
          \bezier{20}(45,18)(45,33)(25,33)
          \bezier{20}(5,18)(5,3)(25,3)
          \bezier{20}(45,18)(45,3)(25,3)
       \end{picture} 
    \end{array} \Bigg\rangle ~.
    \label{vol_on_ngi_3vertex}
\end{eqnarray}
This graphical representation of the vertex is nothing but an 
enlargement of the circle in (\ref{holonomy_action2}), 
rotated around $90^\circ$.
To arrive at (\ref{vol_on_ngi_3vertex}) we took advantage 
of the linearity of the volume operator and the fact that in our 
case both sums in (\ref{volume}) and (\ref{V_v}) reduce to a
single term. The operator $\hat{W}_{[pqc]}$ 
denotes the grasping of the three real edges of the 
non-gauge-invariant 3-vertex, colored $p,\,q$ and $c$ in 
this order. The $3!$ factor in (\ref{V_v}) is canceled out by 
those terms that appear due to permutations of the three grasped 
edges, since they are all equal up a sign.
Recall that admissibility of the vertex restricts the range 
of $c$ to $|r-m|,\,|r-m|+2, \ldots, (r+m)$.

In order to make sense of the absolute value and the square root
in (\ref{vol_on_ngi_3vertex}), we consider first
the `square of the volume', i.e. the action of $\hat{W}$ on the 
vertex for a fixed color $c$.
The action of $\hat{W}$ on a non-gauge-invariant 3-vertex can generally 
be expressed as
\begin{equation}
  \label{grapheqn_for_W}
  \hat{W}_{[pqc]} 
    \;\;\Bigg|\!\!
     \begin{array}{c}\setlength{\unitlength}{1.5 pt}
        \begin{picture}(50,40)
          \put(1,0){${\scriptstyle q}$}\put(1,34){${\scriptstyle p}$}
          \put(36,14){${\scriptstyle m}$}\put(46,34){${\scriptstyle c}$}
          \put(20,18){\line(-1,-1){15}}\put(20,18){\line(-1,1){15}}
          \put(30,18){\line(1,1){15}}\put(30,18){\line(1,-1){8}}
          \put(20,18){\line(1,0){10}}\put(24,21){${\scriptstyle \alpha}$}
          \put(20,18){\circle*{3}}\put(30,18){\circle*{3}}
          \bezier{20}(5,18)( 5,33)(25,33)
          \bezier{20}(45,18)(45,33)(25,33)
          \bezier{20}(5,18)(5,3)(25,3)
          \bezier{20}(45,18)(45,3)(25,3)
       \end{picture}
    \end{array} \Bigg\rangle 
  = \sum_\beta W^{(4)}_{[pqc]}(p,q,m,c){}_\alpha{}^\beta
    \;\;\Bigg|\!\!
    \begin{array}{c}\setlength{\unitlength}{1.5 pt}
       \begin{picture}(50,40)
          \put(1,0){${\scriptstyle q}$}\put(1,34){${\scriptstyle p}$}
          \put(36,14){${\scriptstyle m}$}\put(46,34){${\scriptstyle c}$}
          \put(20,18){\line(-1,-1){15}}\put(20,18){\line(-1,1){15}}
          \put(30,18){\line(1,1){15}}\put(30,18){\line(1,-1){8}}
          \put(20,18){\line(1,0){10}}\put(24,21){${\scriptstyle \beta}$}
          \put(20,18){\circle*{3}}\put(30,18){\circle*{3}}
          \bezier{20}(5,18)( 5,33)(25,33)
          \bezier{20}(45,18)(45,33)(25,33)
          \bezier{20}(5,18)(5,3)(25,3)
          \bezier{20}(45,18)(45,3)(25,3)
       \end{picture}
    \end{array} \Bigg\rangle ~,
\end{equation}
or using a self-explanatory shorthand notation for the state vectors, as
\begin{equation}
  \label{eqn_for_W}
  \hat{W}_{[pqc]} \; \ket{v_\alpha}
   = \sum_\beta W^{(4)}_{[pqc]}(p,q,m,c){}_\alpha{}^\beta \; 
        \ket{v_\beta} ~.
\end{equation}
Here $\alpha$ ($\beta$) is determined by admissibility of the triples 
$\{p, q, \alpha\}$ and $\{c, m, \alpha\}$ ($\{p, q, \beta\}$ and 
$\{c, m, \beta\}$).
The number of these triple determines the dimension of the
intertwiner space on which $\hat{W}_{[pqc]}$ acts.
The matrix $W^{(4)}_{[pqc]}$ is defined in the context of
gauge-invariant 4-valent vertices. It represents the matrix 
elements of the operator $\hat{W}_{[pqc]}$ that acts on the triple of 
edges colored $(p,q,c)$, in a basis of 4-valent vertices.
It has been calculated for the first time in
\cite{DePietriRovelli96}, where the general case concerning the
volume operator acting on $n$-valent vertices is considered.
It has also been shown that in an appropriate basis the operators 
$i \hat{W}$ are represented by antisymmetric, purely 
imaginary, i.e. hermitian matrices, which are diagonalizable 
and have real eigenvalues. Hence the absolute value and the 
square root in (\ref{vol_on_ngi_3vertex}) are well-defined. 

This basis is realized by a 
rescaling, or vertex normalization respectively.
The virtual internal edge is multiplied by $\sqrt{\Delta}$,
and each of the two virtual nodes is divided by an appropriate 
$\sqrt{\Theta}$, giving
\begin{equation}
     \label{vertex_norm}
     \Bigg| \!\!\!\!
     \begin{array}{c}\setlength{\unitlength}{1.0 pt}
        \begin{picture}(50,40)
          \put(5,0){${\scriptstyle q}$}\put(5,33){${\scriptstyle p}$}
          \put(40,0){${\scriptstyle m}$}\put(43,32){${\scriptstyle c}$}
          \put(20,18){\line(-1,-1){12}}\put(20,18){\line(-1,1){12}}
          \put(30,18){\line(1,1){12}}\put(30,18){\line(1,-1){12}}
          \put(20,18){\line(1,0){10}}\put(22.5,20){${\scriptstyle \alpha}$}
          \put(20,18){\circle*{3}}\put(30,18){\circle*{3}}
       \end{picture} 
    \end{array} \!\!\!\Bigg\rangle_{\! N} = 
    \sqrt{\frac{\begin{array}{c}\setlength{\unitlength}{.5 pt}
          \begin{picture}(40,32)
            \put(15,19){$\scriptstyle \alpha$} 
            \put(17,15){\oval(34,30)[l]}        
            \put(23,15){\oval(34,30)[r]} 
            \put(17,-6){\line(0,1){12}}\put(23,-6){\line(0,1){12}}      
            \put(17,-6){\line(1,0){6}}\put(17,6){\line(1,0){6}}
            \put(17,30){\line(1,0){6}}
          \end{picture}\end{array}}
        {\begin{array}{c}\setlength{\unitlength}{.5 pt}
          \begin{picture}(40,44)
            \put(17,35){$\scriptstyle p$}
            \put(16,18){$\scriptstyle \alpha$} 
            \put(17, 5){$\scriptstyle q$} 
            \put(20,15){\oval(40,30)} \put( 0,15){\line(1,0){40}} 
            \put( 0,15){\circle*{3}}  \put(40,15){\circle*{3}}
          \end{picture} ~~
          \begin{picture}(40,44)
            \put(17,32.5){$\scriptstyle c$}
            \put(15,18){$\scriptstyle \alpha$} 
            \put(13, 2){$\scriptstyle m$} 
            \put(20,15){\oval(40,30)} \put( 0,15){\line(1,0){40}} 
            \put( 0,15){\circle*{3}}  \put(40,15){\circle*{3}}
          \end{picture}
        \end{array}}}
     \;\;\; \Bigg| \!\!\!\!
    \begin{array}{c}\setlength{\unitlength}{1.0 pt}
        \begin{picture}(50,40)
          \put(5,0){${\scriptstyle q}$}\put(5,33){${\scriptstyle p}$}
          \put(40,0){${\scriptstyle m}$}\put(43,32){${\scriptstyle c}$}
          \put(20,18){\line(-1,-1){12}}\put(20,18){\line(-1,1){12}}
          \put(30,18){\line(1,1){12}}\put(30,18){\line(1,-1){12}}
          \put(20,18){\line(1,0){10}}\put(22.5,20){${\scriptstyle \alpha}$}
          \put(20,18){\circle*{3}}\put(30,18){\circle*{3}}
       \end{picture}  
    \end{array} \!\!\! \Bigg\rangle ~.
\end{equation}
With this normalization\footnote{It is worth noticing that the recoupling 
theorem (\ref{recoupl_thm}), which relates the two possible distinct 
bases in the virtual decomposition of the 4-vertex, can be considered as a 
unitary transformation in the rescaled basis.}, (\ref{eqn_for_W}) is 
rewritten as
\begin{eqnarray}
  \label{norm_eqn_for_W:1}
  \hat{W}_{[pqc]} \; \ket{v_\alpha}_N
   &=& \sum_\beta \sqrt{\frac{\begin{array}{c}\setlength{\unitlength}{.5 pt}
          \begin{picture}(40,44)
            \put(15,19){$\scriptstyle \alpha$} 
            \put(17,15){\oval(34,30)[l]}        
            \put(23,15){\oval(34,30)[r]} 
            \put(17,-6){\line(0,1){12}}\put(23,-6){\line(0,1){12}}      
            \put(17,-6){\line(1,0){6}}\put(17,6){\line(1,0){6}}
            \put(17,30){\line(1,0){6}}
          \end{picture} ~~
          \begin{picture}(40,44)
            \put(17,35){$\scriptstyle p$}
            \put(16,17){$\scriptstyle \beta$} 
            \put(17, 5){$\scriptstyle q$} 
            \put(20,15){\oval(40,30)} \put( 0,15){\line(1,0){40}} 
            \put( 0,15){\circle*{3}}  \put(40,15){\circle*{3}}
          \end{picture} ~~
          \begin{picture}(40,44)
            \put(17,33){$\scriptstyle c$}
            \put(15,17){$\scriptstyle \beta$} 
            \put(13, 2){$\scriptstyle m$} 
            \put(20,15){\oval(40,30)} \put( 0,15){\line(1,0){40}} 
            \put( 0,15){\circle*{3}}  \put(40,15){\circle*{3}}
          \end{picture}
        \end{array}}
        {\begin{array}{c}\setlength{\unitlength}{.5 pt}
          \begin{picture}(40,44)
            \put(16,17){$\scriptstyle \beta$} 
            \put(17,15){\oval(34,30)[l]}        
            \put(23,15){\oval(34,30)[r]} 
            \put(17,-6){\line(0,1){12}}\put(23,-6){\line(0,1){12}}      
            \put(17,-6){\line(1,0){6}}\put(17,6){\line(1,0){6}}
            \put(17,30){\line(1,0){6}}
          \end{picture} ~~
          \begin{picture}(40,44)
            \put(17,35){$\scriptstyle p$}
            \put(16,18){$\scriptstyle \alpha$} 
            \put(17, 5){$\scriptstyle q$} 
            \put(20,15){\oval(40,30)} \put( 0,15){\line(1,0){40}} 
            \put( 0,15){\circle*{3}}  \put(40,15){\circle*{3}}
          \end{picture} ~~
          \begin{picture}(40,44)
            \put(17,33){$\scriptstyle c$}
            \put(15,18){$\scriptstyle \alpha$} 
            \put(13, 2){$\scriptstyle m$} 
            \put(20,15){\oval(40,30)} \put( 0,15){\line(1,0){40}} 
            \put( 0,15){\circle*{3}}  \put(40,15){\circle*{3}}
          \end{picture}
        \end{array}}}
        \cdot W^{(4)}_{[pqc]}(p,q,m,c){}_\alpha{}^\beta \;\, 
        \ket{v_\beta}_N  \\
     &=& \sum_\beta \tilde{W}^{(4)}_{[pqc]}(p,q,m,c){}_\alpha{}^\beta \; 
        \ket{v_\beta}_N ~,
        \label{norm_eqn_for_W:2}
\end{eqnarray}
where the matrix elements of $\hat{W}_{[pqc]}$ between two normalized 
state vectors are denoted by $\tilde{W}^{(4)}_{[pqc]}{}_\alpha{}^\beta$.
Evaluating (\ref{norm_eqn_for_W:2}) by using the grasping 
operation according to Figure \ref{grasping}, and 
closing the open network with itself, one obtains
\begin{equation}
  \label{W_tilde_matrix}
  \tilde{W}^{(4)}_{[pqc]}(p,q,m,c){}_\alpha{}^\beta ~=~ p q c \,
  \sqrt{\frac{\begin{array}{c}\setlength{\unitlength}{.5 pt}
          \begin{picture}(40,32)
            \put(15,18){$\scriptstyle \alpha$} 
            \put(17,15){\oval(34,30)[l]}        
            \put(23,15){\oval(34,30)[r]} 
            \put(17,-6){\line(0,1){12}}\put(23,-6){\line(0,1){12}}      
            \put(17,-6){\line(1,0){6}}\put(17,6){\line(1,0){6}}
            \put(17,30){\line(1,0){6}}
          \end{picture} ~~
          \begin{picture}(40,32)
            \put(16,16){$\scriptstyle \beta$} 
            \put(17,15){\oval(34,30)[l]}        
            \put(23,15){\oval(34,30)[r]} 
            \put(17,-6){\line(0,1){12}}\put(23,-6){\line(0,1){12}}      
            \put(17,-6){\line(1,0){6}}\put(17,6){\line(1,0){6}}
            \put(17,30){\line(1,0){6}}
          \end{picture}
         \end{array}}
         {\begin{array}{c}\setlength{\unitlength}{.5 pt}
          \begin{picture}(40,44)
            \put(17,35){$\scriptstyle p$}
            \put(15,18){$\scriptstyle \alpha$} 
            \put(17, 5){$\scriptstyle q$} 
            \put(20,15){\oval(40,30)} \put( 0,15){\line(1,0){40}} 
            \put( 0,15){\circle*{3}}  \put(40,15){\circle*{3}}
          \end{picture} ~~
          \begin{picture}(40,44)
            \put(17,35){$\scriptstyle p$}
            \put(16,17){$\scriptstyle \beta$} 
            \put(17, 5){$\scriptstyle q$} 
            \put(20,15){\oval(40,30)} \put( 0,15){\line(1,0){40}} 
            \put( 0,15){\circle*{3}}  \put(40,15){\circle*{3}}
          \end{picture} ~~
          \begin{picture}(40,44)
            \put(17,34){$\scriptstyle c$}
            \put(15,18){$\scriptstyle \alpha$} 
            \put(13, 2){$\scriptstyle m$} 
            \put(20,15){\oval(40,30)} \put( 0,15){\line(1,0){40}} 
            \put( 0,15){\circle*{3}}  \put(40,15){\circle*{3}}
          \end{picture} ~~
          \begin{picture}(40,44)
            \put(17,34){$\scriptstyle c$}
            \put(15,17){$\scriptstyle \beta$} 
            \put(13, 2){$\scriptstyle m$} 
            \put(20,15){\oval(40,30)} \put( 0,15){\line(1,0){40}} 
            \put( 0,15){\circle*{3}}  \put(40,15){\circle*{3}}
          \end{picture}
        \end{array}}}
  \begin{array}{c}\setlength{\unitlength}{1 pt}
     \begin{picture}(110,60)
        \put(15,40){\line(0,1){10}}  \put(18,41){${\scriptstyle q}$}
        \put(15,32){\oval(20,20)[tl]} 
        \put(15,42){\circle*{3}}      
        \put(40,30){\line(0,1){30}}  \put(44,41){${\scriptstyle p}$}
        \put(40,32){\oval(20,20)[tl]} 
        \put(40,42){\circle*{3}}     \put(40,30){\circle*{3}}
        \put(40,60){\circle*{3}}
        \put(65,30){\line(0,1){30}}  \put(69,41){${\scriptstyle c}$}
        \put(65,32){\oval(20,20)[tl]} 
        \put(65,42){\circle*{3}}     \put(65,30){\circle*{3}}
        \put(65,60){\circle*{3}}
        \put(90,40){\line(0,1){10}}  \put(92,41){${\scriptstyle m}$}
        \put( 5,32){\line(0,-1){12}} \put(3,6){${\scriptstyle 2}$}
        \put(30,28){\line(0,-1){18}} \put(24,17){${\scriptstyle 2}$}
        \put(55,28){\line(0,-1){ 8}} \put(53,6){${\scriptstyle 2}$}
        \put(30,20){\oval(50,20)[b]} \put(30,10){\circle*{3}}
        \put(40,40){\oval(50,20)[bl]}\put(65,40){\oval(50,20)[br]}
        \put(40,30){\line(1,0){25}}  \put(46,24){${\scriptstyle \alpha}$} 
        \put(40,50){\oval(50,20)[tl]}\put(65,50){\oval(50,20)[tr]}
        \put(40,60){\line(1,0){25}}  \put(50,53){${\scriptstyle \beta}$}
     \end{picture}
  \end{array}.
\end{equation}
The closed network in this expression is simplified by applying the 
reduction formula 
(\ref{3vertex_reduction}) to the upper right three triangle-like 
vertices $\left\{ (\beta,c,m), (c,2,c), (\alpha,c,m) \right\}$,
reducing it to a $9j-$symbol. Most 
importantly, an interesting structure of the antisymmetric matrix 
$\tilde{W}^{(4)}{}_\alpha{}^\beta$ is revealed. 
Non-zero elements appear only in those entries that are subject 
to \mbox{$|\alpha-\beta| = 2$}. 
This follows from admissibility of the three edges adjacent to
the vertex that appears after performing the just mentioned reduction, 
and the antisymmetry of the matrix. Hence $\tilde{W}^{(4)}{}_\alpha{}^\beta$ 
has only sub- and superdiagonal non-zero entries! 
Evaluating (\ref{W_tilde_matrix}) further until only fundamental, or 
`minimal' closed networks remain, gives
\begin{eqnarray}
  \label{W_tilde_matrix_final}
  \lefteqn{\tilde{W}^{(4)}_{[pqc]}(p,q,m,c){}_\alpha{}^\beta ~=~ 
  \sqrt{\frac{\begin{array}{c}\setlength{\unitlength}{.5 pt}
          \begin{picture}(40,32)
            \put(15,18){$\scriptstyle \alpha$} 
            \put(17,15){\oval(34,30)[l]}        
            \put(23,15){\oval(34,30)[r]} 
            \put(17,-6){\line(0,1){12}}\put(23,-6){\line(0,1){12}}      
            \put(17,-6){\line(1,0){6}}\put(17,6){\line(1,0){6}}
            \put(17,30){\line(1,0){6}}
          \end{picture} ~~
          \begin{picture}(40,32)
            \put(16,16){$\scriptstyle \beta$} 
            \put(17,15){\oval(34,30)[l]}        
            \put(23,15){\oval(34,30)[r]} 
            \put(17,-6){\line(0,1){12}}\put(23,-6){\line(0,1){12}}      
            \put(17,-6){\line(1,0){6}}\put(17,6){\line(1,0){6}}
            \put(17,30){\line(1,0){6}}
          \end{picture}
         \end{array}}
         {\begin{array}{c}\setlength{\unitlength}{.5 pt}
          \begin{picture}(40,44)
            \put(17,35){$\scriptstyle p$}
            \put(15,18){$\scriptstyle \alpha$} 
            \put(17, 5){$\scriptstyle q$} 
            \put(20,15){\oval(40,30)} \put( 0,15){\line(1,0){40}} 
            \put( 0,15){\circle*{3}}  \put(40,15){\circle*{3}}
          \end{picture} ~~
          \begin{picture}(40,44)
            \put(17,35){$\scriptstyle p$}
            \put(16,17){$\scriptstyle \beta$} 
            \put(17, 5){$\scriptstyle q$} 
            \put(20,15){\oval(40,30)} \put( 0,15){\line(1,0){40}} 
            \put( 0,15){\circle*{3}}  \put(40,15){\circle*{3}}
          \end{picture} ~~
          \begin{picture}(40,44)
            \put(17,33){$\scriptstyle c$}
            \put(15,18){$\scriptstyle \alpha$} 
            \put(13, 2){$\scriptstyle m$} 
            \put(20,15){\oval(40,30)} \put( 0,15){\line(1,0){40}} 
            \put( 0,15){\circle*{3}}  \put(40,15){\circle*{3}}
          \end{picture} ~~
          \begin{picture}(40,44)
            \put(17,33){$\scriptstyle c$}
            \put(15,17){$\scriptstyle \beta$} 
            \put(13, 2){$\scriptstyle m$} 
            \put(20,15){\oval(40,30)} \put( 0,15){\line(1,0){40}} 
            \put( 0,15){\circle*{3}}  \put(40,15){\circle*{3}}
          \end{picture}
        \end{array}}}~~
     \begin{array}{c}\setlength{\unitlength}{.6 pt}
       \begin{picture}(40,40)
         \put(16.5,33){$\scriptstyle \alpha$}
         \put(17,17){$\scriptstyle 2$} 
         \put(16, 3){$\scriptstyle \beta$} 
         \put(20,15){\oval(40,30)} \put( 0,15){\line(1,0){40}} 
         \put( 0,15){\circle*{3}}  \put(40,15){\circle*{3}}
       \end{picture}
     \end{array} ~ \times} \hspace{1cm} \nonumber \\
  && \times~ \bigg[\, c~
   \frac{\begin{array}{c}\setlength{\unitlength}{.8 pt}
     \begin{picture}(45,40)
        \put( 0,15){\line(1,-1){15}} \put(-1,21){${\scriptstyle \beta}$}
        \put( 0,15){\line(1, 1){15}} \put(0,3){${\scriptstyle \alpha}$}
        \put( 0,15){\circle*{3}}
        \put(30,15){\line(-1, 1){15}} \put(26,21){${\scriptstyle c}$}
        \put(30,15){\line(-1,-1){15}} \put(26, 4){${\scriptstyle c}$}
        \put(30,15){\circle*{3}}
        \put( 0,15){\line(1,0){30}} \put(12,17){${\scriptstyle 2}$}
        \put(15,30){\line(1,0){25}} \put(15,30){\circle*{3}}
        \put(15, 0){\line(1,0){25}} \put(15, 0){\circle*{3}}
        \put(40, 0){\line(0,1){30}} \put(43,13){${\scriptstyle m}$}
     \end{picture}\end{array}}
  {\begin{array}{c}\setlength{\unitlength}{.5 pt}
     \begin{picture}(40,44)
        \put(16,33){$\scriptstyle \alpha$}
        \put(17,17){$\scriptstyle 2$} 
        \put(17, 2){$\scriptstyle \beta$} 
        \put(20,15){\oval(40,30)} \put( 0,15){\line(1,0){40}} 
        \put( 0,15){\circle*{3}}  \put(40,15){\circle*{3}}
     \end{picture}
   \end{array}} \bigg]
   ~ \bigg[\, p~
   \frac{\begin{array}{c}\setlength{\unitlength}{.8 pt}
     \begin{picture}(45,40)
        \put( 0,15){\line(1,-1){15}} \put(0,23){${\scriptstyle \alpha}$}
        \put( 0,15){\line(1, 1){15}} \put(-1,1.5){${\scriptstyle \beta}$}
        \put( 0,15){\circle*{3}}
        \put(30,15){\line(-1, 1){15}} \put(26,23){${\scriptstyle p}$}
        \put(30,15){\line(-1,-1){15}} \put(26, 4){${\scriptstyle p}$}
        \put(30,15){\circle*{3}}
        \put( 0,15){\line(1,0){30}} \put(12,17){${\scriptstyle 2}$}
        \put(15,30){\line(1,0){25}} \put(15,30){\circle*{3}}
        \put(15, 0){\line(1,0){25}} \put(15, 0){\circle*{3}}
        \put(40, 0){\line(0,1){30}} \put(43,13){${\scriptstyle q}$}
     \end{picture}\end{array}}
  {\begin{array}{c}\setlength{\unitlength}{.5 pt}
     \begin{picture}(40,44)
        \put(16,33){$\scriptstyle \alpha$}
        \put(17,17){$\scriptstyle 2$} 
        \put(17, 2){$\scriptstyle \beta$} 
        \put(20,15){\oval(40,30)} \put( 0,15){\line(1,0){40}} 
        \put( 0,15){\circle*{3}}  \put(40,15){\circle*{3}}
     \end{picture}
   \end{array}} \bigg]
   ~ \bigg[\, p~
   \frac{\begin{array}{c}\setlength{\unitlength}{.8 pt}
     \begin{picture}(45,40)
        \put( 0,15){\line(1,-1){15}} \put(0,22){${\scriptstyle p}$}
        \put( 0,15){\line(1, 1){15}} \put(0,4){${\scriptstyle p}$}
        \put( 0,15){\circle*{3}}
        \put(30,15){\line(-1, 1){15}} \put(26,21){${\scriptstyle 2}$}
        \put(30,15){\line(-1,-1){15}} \put(26, 2.5){${\scriptstyle 2}$}
        \put(30,15){\circle*{3}}
        \put( 0,15){\line(1,0){30}} \put(12,17){${\scriptstyle 2}$}
        \put(15,30){\line(1,0){25}} \put(15,30){\circle*{3}}
        \put(15, 0){\line(1,0){25}} \put(15, 0){\circle*{3}}
        \put(40, 0){\line(0,1){30}} \put(43,13){${\scriptstyle p}$}
     \end{picture}\end{array}}
  {\begin{array}{c}\setlength{\unitlength}{.5 pt}
     \begin{picture}(40,44)
        \put(17,35){$\scriptstyle p$}
        \put(17,17){$\scriptstyle 2$} 
        \put(17,5){$\scriptstyle p$} 
        \put(20,15){\oval(40,30)} \put( 0,15){\line(1,0){40}} 
        \put( 0,15){\circle*{3}}  \put(40,15){\circle*{3}}
     \end{picture}
   \end{array}} 
  ~ - ~b~
   \frac{\begin{array}{c}\setlength{\unitlength}{.8 pt}
     \begin{picture}(45,40)
        \put( 0,15){\line(1,-1){15}} \put(-1,21){${\scriptstyle \beta}$}
        \put( 0,15){\line(1, 1){15}} \put(0,3){${\scriptstyle \alpha}$}
        \put( 0,15){\circle*{3}}
        \put(30,15){\line(-1, 1){15}} \put(26,21){${\scriptstyle 2}$}
        \put(30,15){\line(-1,-1){15}} \put(26,2.5){${\scriptstyle 2}$}
        \put(30,15){\circle*{3}}
        \put( 0,15){\line(1,0){30}} \put(12,17){${\scriptstyle 2}$}
        \put(15,30){\line(1,0){25}} \put(15,30){\circle*{3}}
        \put(15, 0){\line(1,0){25}} \put(15, 0){\circle*{3}}
        \put(40, 0){\line(0,1){30}} \put(43,13){${\scriptstyle \beta}$}
     \end{picture}\end{array}}
  {\begin{array}{c}\setlength{\unitlength}{.5 pt}
     \begin{picture}(40,44)
        \put(16,33){$\scriptstyle \alpha$}
        \put(17,17){$\scriptstyle 2$} 
        \put(17, 2){$\scriptstyle \beta$} 
        \put(20,15){\oval(40,30)} \put( 0,15){\line(1,0){40}} 
        \put( 0,15){\circle*{3}}  \put(40,15){\circle*{3}}
     \end{picture}
   \end{array}} \bigg] ~.
\end{eqnarray}
A more useful algebraic expression which is appropriate 
for further computations can be derived by a complete chromatic 
evaluation using the formulae given in appendix \ref{app:recoupl_theory}.
For the sake of completeness, we state 
the result \cite{Borissovetal97}. Defining $t=(\beta+\alpha)/2$ and 
\mbox{$e=(\beta-\alpha)/2 = \pm 1$}, the non-zero matrix elements are
\begin{eqnarray}
  \label{W_tilde_explicit} 
  \lefteqn{\tilde{W}^{(4)}_{[pqc]}(p,q,m,c){}_{t-e}{}^{t+e}  
    = -e \,(-1)^{\frac{p+q+m+c}{2}} ~~ \times}
                \hspace{2cm} \nonumber\\
    && \times~ \bigg[\, \frac{1}{1024\, t(t+2)} \, 
          (p+q+t+3) (m+c+t+3) ~~ \times \nonumber \\ 
    && \times~~\, (1+p+q-t) (1+p+t-q) (1+q+t-p) ~~\times \nonumber \\
    && \times~~\, (1+m+c-t) (1+m+t-c) (1+c+t-m) \,\bigg]^{1/2}~~.
\end{eqnarray}
Thus we end up with the explicit matrix elements of the
real antisymmetric $\tilde{W}^{(4)}$ in the rescaled basis. Since
the required $i \tilde{W}^{(4)}$ is hermitian and 
obviously normal as well, it can be diagonalized by a 
unitary matrix $U$, leading to $i \tilde{W}_D := U i \tilde{W} U^{-1}$
which has real eigenvalues (in addition, we 
know that if $\lambda$ is an eigenvalue, $-\lambda$ is as well).
It is this diagonal form that allows to take the required absolute 
value and square root in (\ref{vol_on_ngi_3vertex}) in a well-defined way.
However, since two base transformations have been performed to
arrive at the diagonal form, we need to revert them to obtain
the explicit action of the volume operator. More explicitly, 
the first transformation has been the rescaling (\ref{vertex_norm}).
Denoting the normalization factor by $n(\alpha)$, the diagonal matrix 
that changes the basis in the space of 4-valent vertices is denoted by
$\Lambda_\alpha{}^\beta = n(\alpha)\, \delta_\alpha{}^\beta$, giving
$\ket{v_\alpha}_N = \Lambda_\alpha{}^\beta \,
\ket{v_\beta} = n(\alpha)\, \ket{v_\alpha}$. Secondly, the now 
normalized $i \tilde{W}$ is diagonalized with the above introduced 
unitary matrix $U$. 
Taking the absolute value and the square root, the double base
transformation needs to be reverted to return to the original basis
in which the whole calculation is performed.
In all, this is written as
\begin{equation}
   \sqrt{|i W|} = \Lambda^{-1} U^{-1} \sqrt{|i \tilde{W}_D}| \:
        U \Lambda ~,
\end{equation}
or explicitly in terms of the matrix elements,
\begin{equation}
   \label{sqrt_W_matrix}
   \sqrt{|i W|}\: {}_\alpha{}^\beta = \frac{n(\beta)}{n(\alpha)} \:
   U^{-1}{}_\alpha{}^\rho \,\sqrt{|i \tilde{W}_D|}\,{}_\rho{}^\sigma 
   \; U_\sigma{}^\beta ~.
\end{equation}
The action of the volume operator is in general 
\emph{not} diagonal!\footnote{One exception 
turns out to be given by Thiemann's original 
$m=1$ HCO. In this case, the action of the volume is indeed diagonal,
and $\tilde{W}^{(4)}$ is a $(2 \times 2)$ matrix, allowing 
explicit calculations \cite{Borissovetal97}, see also 
appendix \ref{app:ThiemannsHCO}.}
Unfortunately, the complexity of the problem for arbitrary $m$
and colorings of the vertex, keeps us from
calculating (\ref{sqrt_W_matrix}) explicitly. 
Nevertheless, we should mention that there is no difficulty on principle. 
As soon as specific colorings of a vertex are chosen, the complete 
calculation can be performed. It is just the general expression that is 
lacking.

In the following we return to the computation of the Euclidean 
HCO's action, using notation (\ref{sqrt_W_matrix}) for the 
matrix elements.
The relation between the vertex operator $\hat{V}_v$ and 
the square root of the local grasp $i \hat{W}$ reads
in the trivalent case
\begin{equation}
  \label{V_def}
  \sqrt{|i W|}\: {}_\alpha{}^\beta = (V_v){}_\alpha{}^\beta 
        \equiv V{}_\alpha{}^\beta ~.
\end{equation}
Inserting this in (\ref{vol_on_ngi_3vertex}), we obtain
for the non-diagonal action of the volume operator
\begin{eqnarray}
  \label{final_action_of_V}
  \hat{V} \left( \hat{h}^{(m)}[s^{-1}_{k}] \, \ket{v} \right)
        &=&  \frac{l_0^3}{4} \, \sum_c \, 
        \frac{\begin{array}{c}\setlength{\unitlength}{.5 pt}
          \begin{picture}(40,35)
            \put(16,19){$\scriptstyle c$} 
            \put(17,15){\oval(34,30)[l]}        
            \put(23,15){\oval(34,30)[r]} 
            \put(17,-6){\line(0,1){12}}\put(23,-6){\line(0,1){12}}      
            \put(17,-6){\line(1,0){6}}\put(17,6){\line(1,0){6}}
            \put(17,30){\line(1,0){6}}
          \end{picture}\end{array}}
        {\begin{array}{c}\setlength{\unitlength}{.5 pt}
          \begin{picture}(40,42)
            \put(17,32.5){$\scriptstyle r$}
            \put(14,17){$\scriptstyle m$} 
            \put(16,3){$\scriptstyle c$} 
            \put(20,15){\oval(40,30)} \put( 0,15){\line(1,0){40}} 
            \put( 0,15){\circle*{3}}  \put(40,15){\circle*{3}}
          \end{picture}
        \end{array}} 
           \, \sqrt{\left| i \hat{W}_{[pqc]} \right|}
           \;\;\Bigg|\!\!
     \begin{array}{c}\setlength{\unitlength}{1.3 pt}
        \begin{picture}(50,40)
          \put(1,0){${\scriptstyle q}$}\put(1,34){${\scriptstyle p}$}
          \put(36,14){${\scriptstyle m}$}\put(46,34){${\scriptstyle c}$}
          \put(20,18){\line(-1,-1){15}}\put(20,18){\line(-1,1){15}}
          \put(30,18){\line(1,1){15}}\put(30,18){\line(1,-1){8}}
          \put(20,18){\line(1,0){10}}\put(24,20){${\scriptstyle r}$}
          \put(20,18){\circle*{3}}\put(30,18){\circle*{3}}
          \bezier{20}(5,18)( 5,33)(25,33)
          \bezier{20}(45,18)(45,33)(25,33)
          \bezier{20}(5,18)(5,3)(25,3)
          \bezier{20}(45,18)(45,3)(25,3)
       \end{picture} 
    \end{array} \!\Bigg\rangle  \nonumber \\
        &=& \frac{l_0^3}{4} \, \sum_c \,
        \frac{\begin{array}{c}\setlength{\unitlength}{.5 pt}
          \begin{picture}(40,35)
            \put(16,19){$\scriptstyle c$} 
            \put(17,15){\oval(34,30)[l]}        
            \put(23,15){\oval(34,30)[r]} 
            \put(17,-6){\line(0,1){12}}\put(23,-6){\line(0,1){12}}      
            \put(17,-6){\line(1,0){6}}\put(17,6){\line(1,0){6}}
            \put(17,30){\line(1,0){6}}
          \end{picture}\end{array}}
        {\begin{array}{c}\setlength{\unitlength}{.5 pt}
          \begin{picture}(40,42)
            \put(17,32.5){$\scriptstyle r$}
            \put(14,17){$\scriptstyle m$} 
            \put(16,3){$\scriptstyle c$} 
            \put(20,15){\oval(40,30)} \put( 0,15){\line(1,0){40}} 
            \put( 0,15){\circle*{3}}  \put(40,15){\circle*{3}}
          \end{picture}
        \end{array}} 
     ~\sum_{\beta} V{}_r{}^\beta (p,q,m,c) \;\;\Bigg|\!\!
        \begin{array}{c}\setlength{\unitlength}{1.3 pt}
         \begin{picture}(50,40)
          \put(1,0){${\scriptstyle q}$}\put(1,34){${\scriptstyle p}$}
          \put(36,14){${\scriptstyle m}$}\put(46,34){${\scriptstyle c}$}
          \put(20,18){\line(-1,-1){15}}\put(20,18){\line(-1,1){15}}
          \put(30,18){\line(1,1){15}}\put(30,18){\line(1,-1){8}}
          \put(20,18){\line(1,0){10}}\put(24,21){${\scriptstyle \beta}$}
          \put(20,18){\circle*{3}}\put(30,18){\circle*{3}}
          \bezier{20}(5,18)( 5,33)(25,33)
          \bezier{20}(45,18)(45,33)(25,33)
          \bezier{20}(5,18)(5,3)(25,3)
          \bezier{20}(45,18)(45,3)(25,3)
         \end{picture} 
      \end{array} \!\Bigg\rangle \nonumber \\
      &=&  \frac{l_0^3}{4} \, \sum_{c,\beta} \,
        \frac{\begin{array}{c}\setlength{\unitlength}{.5 pt}
          \begin{picture}(40,35)
            \put(16,19){$\scriptstyle c$} 
            \put(17,15){\oval(34,30)[l]}        
            \put(23,15){\oval(34,30)[r]} 
            \put(17,-6){\line(0,1){12}}\put(23,-6){\line(0,1){12}}      
            \put(17,-6){\line(1,0){6}}\put(17,6){\line(1,0){6}}
            \put(17,30){\line(1,0){6}}
          \end{picture}\end{array}}
        {\begin{array}{c}\setlength{\unitlength}{.5 pt}
          \begin{picture}(40,42)
            \put(17,33){$\scriptstyle r$}
            \put(13,17){$\scriptstyle m$} 
            \put(16,3){$\scriptstyle c$} 
            \put(20,15){\oval(40,30)} \put( 0,15){\line(1,0){40}} 
            \put( 0,15){\circle*{3}}  \put(40,15){\circle*{3}}
          \end{picture}
        \end{array}} 
      \: V{}_r{}^\beta (p,q,m,c) ~~\Bigg| \!\!\!
     \begin{array}{c}\mbox{\epsfig{file=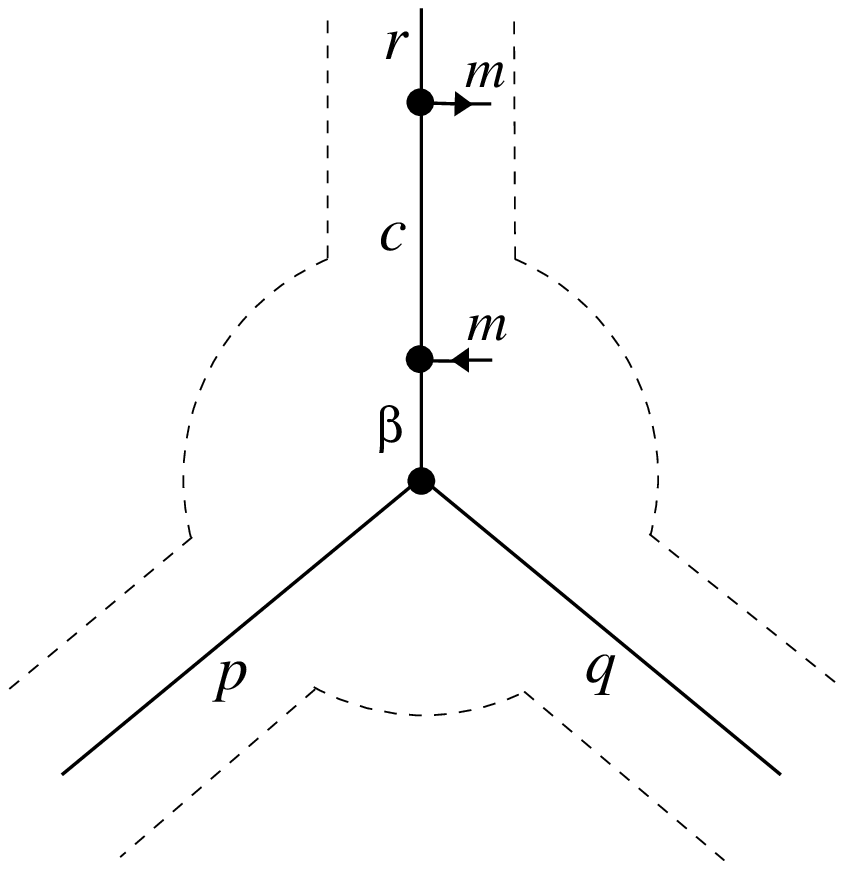,width=3.5cm}} 
    \end{array} \!\!\! \Bigg\rangle ~.
\end{eqnarray}
In the last step we returned to the same diagrammatic representation of 
the state as in the beginning of this section.


\subsubsection{Completing the action of $\hat{\mathcal{H}}^m$}
In this subsection we finish the computation of the action of
the generalized Euclidean HCO $\hat{\mathcal{H}}_{\Delta}^m$.
The relevant part of the operator (\ref{H_m_Delta}) is
\begin{equation}
  \label{hh_action}
  \mbox{Tr}\left(
  \frac{\hat{h}^{(m)}[\alpha_{ij}] - \hat{h}^{(m)}[\alpha_{ji}]}{2} \:
  \hat{h}^{(m)}[s_{k}]\, \hat{V} \, \hat{h}^{(m)}[s^{-1}_{k}] 
  \right) ~.
\end{equation}
The holonomies on the left hand side of the volume remain to be
evaluated acting on (\ref{final_action_of_V}).
Therefore, we consider first the action of
$\hat{h}^{(m)}[\alpha_{ij}]\, \hat{h}^{(m)}[s_{k}]$ on a 
gauge-invariant 3-vertex as in Fig. \ref{3vertexstate}. Similar  
to (\ref{holonomy_action1}), we obtain
\begin{equation}
    \hat{h}^{(m)}[\alpha_{ij}]\, \hat{h}^{(m)}[s_{k}]
    ~ \Bigg| \!\!
    \begin{array}{c}\mbox{\epsfig{file=v1.eps,width=3.5cm}}
    \end{array} \!\!\Bigg\rangle 
   ~=~  \Bigg| \!\!
    \begin{array}{c}\mbox{\epsfig{file=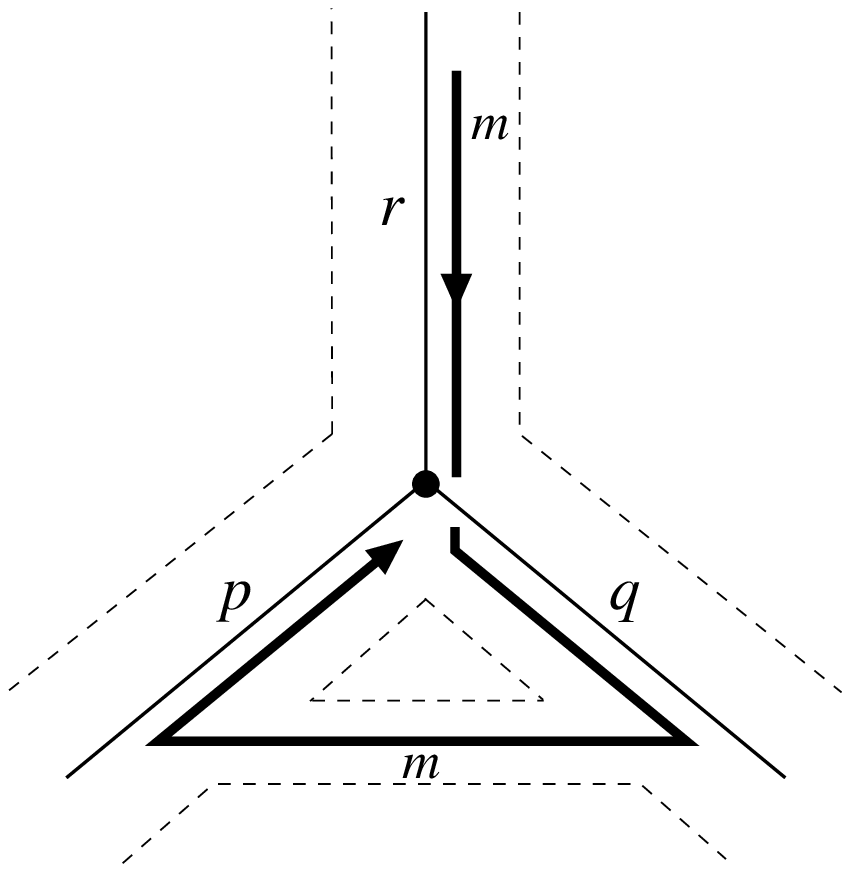,width=3.5cm}}
      \end{array}\!\!\Bigg\rangle  ~,
\end{equation}
where the lower triangular part of the added open loop corresponds
to $\alpha_{ij}$. The second operator  
$\hat{h}^{(m)}[\alpha_{ji}]\, \hat{h}^{(m)}[s_{k}]$ in (\ref{hh_action})
acts by reversing the orientation of the triangular loop.

These partial results are now applied to the relevant non-gauge-invariant 
3-vertex in (\ref{final_action_of_V}). One obtains two terms in which
the open loops are added by attaching their ends in such a way that 
compatibility of orientations is taken into account.
The trace in the color-$m$ representation ensures the connection and
summation of the free matrix indices\footnote{The matrix indices can 
be seen as sitting at the end of edges --- one index at each ending.
Connecting two lines in the graphical representation corresponds to
contracting these indices. Since only closed loops appear in the 
gauge-invariant context, all dummy indices are summed over. 
For this reason they are never explicitly shown.}. Our conventions
give an additional sign factor for tracing. It depends on the
color of the representation in which the indices of the added edges
live \cite{DePietriRovelli96}.
Performing this computation diagrammatically, one obtains

\begin{eqnarray*}
   \label{hh_on_ng3vert}
   \lefteqn{\frac{\hat{h}^{(m)}[\alpha_{ij}] -
        \hat{h}^{(m)}[\alpha_{ji}]}{2} \: \hat{h}^{(m)}[s_{k}] \!
   ~\;\Bigg|\!\!\!
   \begin{array}{c}\mbox{\epsfig{file=v3.eps,width=3.5cm}} 
   \end{array} \!\!\!\Bigg\rangle }  \hspace{2cm} \\
   && =~ \frac{(-1)^m}{2} ~\left[\rule{0cm}{1.3cm}\right. 
        ~\Bigg|\!\!\!
        \begin{array}{c}\mbox{\epsfig{file=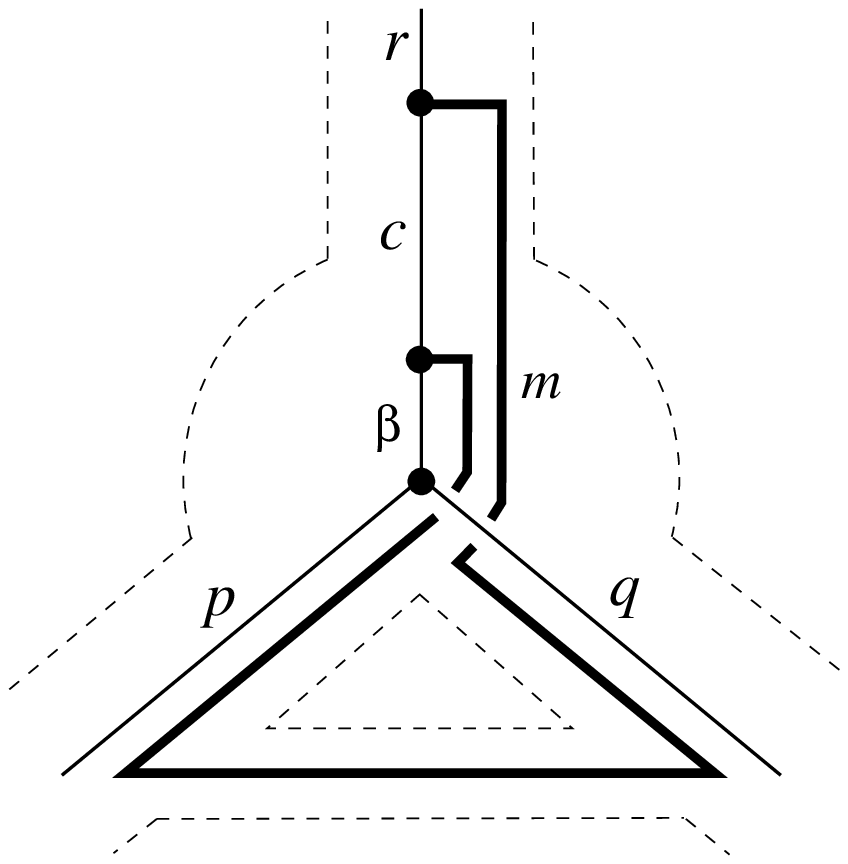,width=3.5cm}}
        \end{array} \!\!\!\Bigg\rangle
        ~-~ \Bigg|\!\!\! 
        \begin{array}{c}\mbox{\epsfig{file=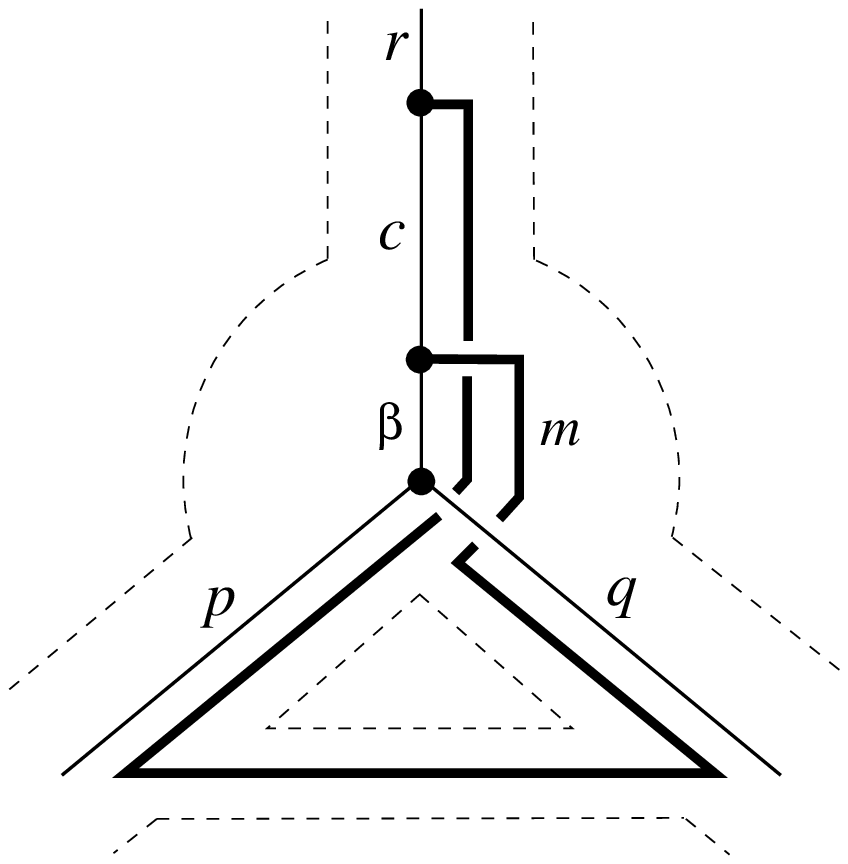,width=3.5cm}}
        \end{array} \!\!\!\Bigg\rangle \left.\rule{0cm}{1.3cm}\right]
         \hspace{1cm} \nonumber \\
\end{eqnarray*}
\newpage
\begin{eqnarray}
   && =~ \frac{(-1)^m}{2} ~\sum_{a,b} \,
        \frac{\begin{array}{c}\setlength{\unitlength}{.5 pt}
          \begin{picture}(40,35)
            \put(16,18){$\scriptstyle a$} 
            \put(17,15){\oval(34,30)[l]}        
            \put(23,15){\oval(34,30)[r]} 
            \put(17,-6){\line(0,1){12}}\put(23,-6){\line(0,1){12}}      
            \put(17,-6){\line(1,0){6}}\put(17,6){\line(1,0){6}}
            \put(17,30){\line(1,0){6}}
          \end{picture}~~
          \begin{picture}(40,35)
            \put(17,17){$\scriptstyle b$} 
            \put(17,15){\oval(34,30)[l]}        
            \put(23,15){\oval(34,30)[r]} 
            \put(17,-6){\line(0,1){12}}\put(23,-6){\line(0,1){12}}      
            \put(17,-6){\line(1,0){6}}\put(17,6){\line(1,0){6}}
            \put(17,30){\line(1,0){6}}
          \end{picture}
         \end{array}}
        {\begin{array}{c}\setlength{\unitlength}{.5 pt}
          \begin{picture}(40,44)
            \put(17,35){$\scriptstyle p$}
            \put(13,17){$\scriptstyle m$} 
            \put(16, 3){$\scriptstyle a$} 
            \put(20,15){\oval(40,30)} \put( 0,15){\line(1,0){40}} 
            \put( 0,15){\circle*{3}}  \put(40,15){\circle*{3}}
          \end{picture}~~
          \begin{picture}(40,44)
            \put(17,35){$\scriptstyle q$}
            \put(13,17){$\scriptstyle m$} 
            \put(17,2){$\scriptstyle b$} 
            \put(20,15){\oval(40,30)} \put( 0,15){\line(1,0){40}} 
            \put( 0,15){\circle*{3}}  \put(40,15){\circle*{3}}
          \end{picture}
        \end{array}} \: \lambda^{mp}_a \lambda^{mq}_b 
        ~ \times \nonumber \\
        && \hspace{1cm} \times \: \left[\rule{0cm}{1.3cm}\right.
        \: \lambda^{m \beta}_c ~~ \Bigg|\!\!\!
        \begin{array}{c}\mbox{\epsfig{file=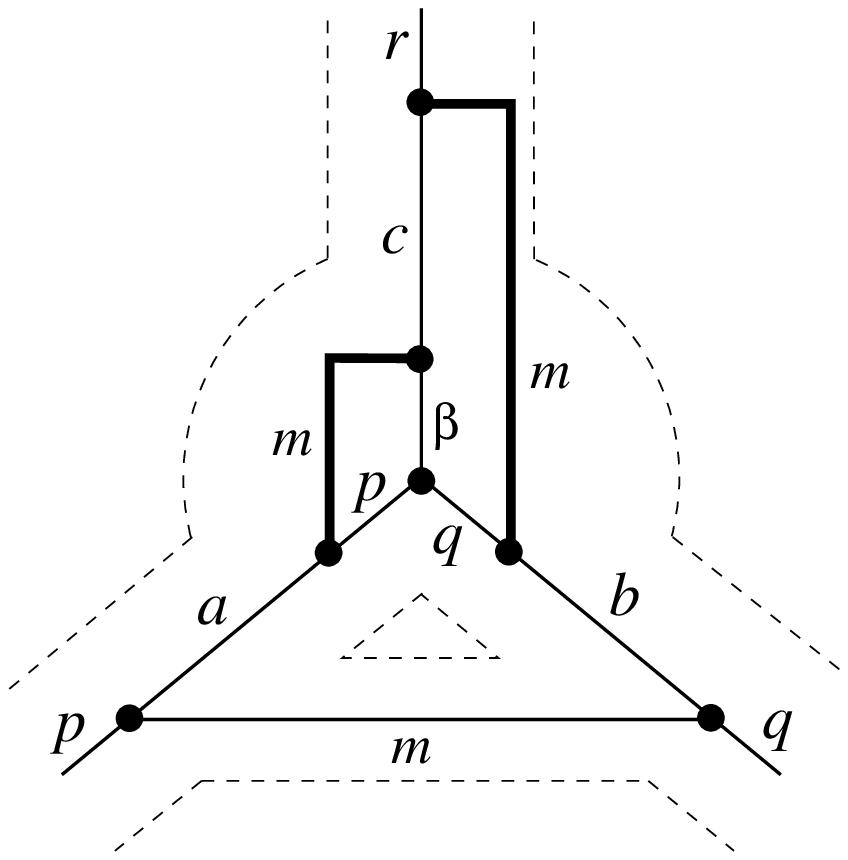,width=3.5cm}}
        \end{array} \!\!\!\Bigg\rangle
         ~-~ \lambda^{mr}_c ~~\Bigg|\!\!\!
          \begin{array}{c}\mbox{\epsfig{file=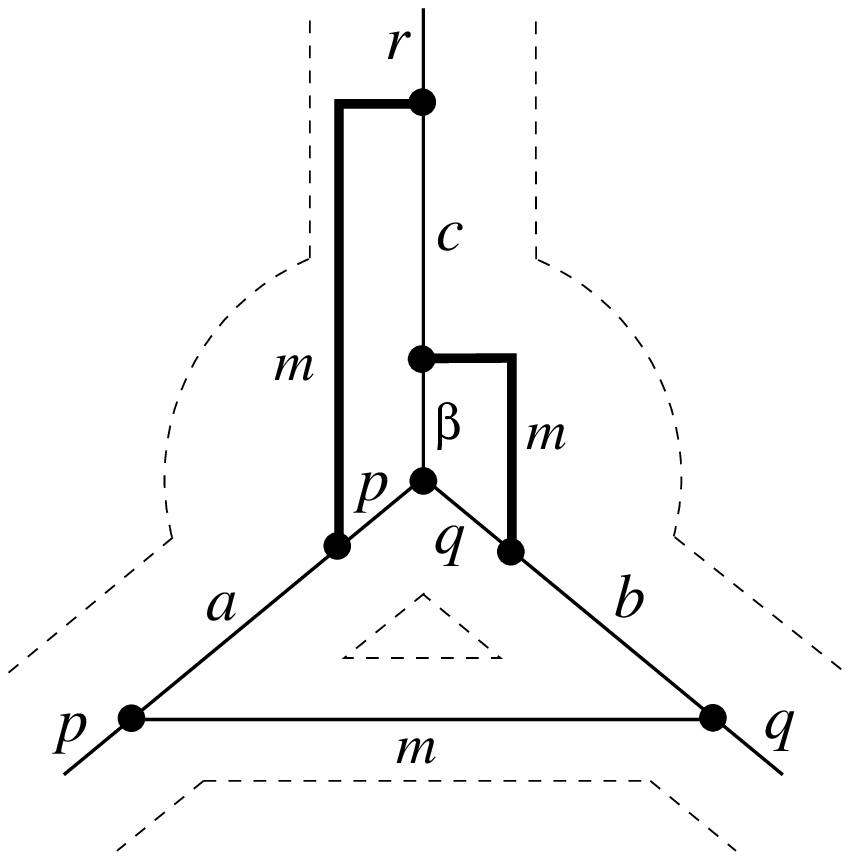,width=3.5cm}}
        \end{array}\!\!\!\Bigg\rangle  \left.\rule{0cm}{1.3cm}\right] ~.
        \label{final_eval_I}
\end{eqnarray}
The first equality stems from a direct application 
of the holonomy operators, while $(-1)^m$ is the mentioned 
tracing factor. Subsequently, the Clebsch-Gordan decomposition is
performed on the color-$p$ and $q$ edges via the edge addition 
formula (\ref{edge_addition}), and the twist property (\ref{twist}) 
is used three times in each of the terms. Admissibility restricts
the ranges of the segments between the newly added vertices
to $a \in \big\{ |p-m|, |p-m|+2, \ldots, p+m \big\}$ and \linebreak
$b \in \big\{ |q-m|, |q-m|+2, \ldots, q+m \big\}$.

The tangles in (\ref{final_eval_I}) can be further evaluated by 
applying the reduction formula (\ref{3vertex_reduction}) two times 
on each of the networks. Consider, for example the first term. 
The three vertices $(c,m,\beta)$, $(p,\beta,q)$ and $(m,p,a)$ are 
reduced to a single one, followed by reducing the remaining three
in the same way, i.e.

\begin{eqnarray}
   \Bigg|\!\!\!
   \begin{array}{c}\mbox{\epsfig{file=v7.eps,width=3.5cm}} 
     \end{array} \!\!\!\Bigg\rangle
    ~&=&~ 
   \frac{\begin{array}{c}\setlength{\unitlength}{.8 pt}
     \begin{picture}(45,40)
        \put( 0,15){\line(1,-1){15}} \put(0,22.5){${\scriptstyle q}$}
        \put( 0,15){\line(1, 1){15}} \put(0,4){${\scriptstyle a}$}
        \put( 0,15){\circle*{3}}
        \put(30,15){\line(-1, 1){15}} \put(26.5,21.5){${\scriptstyle \beta}$}
        \put(30,15){\line(-1,-1){15}} \put(25, 4){${\scriptstyle m}$}
        \put(30,15){\circle*{3}}
        \put( 0,15){\line(1,0){30}} \put(13,17){${\scriptstyle c}$}
        \put(15,30){\line(1,0){25}} \put(15,30){\circle*{3}}
        \put(15, 0){\line(1,0){25}} \put(15, 0){\circle*{3}}
        \put(40, 0){\line(0,1){30}} \put(43,13){${\scriptstyle p}$}
     \end{picture}\end{array}}
  {\begin{array}{c}\setlength{\unitlength}{.5 pt}
     \begin{picture}(40,44)
        \put(17,33){$\scriptstyle c$}
        \put(17,18){$\scriptstyle a$} 
        \put(17,5){$\scriptstyle q$} 
        \put(20,15){\oval(40,30)} \put( 0,15){\line(1,0){40}} 
        \put( 0,15){\circle*{3}}  \put(40,15){\circle*{3}}
     \end{picture}
   \end{array}}
   ~~\Bigg|\!\!\!
   \begin{array}{c}\mbox{\epsfig{file=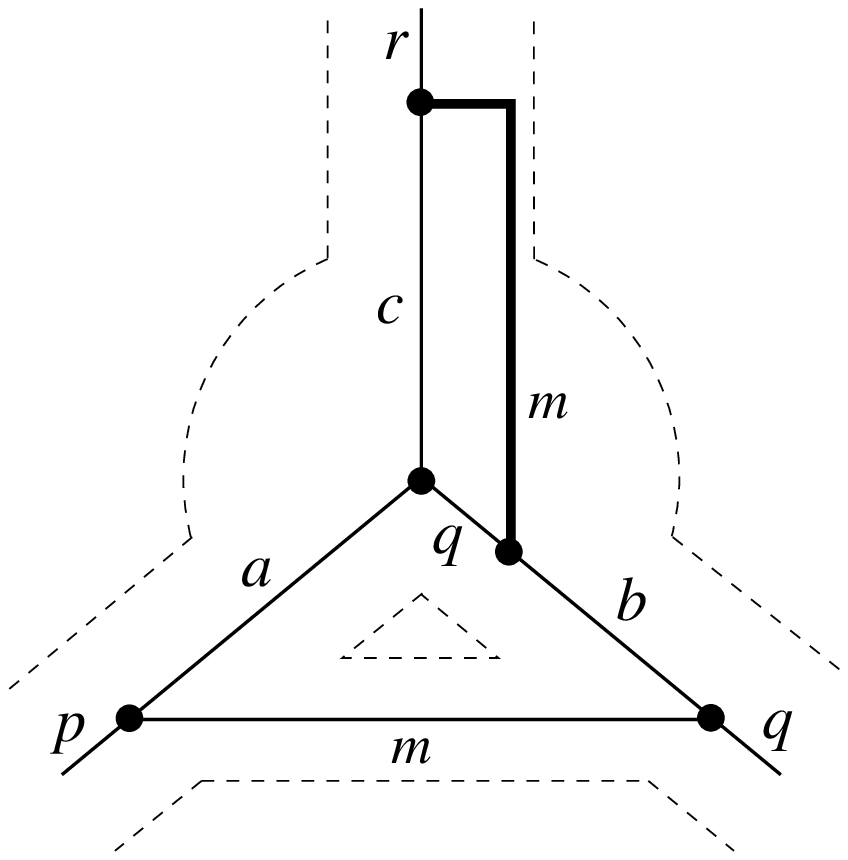,width=3.5cm}}
     \end{array} \!\!\!\Bigg\rangle \nonumber \\
    ~&=&~ 
   \frac{\begin{array}{c}\setlength{\unitlength}{.8 pt}
     \begin{picture}(45,40)
        \put( 0,15){\line(1,-1){15}} \put(-1,22.5){${\scriptstyle q}$}
        \put( 0,15){\line(1, 1){15}} \put(0,4){${\scriptstyle a}$}
        \put( 0,15){\circle*{3}}
        \put(30,15){\line(-1, 1){15}} \put(26.5,21.5){${\scriptstyle \beta}$}
        \put(30,15){\line(-1,-1){15}} \put(25, 4){${\scriptstyle m}$}
        \put(30,15){\circle*{3}}
        \put( 0,15){\line(1,0){30}} \put(13,17){${\scriptstyle c}$}
        \put(15,30){\line(1,0){25}} \put(15,30){\circle*{3}}
        \put(15, 0){\line(1,0){25}} \put(15, 0){\circle*{3}}
        \put(40, 0){\line(0,1){30}} \put(43,13){${\scriptstyle p}$}
     \end{picture} ~~~
     \begin{picture}(45,40)
        \put( 0,15){\line(1,-1){15}} \put(0,22){${\scriptstyle b}$}
        \put( 0,15){\line(1, 1){15}} \put(0,4){${\scriptstyle a}$}
        \put( 0,15){\circle*{3}}
        \put(30,15){\line(-1, 1){15}} \put(26,22){${\scriptstyle m}$}
        \put(30,15){\line(-1,-1){15}} \put(25, 4){${\scriptstyle c}$}
        \put(30,15){\circle*{3}}
        \put( 0,15){\line(1,0){30}} \put(13,17){${\scriptstyle r}$}
        \put(15,30){\line(1,0){25}} \put(15,30){\circle*{3}}
        \put(15, 0){\line(1,0){25}} \put(15, 0){\circle*{3}}
        \put(40, 0){\line(0,1){30}} \put(43,13){${\scriptstyle q}$}
     \end{picture}
   \end{array}}
  {\begin{array}{c}\setlength{\unitlength}{.5 pt}
     \begin{picture}(40,44)
        \put(18,32.5){$\scriptstyle c$}
        \put(17,17.5){$\scriptstyle a$} 
        \put(17,5){$\scriptstyle q$} 
        \put(20,15){\oval(40,30)} \put( 0,15){\line(1,0){40}} 
        \put( 0,15){\circle*{3}}  \put(40,15){\circle*{3}}
     \end{picture}~~
     \begin{picture}(40,44)
        \put(18,32.5){$\scriptstyle r$}
        \put(17,17.5){$\scriptstyle a$} 
        \put(17,2){$\scriptstyle b$} 
        \put(20,15){\oval(40,30)} \put( 0,15){\line(1,0){40}} 
        \put( 0,15){\circle*{3}}  \put(40,15){\circle*{3}}
     \end{picture}
   \end{array}}
   ~~\Bigg|\!\!\!
   \begin{array}{c}\mbox{\epsfig{file=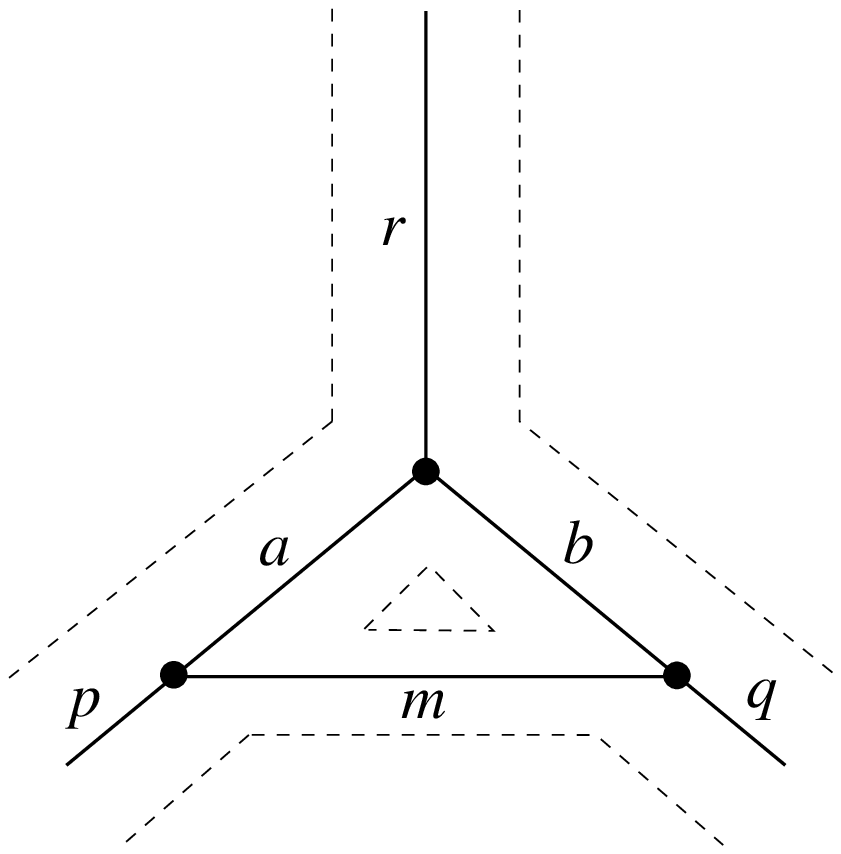,width=3.5cm}}
     \end{array} \!\!\!\Bigg\rangle ~~.
\end{eqnarray}
Analogously, the second term gives
\begin{equation}
   \Bigg|\!\!\!
   \begin{array}{c}\mbox{\epsfig{file=v8.eps,width=3.5cm}} 
     \end{array} \!\!\!\Bigg\rangle
    ~=~ 
   \frac{\begin{array}{c}\setlength{\unitlength}{.8 pt}
     \begin{picture}(45,40)
        \put( 0,15){\line(1,-1){15}} \put(0,22){${\scriptstyle b}$}
        \put( 0,15){\line(1, 1){15}} \put(0,3){${\scriptstyle p}$}
        \put( 0,15){\circle*{3}}
        \put(30,15){\line(-1, 1){15}} \put(25,22){${\scriptstyle m}$}
        \put(30,15){\line(-1,-1){15}} \put(26,4){${\scriptstyle \beta}$}
        \put(30,15){\circle*{3}}
        \put( 0,15){\line(1,0){30}} \put(13,17){${\scriptstyle c}$}
        \put(15,30){\line(1,0){25}} \put(15,30){\circle*{3}}
        \put(15, 0){\line(1,0){25}} \put(15, 0){\circle*{3}}
        \put(40, 0){\line(0,1){30}} \put(43,13){${\scriptstyle q}$}
     \end{picture} ~~~
     \begin{picture}(45,40)
        \put( 0,15){\line(1,-1){15}} \put(0,22){${\scriptstyle b}$}
        \put( 0,15){\line(1, 1){15}} \put(0,3){${\scriptstyle a}$}
        \put( 0,15){\circle*{3}}
        \put(30,15){\line(-1, 1){15}} \put(26,22){${\scriptstyle c}$}
        \put(30,15){\line(-1,-1){15}} \put(25,3){${\scriptstyle m}$}
        \put(30,15){\circle*{3}}
        \put( 0,15){\line(1,0){30}} \put(13,17){${\scriptstyle r}$}
        \put(15,30){\line(1,0){25}} \put(15,30){\circle*{3}}
        \put(15, 0){\line(1,0){25}} \put(15, 0){\circle*{3}}
        \put(40, 0){\line(0,1){30}} \put(43,13){${\scriptstyle p}$}
     \end{picture}
   \end{array}}
  {\begin{array}{c}\setlength{\unitlength}{.5 pt}
     \begin{picture}(40,44)
        \put(18,33){$\scriptstyle c$}
        \put(17,20){$\scriptstyle p$} 
        \put(17,2){$\scriptstyle b$} 
        \put(20,15){\oval(40,30)} \put( 0,15){\line(1,0){40}} 
        \put( 0,15){\circle*{3}}  \put(40,15){\circle*{3}}
     \end{picture}~~
     \begin{picture}(40,44)
        \put(18,33){$\scriptstyle r$}
        \put(16,18){$\scriptstyle a$} 
        \put(17,2){$\scriptstyle b$} 
        \put(20,15){\oval(40,30)} \put( 0,15){\line(1,0){40}} 
        \put( 0,15){\circle*{3}}  \put(40,15){\circle*{3}}
     \end{picture}
   \end{array}}
   ~~\Bigg|\!\!\!
   \begin{array}{c}\mbox{\epsfig{file=v10.eps,width=3.5cm}}
     \end{array}\!\!\!\Bigg\rangle ~~.
\end{equation}
Notice that there is no problem on the left hand side of the last 
equation with the uppermost vertex $(r,c,m)$ 
lying outside the dashed circle. It is situated in a (virtual) ribbon
edge, indicating that the color-$m$ line emanating from this vertex 
and the adjacent color-$c$ line lie in fact on top of each other. 
Therefore we can use recoupling theory and retrace it inside the original 
vertex within the dashed circle.

Thus we have finished calculating the trace part of the generalized 
HCO in (\ref{H_m_Delta}). To summarize, we obtain
\begin{eqnarray}
 \label{trace_part}
 \lefteqn{
  \mbox{Tr}\left(
  \frac{\hat{h}^{(m)}[\alpha_{ij}] - \hat{h}^{(m)}[\alpha_{ji}]}{2} \:
  \hat{h}^{(m)}[s_{k}]\, \hat{V} \, \hat{h}^{(m)}[s^{-1}_{k}] 
  \right) \, |v (p,q,r) \rangle} \hspace{2.5cm} \nonumber \\
  & &=~ (-1)^m \: \frac{l_0^3}{8} \, \sum_{a,b} 
        \, A^{(m)}(p,a|q,b|r\,\cdot) 
   ~\Bigg| \!\!\!
   \begin{array}{c}\mbox{\epsfig{file=v10.eps,width=3.5cm}}
     \end{array} \!\!\!\Bigg\rangle ~.
\end{eqnarray}
To simplify the notation, we have introduced the 
amplitude
\begin{eqnarray}
  \lefteqn{
  A^{(m)}(p,a|q,b|r\,\cdot) := \sum_c\; \lambda^{mp}_a 
        \lambda^{mq}_b \lambda^{mr}_c \:
     \frac{\begin{array}{c}\setlength{\unitlength}{.5 pt}
          \begin{picture}(40,32)
            \put(16,19){$\scriptstyle a$} 
            \put(17,15){\oval(34,30)[l]}        
            \put(23,15){\oval(34,30)[r]} 
            \put(17,-6){\line(0,1){12}}\put(23,-6){\line(0,1){12}}      
            \put(17,-6){\line(1,0){6}}\put(17,6){\line(1,0){6}}
            \put(17,30){\line(1,0){6}}
          \end{picture} ~~
          \begin{picture}(40,32)
            \put(16,17){$\scriptstyle b$} 
            \put(17,15){\oval(34,30)[l]}        
            \put(23,15){\oval(34,30)[r]} 
            \put(17,-6){\line(0,1){12}}\put(23,-6){\line(0,1){12}}      
            \put(17,-6){\line(1,0){6}}\put(17,6){\line(1,0){6}}
            \put(17,30){\line(1,0){6}}
          \end{picture} ~~
          \begin{picture}(40,32)
            \put(16,19){$\scriptstyle c$} 
            \put(17,15){\oval(34,30)[l]}        
            \put(23,15){\oval(34,30)[r]} 
            \put(17,-6){\line(0,1){12}}\put(23,-6){\line(0,1){12}}      
            \put(17,-6){\line(1,0){6}}\put(17,6){\line(1,0){6}}
            \put(17,30){\line(1,0){6}}
          \end{picture}
         \end{array}}
         {\begin{array}{c}\setlength{\unitlength}{.5 pt}
          \begin{picture}(40,44)
            \put(17,35){$\scriptstyle p$}
            \put(14,18){$\scriptstyle m$} 
            \put(17, 3){$\scriptstyle a$} 
            \put(20,15){\oval(40,30)} \put( 0,15){\line(1,0){40}} 
            \put( 0,15){\circle*{3}}  \put(40,15){\circle*{3}}
          \end{picture} ~~
          \begin{picture}(40,44)
            \put(17,35){$\scriptstyle q$}
            \put(14,18){$\scriptstyle m$} 
            \put(17, 2.5){$\scriptstyle b$} 
            \put(20,15){\oval(40,30)} \put( 0,15){\line(1,0){40}} 
            \put( 0,15){\circle*{3}}  \put(40,15){\circle*{3}}
          \end{picture} ~~
          \begin{picture}(40,44)
            \put(17.5,33){$\scriptstyle r$}
            \put(14,18){$\scriptstyle m$} 
            \put(17, 3){$\scriptstyle c$} 
            \put(20,15){\oval(40,30)} \put( 0,15){\line(1,0){40}} 
            \put( 0,15){\circle*{3}}  \put(40,15){\circle*{3}}
          \end{picture} 
        \end{array}} ~~\times} \nonumber \\
   &&\hspace{-0.5cm} \times \sum_{\beta(p,q,m,c)} 
     \!\!\!\! V{}_r{}^\beta (p,q,m,c) 
 \left[\rule{0cm}{1.2cm}\right.
   \lambda^{0 \beta}_r \;
   \frac{\begin{array}{c}\setlength{\unitlength}{.8 pt}
     \begin{picture}(45,40)
        \put( 0,15){\line(1,-1){15}} \put(-1,22.5){${\scriptstyle q}$}
        \put( 0,15){\line(1, 1){15}} \put(0,4){${\scriptstyle a}$}
        \put( 0,15){\circle*{3}}
        \put(30,15){\line(-1, 1){15}} \put(26.5,21.5){${\scriptstyle \beta}$}
        \put(30,15){\line(-1,-1){15}} \put(25, 4){${\scriptstyle m}$}
        \put(30,15){\circle*{3}}
        \put( 0,15){\line(1,0){30}} \put(13,17){${\scriptstyle c}$}
        \put(15,30){\line(1,0){25}} \put(15,30){\circle*{3}}
        \put(15, 0){\line(1,0){25}} \put(15, 0){\circle*{3}}
        \put(40, 0){\line(0,1){30}} \put(43,13){${\scriptstyle p}$}
     \end{picture} ~~~
     \begin{picture}(45,40)
        \put( 0,15){\line(1,-1){15}} \put(0,21){${\scriptstyle b}$}
        \put( 0,15){\line(1, 1){15}} \put(0,4){${\scriptstyle a}$}
        \put( 0,15){\circle*{3}}
        \put(30,15){\line(-1, 1){15}} \put(26,21){${\scriptstyle m}$}
        \put(30,15){\line(-1,-1){15}} \put(25, 4){${\scriptstyle c}$}
        \put(30,15){\circle*{3}}
        \put( 0,15){\line(1,0){30}} \put(13,17){${\scriptstyle r}$}
        \put(15,30){\line(1,0){25}} \put(15,30){\circle*{3}}
        \put(15, 0){\line(1,0){25}} \put(15, 0){\circle*{3}}
        \put(40, 0){\line(0,1){30}} \put(43,13){${\scriptstyle q}$}
     \end{picture}
   \end{array}}
  {\begin{array}{c}\setlength{\unitlength}{.5 pt}
     \begin{picture}(40,44)
        \put(18,35.5){$\scriptstyle q$}
        \put(17,17){$\scriptstyle c$} 
        \put(17,3){$\scriptstyle a$} 
        \put(20,15){\oval(40,30)} \put( 0,15){\line(1,0){40}} 
        \put( 0,15){\circle*{3}}  \put(40,15){\circle*{3}}
     \end{picture}~~
     \begin{picture}(40,44)
        \put(18,33){$\scriptstyle r$}
        \put(17,17){$\scriptstyle b$} 
        \put(17,3){$\scriptstyle a$} 
        \put(20,15){\oval(40,30)} \put( 0,15){\line(1,0){40}} 
        \put( 0,15){\circle*{3}}  \put(40,15){\circle*{3}}
     \end{picture}
   \end{array}}
  ~-~ 
   \frac{\begin{array}{c}\setlength{\unitlength}{.8 pt}
     \begin{picture}(45,40)
        \put( 0,15){\line(1,-1){15}} \put(0,21){${\scriptstyle b}$}
        \put( 0,15){\line(1, 1){15}} \put(-1,4){${\scriptstyle p}$}
        \put( 0,15){\circle*{3}}
        \put(30,15){\line(-1, 1){15}} \put(26,21){${\scriptstyle m}$}
        \put(30,15){\line(-1,-1){15}} \put(26,4){${\scriptstyle \beta}$}
        \put(30,15){\circle*{3}}
        \put( 0,15){\line(1,0){30}} \put(13,17){${\scriptstyle c}$}
        \put(15,30){\line(1,0){25}} \put(15,30){\circle*{3}}
        \put(15, 0){\line(1,0){25}} \put(15, 0){\circle*{3}}
        \put(40, 0){\line(0,1){30}} \put(43,13){${\scriptstyle q}$}
     \end{picture} ~~~
     \begin{picture}(45,40)
        \put( 0,15){\line(1,-1){15}} \put(0,22){${\scriptstyle b}$}
        \put( 0,15){\line(1, 1){15}} \put(0,3){${\scriptstyle a}$}
        \put( 0,15){\circle*{3}}
        \put(30,15){\line(-1, 1){15}} \put(26,22){${\scriptstyle c}$}
        \put(30,15){\line(-1,-1){15}} \put(25,3){${\scriptstyle m}$}
        \put(30,15){\circle*{3}}
        \put( 0,15){\line(1,0){30}} \put(13,17){${\scriptstyle r}$}
        \put(15,30){\line(1,0){25}} \put(15,30){\circle*{3}}
        \put(15, 0){\line(1,0){25}} \put(15, 0){\circle*{3}}
        \put(40, 0){\line(0,1){30}} \put(43,13){${\scriptstyle p}$}
     \end{picture}
   \end{array}}
  {\begin{array}{c}\setlength{\unitlength}{.5 pt}
     \begin{picture}(40,44)
        \put(18,35.5){$\scriptstyle p$}
        \put(17,17.5){$\scriptstyle c$} 
        \put(17,2){$\scriptstyle b$} 
        \put(20,15){\oval(40,30)} \put( 0,15){\line(1,0){40}} 
        \put( 0,15){\circle*{3}}  \put(40,15){\circle*{3}}
     \end{picture}~~
     \begin{picture}(40,44)
        \put(18,33){$\scriptstyle r$}
        \put(17,17.5){$\scriptstyle a$} 
        \put(17,2){$\scriptstyle b$} 
        \put(20,15){\oval(40,30)} \put( 0,15){\line(1,0){40}} 
        \put( 0,15){\circle*{3}}  \put(40,15){\circle*{3}}
     \end{picture}
   \end{array}} \left.\rule{0cm}{1.2cm}\right],
   \label{amplitude}
\end{eqnarray}
where $A^{(m)}(p,a|q,b|r\,\cdot)$ indicates the dependence on only five of
the six colors $\{p,a,q,b,r,c\}$ adjacent to the vertex. The remaining
color is 'internally' summed over, i.e. the corresponding state 
does not depend on it, see (\ref{trace_part}).
Note furthermore that the order of arguments in $A^{(m)}$ is relevant. 
The summation index $\beta = \beta(p,q,m,c)$ which appears due to 
the non-diagonal action of the volume operator, takes $d(q,p,c,m)$
different values which are determined by the simultaneous
admissibility of the triples $\{p,q,\beta \}$ and 
$\{m,c,\beta\}$. The set 
$(q,p,c,m)$ denotes the edge colorings that fix the 
dimension $d(q,p,c,m)$ of the intertwiner space on which 
the $\hat{W}$ operator acts, see Section \ref{subsubsec:volumeaction}.
The sign factor $\lambda^{0 \beta}_r$ is nothing but 
the usual $\lambda$ in the twist property (\ref{twist}) extended to
zero valued indices. 
We refrain from a complete chromatical evaluation of the 
amplitude $A^{(m)}$. For general $m$, the expression would be 
useless due to its complexity which arises because of large
allowed color ranges of the representations involved.

Having obtained this partial result, one immediately deduces
the full action of the generalized HCO on  $|v (p,q,r) \rangle$.
According to (\ref{H_m_Delta}), it is given (up to constant factors) 
by contracting the trace part (\ref{trace_part}) with 
$\epsilon^{ijk}$. In all, this leads in the complete action of 
$\hat{\mathcal{H}}^m_{\Delta}$ to a sum of three terms. They are 
distinguished from each other by the assignment of color-$m$ 
segments between mutually distinct pairs of edges adjacent to the 
vertex. The corresponding amplitudes are determined from
(\ref{amplitude}) by cyclic permutations of argument pairs. 
Thus we obtain for a generic 3-vertex $|v (p,q,r) \rangle$,
\begin{eqnarray}
  \label{final_action_of_H}
  \lefteqn{
  \hat{\mathcal{H}}^m_{\Delta} \, \big| v (p,q,r) \big\rangle 
    = (-1)^{m} \: \frac{i l_0}{12 C(m)} \, 
      \left[\rule{0cm}{1.1cm}\right. \sum_{a,b} 
      A^{(m)}(p,a|q,b|r\, \cdot) ~\Bigg|\!\!\!  
     \begin{array}{c} \mbox{\epsfig{file=v10.eps,width=2.8cm}} 
       \end{array} \!\!\!\Bigg\rangle } 
        \hspace{3cm} \nonumber \\
   && +~ \sum_{b,c} A^{(m)}(q,b|r,c|p\, \cdot) 
      ~\Bigg|\!\!\!
   \begin{array}{c} \mbox{\epsfig{file=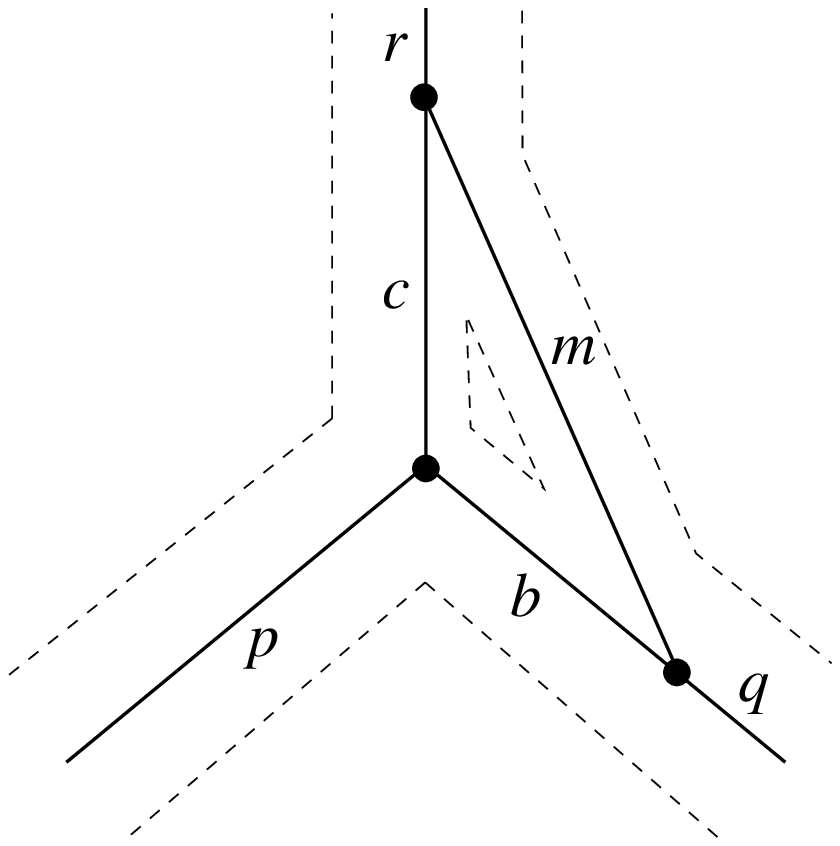,width=2.8cm}} 
     \end{array} \!\!\!\Bigg\rangle \nonumber \\
   && +~ \sum_{a,c} A^{(m)}(r,c|p,a|q\, \cdot) ~\Bigg|\!\!\! 
   \begin{array}{c} \mbox{\epsfig{file=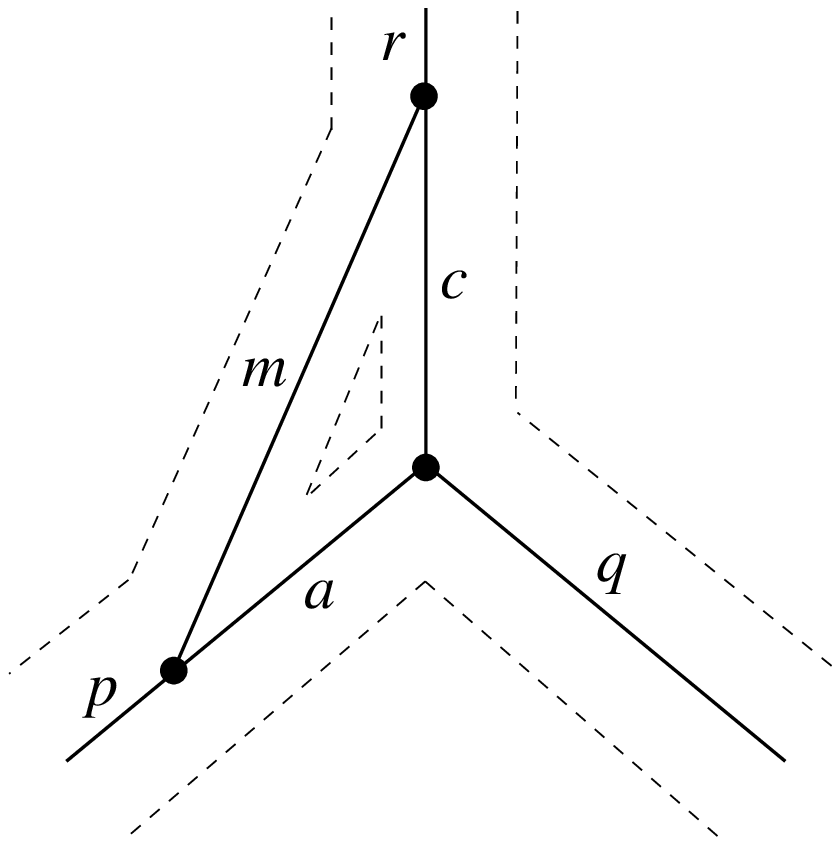,width=2.8cm}} 
     \end{array} \!\!\!\Bigg\rangle 
        \left.\rule{0cm}{1.1cm}\right] ~.
\end{eqnarray}
The above result is useful in order to recognize easily where  
new color-$m$ segments are being added around the
3-vertex. However, rewriting it in a compact way will 
exhibit more symmetries than $A^{(m)}$ alone, as we will 
show in a moment.

We should finally mention that the second possible 
ordering choice for the local Euclidean HCO as given in 
(\ref{ordering2b}) is not considered in this article.
The explicit calculations are performed in \cite{GaulPhD}.


\subsubsection{The final result for the action of 
$\hat{\mathcal{H}}^m_{\Delta}$}
We take advantage of the close connection between
the three amplitudes in (\ref{final_action_of_H}) to rewrite
the action of $\hat{\mathcal{H}}^m_{\Delta}$.
Instead of considering added color-$m$ segments for each of
the terms differently, we treat them on the same footing.
This leads directly to the compact final form for the action of the 
generalized local HCO $\hat{\mathcal{H}}^m_{\Delta}$
on a 3-valent vertex, namely
\begin{equation}
  \label{complete_action}
  \hat{\mathcal{H}}^m_{\Delta} 
  ~\Bigg| \!  
  \begin{array}{c} \mbox{\epsfig{file=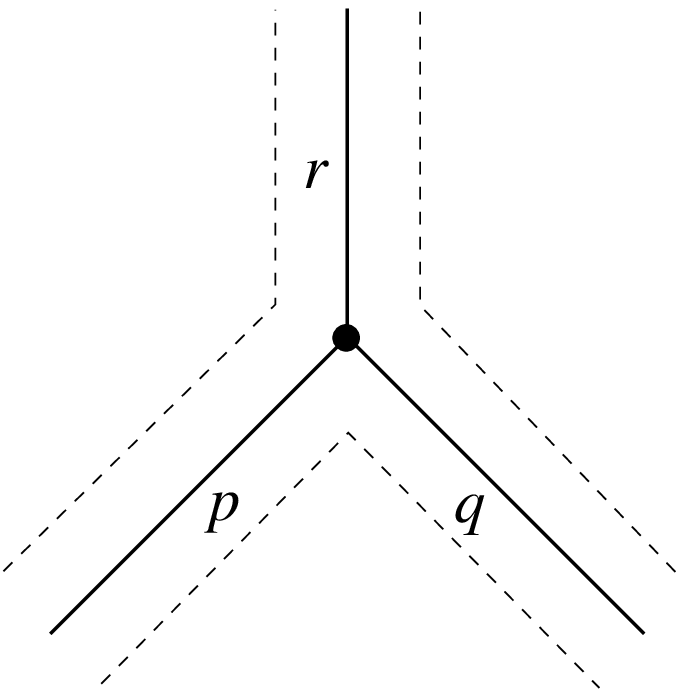,width=2.5cm}} 
    \end{array} \!\Bigg\rangle
  ~=~ (-1)^{m} \, \frac{i l_0}{12 C(m)} \, \sum_{\alpha,\beta,\gamma}
        H^m_{\Delta} (p,\alpha | q,\beta|r,\gamma) ~
   \Bigg|\!\!\! 
   \begin{array}{c} \mbox{\epsfig{file=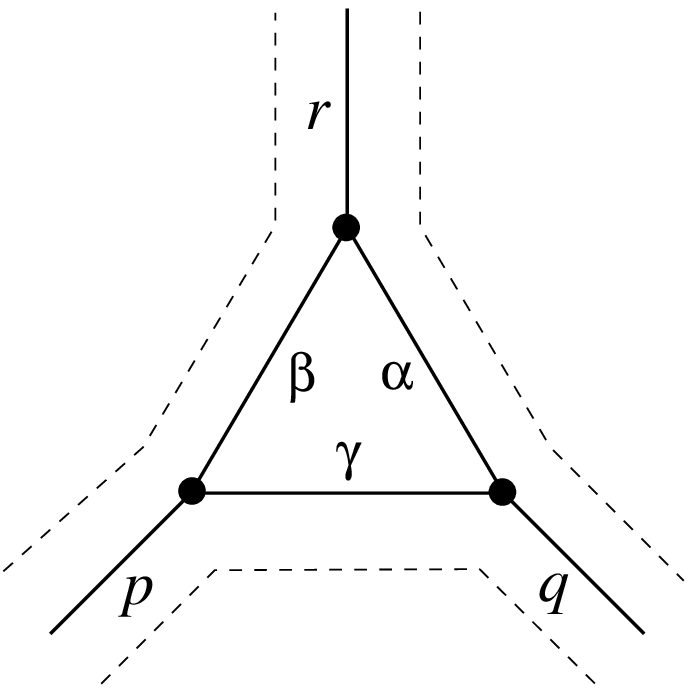,width=2.5cm}} 
     \end{array} \!\!\Bigg\rangle ~.
\end{equation}
Here $\alpha$ denotes the color of the edge opposite to $p$, 
$\beta$ the one opposite to $q$, and $\gamma$ that of $r$.
Note that $\beta$ should not be confused with the index
in (\ref{amplitude}), where it represents an internal summation in
$A^{(m)}$ due to the non-diagonal action of
the volume operator.
The matrix elements of the HCO are denoted by 
$H^m_{\Delta} (p,\alpha | q,\beta|r,\gamma)$, i.e.
\begin{equation}
  H^m_{\Delta} (p,\alpha | q,\beta|r,\gamma) ~\equiv~
    (-1)^{m} \, \frac{12 C(m)}{i l_0} \:
    \Bigg\langle\! 
     \begin{array}{c} \mbox{\epsfig{file=v13-2.eps,width=2.3cm}} 
       \end{array} \!\Bigg|~ \hat{\mathcal{H}}^m_{\Delta}
     ~\Bigg| \!  
     \begin{array}{c} \mbox{\epsfig{file=v14.eps,width=2.3cm}} 
       \end{array} \!\Bigg\rangle~. 
\end{equation}
Their relation to the amplitudes in 
the previous expression (\ref{final_action_of_H}) is
\begin{eqnarray}
  \label{HmMatrixElement}
  \lefteqn{H^m_{\Delta} (p,\alpha | q,\beta|r,\gamma) 
     ~=~ A^{(\gamma)} (p,\beta|q,\alpha|r\,\cdot) 
        \,\delta_m^\gamma} \hspace{2.6cm} \nonumber \\[1mm]
  && +~ A^{(\alpha)}(q,\gamma|r,\beta|p\,\cdot) \,\delta_m^\alpha
    + A^{(\beta)}(r,\alpha|p,\gamma|q\,\cdot) \,\delta_m^\beta ~.
\end{eqnarray}
An important property is their
invariance under cyclic permutations of argument pairs
accord\-ing to the scheme
\begin{displaymath}
  \begin{array}{c}\setlength{\unitlength}{1 pt}
    \begin{picture}(90,80) 
      \put(40,62){${\scriptstyle p}$}
      \put(39,50){${\scriptstyle \alpha}$}
      \put(37,60){\vector(-2,-3){30}}\put(37,47){\vector(-2,-3){16}}
      \put(2,8){${\scriptstyle q}$}
      \put(14,15){${\scriptstyle \beta}$}
      \put(10,9){\vector(1,0){64}}\put(22,17){\vector(1,0){40}}
      \put(78,7){${\scriptstyle r}$}
      \put(65,16){${\scriptstyle \gamma}$}
      \put(77,15){\vector(-2,3){30}}\put(63,23){\vector(-2,3){16}}
      \put(93,17) ,
    \end{picture}
  \end{array} 
\end{displaymath}
i.e.
\begin{equation}
  H^m_{\Delta} (p,\alpha|q,\beta|r,\gamma) =
  H^m_{\Delta} (q,\beta|r,\gamma|p,\alpha) = 
  H^m_{\Delta} (r,\gamma|p,\alpha|q,\beta) ~.
\end{equation}
This symmetry might play a role in subsequent considerations 
concerning crossing symmetry. For closer examinations 
thereof we refer to \cite{GaulRovelli00}.


\section{Conclusions and Outlook}

We have generalized Thiemann's Hamiltonian constraint operator to a
family of classically equivalent constraint operators.  From this
point of view, Thiemann's original color choice is just one of
the possible choices.  The calculation of the generalized matrix
elements shows that computations get complicated, but nevertheless
they can in principle be performed.  In this respect, we compute in
appendix \ref{app:ThiemannsHCO} the particular case of $m=1$.  As
expected, we obtain the same result that has been computed for
Thiemann's Hamiltonian in \cite{Borissovetal97}, providing a nice
consistency check of our evaluations.

As described in the Introduction, the motivation for this paper came
from crossing symmetry.  More precisely, we would like to find a
crossing symmetric HCO. As demonstrated in \cite{GaulRovelli00},
Thiemann's HCO (in fact, any constraint with fixed $m$) can not be
crossing symmetric.  The interesting question is whether there exists
a linear combination of HCO's with different $m$'s which is crossing
symmetric. More precisely whether we can define the physical HCO as 
\begin{equation}
\hat{\mathcal{H}}=\sum_m c_{m} \hat{\mathcal{H}}^m,    
\end{equation}
and fix the coefficients $c_{m}$ by requiring crossing symmetry. We
do not know if this problem has a solution.

The generalization we have studied might also be useful for investigations 
of semi-classical quantum gravity, in particular with regard to the
recently introduced coherent states \cite{Thiemann00a}.


\paragraph{Acknowledgements.}
This work was partially supported by NSF grant PHY-9900791.

\begin{appendix}

\section{The $m=1$ case: Thiemann's Hamiltonian}
\label{app:ThiemannsHCO}
The action of the original Hamiltonian constraint on 3-valent vertices
has already been evaluated in \cite{Borissovetal97} allowing to perform a 
consistency check of our results. Therefore we have to evaluate
(\ref{final_action_of_H}) or (\ref{complete_action}) respectively,
for the particular color $m=1$ and compare it to the result in the 
literature.

At this point, a comment on conventions is appropriate.
The classical expression (\ref{BarberoHamiltonian}) for the Hamiltonian 
constraint is exact. However, one often finds in the literature 
that a factor of $(-2)$ is missing, which is justified by the classical 
identity $\mathcal{C} = 0$.
But since we investigate the case of arbitrary
colors, we have to keep track of this factor, which is nothing but 
$-1 / C(m)$. It emerges in the generalization and ensures the correct 
classical limit. Hence the evaluation for $m=1$ that has to be performed in 
this appendix would differ from \cite{Borissovetal97} by $-1 / C(1) = -2$, 
i.e. the above factor! 

According to (\ref{final_action_of_H}), the Hamiltonian constraint operator 
produces a sum of three terms on generic 3-vertices $\ket{v(p,q,r)}$.
They are determined by the corresponding amplitudes $A^{(m)}_i$, 
$i=1,2,3$, which are invariant under cyclic permutations of the
arguments. Thus it suffices to show the $m=1$ correspondence
of our result and \cite{Borissovetal97} for just one of the three terms.
We consider in the following the first term in (\ref{final_action_of_H}).
In order to avoid the mentioned factor problem, we write it as
\begin{eqnarray}
  \frac{-1}{C(m)} \, \sum_{a,b} \, 
  \tilde{A}^{(m)}(p,a|q,b|r\,\cdot) \: \Bigg|\!\!\!
  \begin{array}{c} \mbox{\epsfig{file=v10.eps,width=2.8cm}} 
    \end{array} \!\!\!\Bigg\rangle ~,
\end{eqnarray}

\newpage\noindent
where
\begin{equation}
  \tilde{A}^{(m)}(p,a|q,b|r\,\cdot) = (-1)^{m-1}\, \frac{i l_0}{12}\,
    A^{(m)}(p,a|q,b|r\,\cdot)  ~.
\end{equation}
Equivalence to the literature will thus be shown by comparing 
with the evaluation of
\begin{eqnarray}
  \lefteqn{
    \tilde{A}^{(m)}(p,a|q,b|r\,\cdot) :=} \hspace{6mm} \nonumber \\[2mm]
   &&  \hspace{-1cm} \sum_c \; (-1)^{m-1} \: \frac{i l_0}{12} \,
     \lambda^{mp}_a \lambda^{mq}_b \lambda^{mr}_c \:
     \frac{\Delta_a \, \Delta_b \, \Delta_c}{\Theta\big(p,m,a \big) \, 
       \Theta\big(q,m,b \big) \, \Theta\big(r,m,c \big)} \, 
     \sum_{\beta(p,q,m,c)} 
     \!\!\!\! V{}_r{}^\beta (p,q,m,c) ~\times \nonumber \\[2mm]
   && \hspace{-10mm} \times \left[\rule{0cm}{1.2cm}\right.
   \lambda^{0 \beta}_r \;
   \frac{{\rule[-4mm]{0mm}{5mm}\rm Tet} \left[
      \begin{array}{ccc}
             a & q & p \\
             \beta & m & c  
      \end{array}\right] \, {\rm Tet}\left[
      \begin{array}{ccc} 
            a & b & q \\
            m & c & r  
      \end{array}\right]}{\rule[0mm]{0mm}{4mm} \Theta\big(q, c, a \big)
    \Theta\big(r, a, b \big)}
  - \frac{{\rule[-4mm]{0mm}{5mm}\rm Tet}\left[
      \begin{array}{ccc} 
            p & b & q \\
            m & \beta & c  
      \end{array}\right] \, {\rm Tet}\left[
      \begin{array}{ccc} 
            a & b & p \\
            c & m & r  
      \end{array}\right]}{\rule[0mm]{0mm}{4mm} \Theta\big(p, c, b \big)
    \Theta\big(r, a, b \big)}
     \left.\rule{0cm}{1.2cm}\right] \hspace{-1mm}.
   \label{amplitude2}
\end{eqnarray}

\noindent
In the general case of arbitrary $m$, the action of the volume operator is
the main obstacle for the explicit computation of a closed expression
for the trivalent vertex amplitude. In the $m=1$ case, however, its 
action is diagonal, allowing a complete evaluation.


\subsection{The volume operator for $m=1$}
We start the evaluation of (\ref{amplitude2}) with the matrix elements 
$V{}_r{}^\beta (p,q,m,c)$ of the volume operator. They were studied for 
general $m$ in Section \ref{subsubsec:volumeaction}, finally obtaining 
(\ref{final_action_of_V}). 
From now on, we will restrict $m$ to fundamental representations of 
$SU(2)$, i.e. $m=1$. Thus the internal 
color $c$, labelling representations in the decomposition of 
$r \otimes m$, is determined to $r \pm 1 \equiv r + \epsilon$, with 
$\epsilon = \pm 1$. 

Recall definition (\ref{V_def}) of the matrix elements 
of the volume operator,
\[ 
  V{}_\alpha{}^\beta = \sqrt{|i W|}\: {}_\alpha{}^\beta ~.
\]
$W$ stems from the action of $\hat{W}$, the `square of the volume',
which has been dealt with above, resulting in (\ref{grapheqn_for_W}), 
\begin{displaymath}
  \hat{W}_{[pqc]} 
    \;\;\Bigg|\!\!
     \begin{array}{c}\setlength{\unitlength}{1.5 pt}
        \begin{picture}(50,40)
          \put(1,0){${\scriptstyle q}$}\put(1,34){${\scriptstyle p}$}
          \put(36,14){${\scriptstyle 1}$}\put(46,34){${\scriptstyle c}$}
          \put(20,18){\line(-1,-1){15}}\put(20,18){\line(-1,1){15}}
          \put(30,18){\line(1,1){15}}\put(30,18){\line(1,-1){8}}
          \put(20,18){\line(1,0){10}}\put(24,21){${\scriptstyle \alpha}$}
          \put(20,18){\circle*{3}}\put(30,18){\circle*{3}}
          \bezier{20}(5,18)( 5,33)(25,33)
          \bezier{20}(45,18)(45,33)(25,33)
          \bezier{20}(5,18)(5,3)(25,3)
          \bezier{20}(45,18)(45,3)(25,3)
       \end{picture}
    \end{array} \Bigg\rangle 
  = \sum_\beta W^{(4)}_{[pqc]}(p,q,1,c){}_\alpha{}^\beta
    \;\;\Bigg|\!\!
    \begin{array}{c}\setlength{\unitlength}{1.5 pt}
       \begin{picture}(50,40)
          \put(1,0){${\scriptstyle q}$}\put(1,34){${\scriptstyle p}$}
          \put(36,14){${\scriptstyle 1}$}\put(46,34){${\scriptstyle c}$}
          \put(20,18){\line(-1,-1){15}}\put(20,18){\line(-1,1){15}}
          \put(30,18){\line(1,1){15}}\put(30,18){\line(1,-1){8}}
          \put(20,18){\line(1,0){10}}\put(24,21){${\scriptstyle \beta}$}
          \put(20,18){\circle*{3}}\put(30,18){\circle*{3}}
          \bezier{20}(5,18)( 5,33)(25,33)
          \bezier{20}(45,18)(45,33)(25,33)
          \bezier{20}(5,18)(5,3)(25,3)
          \bezier{20}(45,18)(45,3)(25,3)
       \end{picture}
    \end{array} \Bigg\rangle ~.
\end{displaymath}
The admissibility conditions applied to the rightmost virtual 3-vertices on 
both sides of the equation, give a bound on $\alpha$ and $\beta$.
Using $c=r \pm 1$, we obtain the two cases
\begin{eqnarray*}
  c = r-1 ~&:&~ \alpha, \beta = r-2,\, r \\
  c = r+1 ~&:&~ \alpha, \beta = r,\, r+2 ~,
\end{eqnarray*}
i.e. the intertwiner space is always 2-dimensional. Normalizing the 
vertex as explained in Section \ref{subsubsec:volumeaction}, 
we end up with real antisymmetric $(2 \times 2)$--matrices
\mbox{$\tilde{W}^{(4)}_{[pq r \pm 1]}(p,q,1,r \pm 1){}_\alpha{}^\beta$}.
Obviously, the diagonal elements equal zero, and 
only one degree of freedom remains for each value of $c$. 
The corresponding matrices can thus be written as

\[
  \Big( \tilde{W}^{(4)}_{[pq r\pm 1]}(p,q,1,r \pm 1){}_\alpha{}^\beta \Big) 
  = \left( \begin{array}{c c}
      0 & \tilde{W}_r{}^{r \pm 2} \\
      - \tilde{W}_r{}^{r \pm 2} & 0 \end{array} \right) ~~.
\] 
They can be diagonalized, leading to $\tilde{W}_D$ with purely 
imaginary eigenvalues $\lambda_{1,2}^{(c)}$. Defining 

\begin{equation}
  \label{w_def}
  w(p,q,1,r \pm 1) =
  | \tilde{W}^{(4)}_{[pq r \pm 1]}(p,q,1,r \pm 1)_r{}^{r \pm 2} | 
  ~ \in \mathbb{R} ~~,
\end{equation}
where we use the notation as in \cite{Borissovetal97}\footnote{We would like 
to point out a mistake in \cite{Borissovetal97}. Equation (A.10), which 
should be the same one as (\ref{w_def}) above, has a wrong square root in it.}, 
they are $\lambda_{1,2}^{(r \pm 1)} = \pm i w(p,q,1,r \pm 1)$.
Using (\ref{W_tilde_explicit}), the matrix elements, and thus the eigenvalues 
of $\tilde{W}^{(4)}$ are explicitly computable for any admissible triple 
$\{ p,q,r \}$. 

Having obtained these eigenvalues, the matrix elements of the volume operator 
can be given immediately. Equation (\ref{sqrt_W_matrix}) provides the 
link between the diagonal form $\tilde{W}_D$ after two base transformations, and 
$W$ in the original basis according to
 
\begin{equation}
   \label{sqrt_W_matrix_again}
   \sqrt{|i W|}\: {}_\alpha{}^\beta = \frac{n(\beta)}{n(\alpha)} \:
   U^{-1}{}_\alpha{}^\rho \,\sqrt{|i \tilde{W}_D|}\,{}_\rho{}^\sigma 
   \; U_\sigma{}^\beta ~.
\end{equation}
The required absolute value of $i \tilde{W}_D$, and therefore its square root,
turn out to be proportional to the identity. Thus  
(\ref{sqrt_W_matrix_again}) is trivial, since $U$ now commutes with the 
square root, giving

\begin{equation}
  \sqrt{|i W|}\: {}_\alpha{}^\beta = \sqrt{w(p,q,1,r + \epsilon)} \:
  \delta {}_\alpha{}^\beta ~.
\end{equation} 
Switching back to the notation of (\ref{amplitude2}), we finally obtain for
the diagonal volume operator

\begin{eqnarray}
  V(p,q,1,r+\epsilon){}_r{}^\beta &=& \sqrt{w(p,q,1,r + \epsilon)} \:
  \delta {}_r{}^\beta \nonumber \\[2mm] 
  &=:& V(p,q,1,r+\epsilon) \; \delta {}_r{}^\beta ~,
  \label{diagonal_volume}
\end{eqnarray} 
where $\epsilon = \pm 1$. 
This result is precisely the same as in \cite{Borissovetal97}, providing the 
first step in the consistency check.


\subsection{The chromatic evaluation of the amplitude}
We may simplify the $m=1$ amplitude (\ref{amplitude2}) using
(\ref{diagonal_volume}) as well the restriction for the internal color $c$.
The inspection of (\ref{final_eval_I}) reveals that the remaining two 
colors $a$ and $b$ are restricted in a similar way.
Namely, we get $a=p+\bar{\epsilon}$, $b=q+\tilde{\epsilon}$ and $c=r+\epsilon$,
whereas $\bar{\epsilon}, \tilde{\epsilon}, \epsilon = \pm 1$.
Note that we attach importance to the use of the same
parameters as in \cite{Borissovetal97} (at least wherever this is possible 
without 
notation ambiguities). Inserting everything in (\ref{amplitude2}), and using
$\lambda^{0 r}_r = +1$, it remains to calculate
\begin{eqnarray}
  \lefteqn{
  \tilde{A}^{(1)}(p,p+\bar{\epsilon}\,|\,q,q+\tilde{\epsilon}\, |\, 
        r\,\cdot)} 
       \hspace{1.5cm} \nonumber \\[2mm]
   && \hspace{-1cm}  =~ \sum_{\epsilon= \pm 1} \frac{i l_0}{12} \,
     \lambda^{1p}_{p+\bar{\epsilon}} \, \lambda^{1q}_{q+\tilde{\epsilon}} \,
     \lambda^{1r}_{r+\epsilon} \:
     \frac{\Delta_{p+\bar{\epsilon}} \, \Delta_{q+\tilde{\epsilon}} \, 
         \Delta_{r+\epsilon}}{\Theta\big(p,1,p+\bar{\epsilon} \big) \, 
       \Theta\big(q,1,q+\tilde{\epsilon} \big) \, \Theta\big(r,1,r+\epsilon 
         \big)}  ~\times \nonumber \\[2mm]
   && \hspace{-0.5cm} \times~ V(p,q,1,r+\epsilon) 
        \left[\rule{0cm}{1.2cm}\right.
   \frac{{\rule[-4mm]{0mm}{5mm}\rm Tet} \left[
      \begin{array}{ccc}
             p+\bar{\epsilon} & q & p \\
             r & 1 & r+\epsilon 
      \end{array}\right] \, {\rm Tet}\left[
      \begin{array}{ccc} 
            p+\bar{\epsilon} & q+\tilde{\epsilon} & q \\
            1 & r+\epsilon & r  
      \end{array}\right]}{\rule[0mm]{0mm}{4mm} 
    \Theta\big(q, r+\epsilon, p+\bar{\epsilon} \big)
    \Theta\big(r, p+\bar{\epsilon}, q+\tilde{\epsilon} \big)} 
      \nonumber \\[2mm]
   && \hspace{2.5cm} ~-~ \frac{{\rule[-4mm]{0mm}{5mm}\rm Tet}\left[
      \begin{array}{ccc} 
            p & q+\tilde{\epsilon} & q \\
            1 & r & r+\epsilon  
      \end{array}\right] \, {\rm Tet}\left[
      \begin{array}{ccc} 
            p+\bar{\epsilon} & q+\tilde{\epsilon} & p \\
            r+\epsilon & 1 & r  
      \end{array}\right]}{\rule[0mm]{0mm}{4mm} 
    \Theta\big(p, r+\epsilon, q+\tilde{\epsilon} \big)
    \Theta\big(r, p+\bar{\epsilon}, q+\tilde{\epsilon} \big)}
     \left.\rule{0cm}{1.2cm}\right] \hspace{-1mm}.
   \label{amplitude_m=1}
\end{eqnarray}
In the following, we will not calculate the sum over $\epsilon$
but rather investigate every single term. 
Therefore we write 
\begin{equation}
  \tilde{A}^{(1)}(p,p+\bar{\epsilon}\,|\,q,q+\tilde{\epsilon}\, |\, 
        r\,\cdot) = \sum_{\epsilon = \pm 1} 
  \tilde{A}^{(1)}(p,p+\bar{\epsilon}\,|\,q,q+\tilde{\epsilon}\, |\,   
        r, r+\epsilon) ~,
\end{equation}
and consider the 8 different terms
$\tilde{A}^{(1)}(p,p+\bar{\epsilon}\,|\,q,q+\tilde{\epsilon}\, |\,   
r, r+\epsilon)$ that are distinguished by 
the possible combinations of $\bar{\epsilon}, \tilde{\epsilon}$ 
and $\epsilon$. They can be compared 
directly with the results of \cite{Borissovetal97}.

First of all, we get the following simple identities from 
the formulae in appendix \ref{app:recoupl_theory},
\begin{equation}
  \lambda^{1 n}_{n+\sigma} ~=~ \left\{ 
        \begin{array}{c @{\quad}l}
                (-1)^n & : \quad \sigma = +1 \\
                \vspace{0mm} & \\
                 (-1)^{n+1} & : \quad \sigma = -1 ~,
        \end{array} \right.   
\end{equation}
and
\begin{equation}
  \frac{\Delta_{n+\sigma}}{\Theta\big(n, n+\sigma, 1 \big)} =
  \frac{\Delta_{n+\sigma}}{\Delta_{n+\frac{\sigma+1}{2}}} = \left\{ 
        \begin{array}{c @{\quad}l}
                +1 & : \quad \sigma = +1 \\
                \vspace{0mm} & \\
                 - \frac{n}{n+1} & : \quad \sigma = -1 ~.
        \end{array} \right.   
\end{equation}
The computationally most expensive and intriguing parts in 
(\ref{amplitude_m=1}) are the tetrahedral networks, 
see (\ref{Tet}) for the general definition.
The 8 different amplitudes that have to be evaluated contain 32 Tets in all.
However, algebraic manipulations show that the expressions simplify 
greatly when considering quotients of Tet and Theta networks as they appear 
in (\ref{amplitude_m=1}). In addition, a detailed analysis reveals that  
these 32 quotients are not all independent, they can rather be traced back 
to only 4 different fundamental evaluations.
Introducing $\tau=0, \pm 1$ and $\sigma = \pm 1$, one gets
for an admissible triple of colors $\{ a, b, c\}$  (not to be mixed up
with the above used notation), 

\begin{eqnarray}
  \frac{{\rule[-4mm]{0mm}{5mm}\rm Tet}\left[
      \begin{array}{ccc} 
            a+1 & b+\tau & a \\
            c-\tau^2 & 1 & c+1-\tau^2  
      \end{array}\right]}{\rule[0mm]{0mm}{4mm}\Theta 
                     \big(a+1, b+ \tau, c+1 - \tau^2 \big)}
  &=& 1\\[3mm]
  \frac{{\rule[-4mm]{0mm}{5mm}\rm Tet}\left[
      \begin{array}{ccc} 
            a-1 & b+\tau & a \\
            c-\tau^2 & 1 & c+1-\tau^2  
      \end{array}\right]}{\rule[0mm]{0mm}{4mm}\Theta 
                     \big(a-1, b+ \tau, c+1 - \tau^2 \big)}
  &=& \frac{\tau(\tau +1) +a +b -c}{2a} ~~, 
\end{eqnarray}
\begin{eqnarray}
  \frac{{\rule[-4mm]{0mm}{5mm}\rm Tet}\left[
      \begin{array}{ccc} 
            a+1 & b+\sigma & a \\
            c+1 & 1 & c  
      \end{array}\right]}{\rule[0mm]{0mm}{4mm}\Theta 
                               \big(a+1, b+ \sigma, c \big)} 
    &=& \frac{\sigma+1+b-a+c}{2+2c} ~~,
\end{eqnarray}
and
\begin{eqnarray}
  \lefteqn{\frac{{\rule[-4mm]{0mm}{5mm}\rm Tet}\left[
      \begin{array}{ccc} 
            a-1 & b+\tau & a \\
            c+\tau^2 & 1 & c-1+\tau^2  
      \end{array}\right]}{\rule[0mm]{0mm}{4mm}\Theta 
               \big(a-1, b+ \tau, c-1+\tau^2 \big)} ~=} \hspace{2cm} 
           \nonumber \\[3mm]
   & & -\, \frac{\big(\tau(\tau -1)+a -b +c 
     \big)\big(\tau(\tau +1)+2+a +b +c \big)}{4a (\tau^2+c)} ~~.
\end{eqnarray}
Using these results, and introducing the abbreviation 
$V(r+\epsilon):= V(p,q,1,r+\epsilon)$, we finally obtain for the different
amplitudes $\tilde{A}$
in (\ref{amplitude_m=1}) the following expressions distinguished only by the 
values of $\epsilon, \bar{\epsilon}$ and $\tilde{\epsilon}$,
\begin{eqnarray*}
  \begin{array}{r @{~~} l}
    i.)& \epsilon=+1,\, \bar{\epsilon}=+1,\: \tilde{\epsilon}=+1 \\
       & {\D \tilde{A}^{(1)}(p,p+1\,|\,q,q+1\, |\, r,r+1) = 
           V(r+1)\, \frac{i l_0 (p-q)}{12 (r+1)}} \\[5mm]
    ii.) \rule[-3mm]{0mm}{8mm} & 
            \epsilon=+1,\, \bar{\epsilon}=-1,\: \tilde{\epsilon}=+1 \\
       & {\D \tilde{A}^{(1)}(p,p-1\,|\,q,q+1\, |\, r,r+1) = 
           V(r+1)\, \frac{i l_0 (p-q+r)(2+p+q)}{24 (r+1)(p+1)}} \\[5mm]
    iii.) \rule[-3mm]{0mm}{8mm} & 
            \epsilon=+1,\, \bar{\epsilon}=+1,\: \tilde{\epsilon}=-1 \\
       & {\D \tilde{A}^{(1)}(p,p+1\,|\,q,q-1\, |\, r,r+1) = 
           - V(r+1)\, \frac{i l_0 (q-p+r)(2+p+q)}{24 (r+1)(q+1)}} \\[5mm]
    iv.) \rule[-3mm]{0mm}{8mm} & 
            \epsilon=+1,\, \bar{\epsilon}=-1,\: \tilde{\epsilon}=-1 \\
       & {\D \tilde{A}^{(1)}(p,p-1\,|\,q,q-1\, |\, r,r+1) = 
           V(r+1)\, \frac{i l_0 (p+q-r)(p-q)(2+p+q+r)}{48 (r+1)(p+1)(q+1)}} 
         \\[5mm]
    v.) \rule[-3mm]{0mm}{8mm} & 
            \epsilon=-1,\, \bar{\epsilon}=+1,\: \tilde{\epsilon}=+1 \\
       & {\D \tilde{A}^{(1)}(p,p+1\,|\,q,q+1\, |\, r,r-1) = 
           V(r-1)\, \frac{i l_0 (q-p)}{12 (r+1)}} \\[5mm]
  \end{array}
\end{eqnarray*}
\begin{eqnarray*}
  \begin{array}{r @{~~} l}
    vi.) \rule[-3mm]{0mm}{8mm} & 
            \epsilon=-1,\, \bar{\epsilon}=-1,\: \tilde{\epsilon}=+1 \\
       & {\D \tilde{A}^{(1)}(p,p-1\,|\,q,q+1\, |\, r,r-1) = 
          - V(r-1)\, \frac{i l_0 (p-q+r)(2+p+q)}{24 (r+1)(p+1)}} \\[5mm]
    vii.) \rule[-3mm]{0mm}{8mm} & 
            \epsilon=-1,\, \bar{\epsilon}=+1,\: \tilde{\epsilon}=-1 \\
       & {\D \tilde{A}^{(1)}(p,p+1\,|\,q,q-1\, |\, r,r-1) = 
           V(r-1)\, \frac{i l_0 (q-p+r)(2+p+q)}{24 (r+1)(q+1)}} \\[5mm]
    viii.) \rule[-3mm]{0mm}{8mm} & 
            \epsilon=-1,\, \bar{\epsilon}=-1,\: \tilde{\epsilon}=-1 \\
       & {\D \tilde{A}^{(1)}(p,p-1\,|\,q,q-1\, |\, r,r-1) = 
           V(r-1)\, \frac{i l_0 (p+q-r)(q-p)(2+p+q+r)}{48 (r+1)(p+1)(q+1)}}
         \\[5mm]
  \end{array}
\end{eqnarray*}
Comparing $i.) - viii.)$ with the corresponding amplitudes in 
\cite{Borissovetal97}\footnote{We should point out another 
mistake in \cite{Borissovetal97} that needs to be corrected 
before comparing the results. In Equation (5.9), which is a 
distinction of four cases, the two expressions in the middle were  
printed in reversed order, i.e. the result associated to 
$\bar{\epsilon}=-1 , \tilde{\epsilon} = +1$ actually corresponds
to $\bar{\epsilon}=+1 , \tilde{\epsilon} = -1$, and vice versa.},
we find that they are identical. This demonstrates
the consistency of the generalized Hamiltonian constraint operator 
at the color $m=1$ level.


\section{Essentials from Recoupling Theory}
\label{app:recoupl_theory}
This appendix is mainly a collection of the basic relations of
diagrammatic recoupling theory. We use the conventions of
\cite{KauffmanLins94} where the general framework is developed in
the context of Temperley-Lieb algebras.
Relevant for the computations in this paper 
is only the classical case of a deformation parameter $A=-1$, 
which corresponds to standard $SU(2)$ re\-coup\-ling theory.

For the simplest closed tangles we give the explicit results of
the chromatic evaluations of the networks, i.e. of the associated 
trace in the Temperley-Lieb algebra or, equivalently, the Kauffman bracket.
Furthermore, it is required that vertices are admissible to avoid
triviality.
A triple $\{a,b,c \}$ of arbitrary colors associated to edges meeting at a 
vertex is said to be admissible, if it satisfies
\begin{eqnarray*}
  &(i)& a+b+c \equiv 0 \pmod{2} \\
  &(ii)& a+b-c \geq 0\\
  && b+c-a \geq 0 \\
  && c+a-b \geq 0 ~~.
\end{eqnarray*}
One easily sees that the admissibility conditions are completely 
equivalent to the triangular inequalities or Clebsch-Gordan relations 
for the decomposition
of a tensor product of two irreducible $SU(2)$ representations, which
are labelled in this context by colors (i.e. twice the spin).

\paragraph{Recoupling Theorem.}
Certainly the most important result is the so-called 
\emph{recoupling theorem}, which tells us (in familiar terms) 
how different couplings of four angular momenta are 
related to each other,
\begin{equation}
\label{recoupl_thm}
\begin{array}{c}\setlength{\unitlength}{1 pt}
\begin{picture}(50,40)
          \put(4,3){${\scriptstyle a}$}\put( 4,30){${\scriptstyle b}$}
          \put(42,3){${\scriptstyle d}$}\put(42,30){${\scriptstyle c}$}
          \put(10,8){\line(1,1){10}}\put(10,28){\line(1,-1){10}}
          \put(30,18){\line(1,1){10}}\put(30,18){\line(1,-1){10}}
          \put(20,18){\line(1,0){10}}\put(23,23){${\scriptstyle j}$}
          \put(20,18){\circle*{3}}\put(30,18){\circle*{3}}
\end{picture}\end{array}
    = ~\sum_i  \left\{\begin{array}{ccc}
                      a  & b & i \\
                      c  & d & j  
                \end{array}\right\}
\begin{array}{c}\setlength{\unitlength}{1 pt}
\begin{picture}(40,40)
      \put(4,-3){${\scriptstyle a}$}\put(4,34){${\scriptstyle b}$}
      \put(32,-3){${\scriptstyle d}$}\put(32,34){${\scriptstyle c}$}
      \put(10,3){\line(1,1){10}}\put(10,33){\line(1,-1){10}}
      \put(20,23){\line(1,1){10}}\put(20,13){\line(1,-1){10}}
      \put(20,13){\line(0,1){10}}\put(24,16){${\scriptstyle i}$}
      \put(20,13){\circle*{3}}\put(20,23){\circle*{3}} 
      \put(45,12){~.}
\end{picture}\end{array}
\end{equation}
The recoupling coefficients in the theorem are $6j-$symbols that 
are given by
\begin{equation}
  \label{6jsymbol}
  \left\{\begin{array}{ccc}
      a  & b & i \\
      c  & d & j  
  \end{array}\right\}
 ~=~ \frac{\displaystyle \Delta_i \;
        {\rm Tet}\left[\begin{array}{ccc}
                          a & b & i \\
                          c & d & j  
                       \end{array}\right]}
        {\Theta(a,d,i) \, \Theta(b,c,i)} ~.
\end{equation}
The networks emerging in this definition are explained below.

\paragraph{The Symmetrizer.}
The symmetrizer is the simplest closed $n$-tangle. 
It is defined and evaluated as
\begin{equation}
\Delta_n:= 
\begin{array}{c}\setlength{\unitlength}{1 pt}
   \begin{picture}(35,30)
        \put(15,-2){\line(0,1){10}}\put(20,-2){\line(0,1){10}}
        \put(15,-2){\line(1,0){5}}\put(15,8){\line(1,0){5}}
        \put(15,23){\line(1,0){5}}  \put(15,17){${\scriptstyle n}$}
        \put(15,13){\oval(30,20)[l]}\put(20,13){\oval(30,20)[r]}
   \end{picture}
\end{array} 
= \; (-1)^n (n+1) ~. 
\label{symmetrizer}
\end{equation}

\paragraph{Theta Net.}
The $\Theta$-net is obtained by closing a trivalent vertex, or
more precisely, by joining and closing the trivalent network with itself.
It is
\begin{eqnarray} 
\label{theta_net}
  \Theta(a,b,c) ~:=~ 
    \begin{array}{c}\setlength{\unitlength}{1 pt}
      \begin{picture}(40,40)
        \put(18,35){${\scriptstyle a}$}
        \put(18,20){${\scriptstyle b}$} 
        \put(18, 5){${\scriptstyle c}$} 
        \put(20,18){\oval(40,30)} \put( 0,18){\line(1,0){40}} 
        \put( 0,18){\circle*{3}}  \put(40,18){\circle*{3}}
      \end{picture}
    \end{array}
  ~=~ (-1)^{(m+n+p)}\; \frac{(m+n+p+1)! ~m!~n!~p!}{a! ~b! ~c!} ~,
\end{eqnarray}
where $m=(a+b-c)/2$, $n=(b+c-a)/2$, $p=(c+a-b)/2$. 

\paragraph{Tetrahedral Net.}
Another important closed tangle that is frequently used, is 
the \emph{tetrahedral net}, or \emph{Tet} for short.
As the naming suggests, this tangle possesses
tetrahedral symmetry. Graphically, it might be represented as usual
in the standard form (the first drawing) or a little bit more suggestive
concerning the symmetry
(of course there exist a lot more equivalent ways of drawing this
and other nets as well, see e.g. \cite{BrinkSatchler68}),
\begin{equation}
  \label{Tet}
  {{\rm Tet}\left[
        \begin{array}{ccc} 
            a & b & e \\
            c & d & f  
        \end{array}\right]}
  ~=~ \begin{array}{c}\setlength{\unitlength}{1 pt}
        \begin{picture}(50,40)
          \put(0,18){\line(1,-1){15}} \put(2,26){${\scriptstyle b}$}
          \put(0,18){\line(1, 1){15}} \put(2,6){${\scriptstyle a}$}
          \put(0,18){\circle*{3}}
          \put(30,18){\line(-1, 1){15}} \put(24,26){${\scriptstyle c}$}
          \put(30,18){\line(-1,-1){15}} \put(23,6){${\scriptstyle d}$}
          \put(30,18){\circle*{3}}
          \put(0,18){\line(1,0){30}} \put(13,21){${\scriptstyle f}$}
          \put(15,33){\line(1,0){25}} \put(15,33){\circle*{3}}
          \put(15,3){\line(1,0){25}} \put(15,3){\circle*{3}}
          \put(40,3){\line(0,1){30}} \put(42,17){${\scriptstyle e}$}
        \end{picture}
    \end{array}  
  \equiv \begin{array}{c}\setlength{\unitlength}{1 pt}
        \begin{picture}(50,40)
          \put(26,18){\circle{37}}
          \put(26,18){\circle*{3}}
          \put(26,18){\line(0,1){18}}\put(26,36){\circle*{3}}
          \put(26,18){\line(1,-1){12.5}}\put(38.5,5.5){\circle*{3}}
          \put(26,18){\line(-1,-1){12.5}}\put(13.5,5.5){\circle*{3}}
          \put(5,25){${\scriptstyle b}$}
          \put(24,-6){${\scriptstyle a}$}
          \put(14,12){${\scriptstyle f}$}
          \put(33,12){${\scriptstyle d}$}
          \put(44,25){${\scriptstyle e}$}
          \put(20,25){${\scriptstyle c}$}
        \end{picture}
    \end{array}  .
\end{equation}
The tetrahedral symmetry is reflected 
in the invariance under permutations of the set
\begin{displaymath}
  \Big\{ \{a,e,d\}, \{b,a,f\}, \{f,d,c\}, \{c,b,e\} \Big\}
\end{displaymath}
of vertices of the tetrahedron. Any element in this set represents
an admissible triple of colored edges. A more practicable way
of formulating the symmetry, which is of course equivalent
to the above one, is the following. 
${\rm Tet}\left[{a\atop c}{b\atop d}{e\atop f}\right]$
is invariant under all 
permutations of its columns and under exchange of any pair of
elements in the upper row with the corresponding pair in the lower 
row. This invariance is often used throughout the paper.

The chromatic evaluation of (\ref{Tet}) is performed in
\cite{KauffmanLins94}, giving
\begin{equation}
  \label{tet_net}
  {{\rm Tet}\left[
        \begin{array}{ccc} 
            a & b & e \\
            c & d & f  
        \end{array}\right]}
  ~=~ \frac{{\cal I}}{{\cal E}} \sum_{m \leq s \leq M}
        \frac{(-1)^{s} (s+1)!}{\prod_i (s-a_i)! ~ \prod_j (b_j-s)!}~~,
\end{equation}
where
\begin{displaymath}
  \begin{array}{rclcrcl} 
    {\cal E} &=& a!\, b!\, c!\, d!\, e!\, f! ~, & \qquad & 
    {\cal I} &=& \prod_{i,j} (b_j-a_i)! ~,\\[3mm]
    a_1 &=& \frac{1}{2}(a+d+e) ~, & \qquad &
        b_1 &=& \frac{1}{2}(b+d+e+f) ~, \\[3mm]
    a_2 &=& \frac{1}{2}(b+c+e) ~, & \qquad &
        b_2 &=& \frac{1}{2}(a+c+e+f) ~, \\[3mm]
    a_3 &=& \frac{1}{2}(a+b+f) ~, & \qquad &
        b_3 &=& \frac{1}{2}(a+b+c+d) ~, \\[3mm]
    a_4 &=& \frac{1}{2}(c+d+f) ~, & \qquad &
        \multicolumn{3}{l}{m = {\rm max} \{ a_i \} ~,  
           \hspace*{3mm} M = {\rm min} \{ b_j \}} ~.
  \end{array}
\end{displaymath}

\paragraph{Edge Addition Formula.}
The next ingredient that is needed
is the \emph{edge addition formula}, which can be viewed as 
the Clebsch-Gordan decomposition of the tensor product of two arbitrary 
irreducible $SU(2)$ representations, i.e. 
\begin{equation}
  \label{edge_add1}
  \begin{array}{c}\setlength{\unitlength}{1 pt}
     \begin{picture}(30,40)
        \put(1,35){${\scriptstyle n}$}\put(24,35){${\scriptstyle m}$}
        \put(10,1){\line(0,1){35}}
        \put(20,1){\line(0,1){35}}
     \end{picture}
  \end{array}
 =~ \sum_i \,  c_i(n,m)
  \begin{array}{c}\setlength{\unitlength}{1 pt}
     \begin{picture}(40,40)
        \put(4,-3){${\scriptstyle n}$}\put(4,34){${\scriptstyle n}$}
        \put(32,-3){${\scriptstyle m}$}\put(32,34){${\scriptstyle m}$}
        \put(10,3){\line(1,1){10}}\put(10,33){\line(1,-1){10}}
        \put(20,23){\line(1,1){10}}\put(20,13){\line(1,-1){10}}
        \put(20,13){\line(0,1){10}}\put(24,16){${\scriptstyle i}$}
        \put(20,13){\circle*{3}}\put(20,23){\circle*{3}} 
        \put(45,12){~.}
      \end{picture}
  \end{array}
\end{equation}
The coefficients $c_i(n,m)$ in the decomposition can either 
be obtained directly by projecting them out, or by considering 
(\ref{edge_add1}) as a special case of the recoupling theorem
(\ref{recoupl_thm}). Admitting empty, i.e. color-$0$ 
edges, and evaluating the appropriate $6j$-symbol, one
gets
\begin{equation}
\label{edge_addition}
  \begin{array}{c}\setlength{\unitlength}{1 pt}
    \begin{picture}(30,40)
          \put(1,35){${\scriptstyle n}$}\put(24,35){${\scriptstyle m}$}
          \put(10,1){\line(0,1){35}}
          \put(20,1){\line(0,1){35}}
    \end{picture}
  \end{array}
  =~ \sum_i \, \
   \frac{\begin{array}{c}\setlength{\unitlength}{.5 pt}
          \begin{picture}(40,32)
            \put(17,17){$\scriptstyle i$} 
            \put(17,15){\oval(34,30)[l]}        
            \put(23,15){\oval(34,30)[r]} 
            \put(17,-6){\line(0,1){12}}\put(23,-6){\line(0,1){12}}      
            \put(17,-6){\line(1,0){6}}\put(17,6){\line(1,0){6}}
            \put(17,30){\line(1,0){6}}
          \end{picture}
        \end{array}}
       {\begin{array}{c}\setlength{\unitlength}{.5 pt}
          \begin{picture}(40,42)
            \put(16,33){$\scriptstyle n$}
            \put(14,17){$\scriptstyle m$} 
            \put(17, 2){$\scriptstyle i$} 
            \put(20,15){\oval(40,30)} \put( 0,15){\line(1,0){40}} 
            \put( 0,15){\circle*{3}}  \put(40,15){\circle*{3}}
          \end{picture}
        \end{array}}
  \begin{array}{c}\setlength{\unitlength}{1 pt}
    \begin{picture}(40,40)
      \put(4,-3){${\scriptstyle n}$}\put(4,34){${\scriptstyle n}$}
      \put(32,-3){${\scriptstyle m}$}\put(32,34){${\scriptstyle m}$}
      \put(10,3){\line(1,1){10}}\put(10,33){\line(1,-1){10}}
      \put(20,23){\line(1,1){10}}\put(20,13){\line(1,-1){10}}
      \put(20,13){\line(0,1){10}}\put(24,16){${\scriptstyle i}$}
      \put(20,13){\circle*{3}}\put(20,23){\circle*{3}} 
      \put(45,12){~,}
    \end{picture}
  \end{array}
\end{equation}
where the internal color $i$ takes values within the range
$|n-m| \leq i \leq (n+m)$ in steps of two.
Furthermore, the empty edge on the left hand side of (\ref{edge_addition})
is omitted by identifying
\begin{displaymath}
  \begin{array}{c}\setlength{\unitlength}{1 pt}
     \begin{picture}(30,40)
        \put(1,35){${\scriptstyle n}$}\put(24,35){${\scriptstyle m}$}
        \put(10,1){\line(0,1){35}}
        \put(20,1){\line(0,1){35}}
     \end{picture}
  \end{array}
\equiv
  \begin{array}{c}\setlength{\unitlength}{1 pt}
     \begin{picture}(50,40)
        \put(4,3){${\scriptstyle n}$}\put( 4,30){${\scriptstyle n}$}
        \put(42,3){${\scriptstyle m}$}\put(42,30){${\scriptstyle m}$}
        \put(10,8){\line(1,1){10}}\put(10,28){\line(1,-1){10}}
        \put(30,18){\line(1,1){10}}\put(30,18){\line(1,-1){10}}
        \put(20,18){\line(1,0){10}}\put(23,23){${\scriptstyle 0}$}
        \put(20,18){\circle*{3}}\put(30,18){\circle*{3}}
     \end{picture}
  \end{array} ~.
\end{displaymath}

\paragraph{Reduction Formulae.}
One can easily derive reductions of various networks.
A frequently used one is the 3-vertex reduction
\begin{equation}
  \label{3vertex_reduction}
  \begin{array}{c}\setlength{\unitlength}{1 pt}
     \begin{picture}(40,40)  
        \put(10, 3){\line(1,0){10}} \put(3, 1){$\scriptstyle c$}
        \put(10,18){\line(1,0){10}} \put(2,16){$\scriptstyle b$}
        \put(10,33){\line(1,0){10}} \put(2,31){$\scriptstyle a$}
        \put(20, 3){\circle*{3}}      
        \put(20,18){\circle*{3}}      
        \put(20,33){\circle*{3}}      
        \put(20, 3){\line(0,1){15}} \put(22,9){$\scriptstyle s$}
        \put(20,18){\line(0,1){15}} \put(22,24){$\scriptstyle r$}
        \put(20,18){\oval(20,30)[r]}\put(32,16){$\scriptstyle t$}
     \end{picture}
  \end{array}  
 = ~\frac{\begin{array}{c}\setlength{\unitlength}{.8 pt}
     \begin{picture}(45,40)
        \put( 0,15){\line(1,-1){15}} \put(-1,21){${\scriptstyle a}$}
        \put( 0,15){\line(1, 1){15}} \put(-1,4){${\scriptstyle c}$}
        \put( 0,15){\circle*{3}}
        \put(30,15){\line(-1, 1){15}} \put(26,21){${\scriptstyle r}$}
        \put(30,15){\line(-1,-1){15}} \put(26, 4){${\scriptstyle s}$}
        \put(30,15){\circle*{3}}
        \put( 0,15){\line(1,0){30}} \put(12,17){${\scriptstyle b}$}
        \put(15,30){\line(1,0){25}} \put(15,30){\circle*{3}}
        \put(15, 0){\line(1,0){25}} \put(15, 0){\circle*{3}}
        \put(40, 0){\line(0,1){30}} \put(43,13){${\scriptstyle t}$}
     \end{picture}\end{array}}
  {\begin{array}{c}\setlength{\unitlength}{.5 pt}
     \begin{picture}(40,40)
        \put(17,32.5){$\scriptstyle a$}
        \put(17,17){$\scriptstyle b$} 
        \put(17, 2){$\scriptstyle c$} 
        \put(20,15){\oval(40,30)} \put( 0,15){\line(1,0){40}} 
        \put( 0,15){\circle*{3}}  \put(40,15){\circle*{3}}
     \end{picture}
   \end{array}} 
 ~\cdot
   \begin{array}{c}\setlength{\unitlength}{1 pt}
     \begin{picture}(20,40)  
        \put(3,1){$\scriptstyle c$}
        \put(10,18){\line(1,0){15}}   
        \put(2,16){$\scriptstyle b$}
        \put(2,31){$\scriptstyle a$}
        \put(25,18){\circle*{3}}      
        \put(10,18){\oval(30,30)[r]} 
     \end{picture}
   \end{array} ~~.
\end{equation}
Another familiar way of drawing it is 
\begin{equation}
  \label{3vertex_reduction2}
  \begin{array}{c}\setlength{\unitlength}{1 pt}
     \begin{picture}(50,40)
        \put(8,11){\line(1, 1){15}}\put(11,19){${\scriptstyle r}$}
        \put(38,11){\line(-1,1){15}}\put(32,19){${\scriptstyle t}$}
        \put(8,11){\line(1,0){30}}\put(21,6){${\scriptstyle s}$}
        \put(23,26){\line(0,1){14}}\put(21,43){${\scriptstyle a}$}
        \put(8,11){\line(-3,-5){8}}\put(-5,-2){${\scriptstyle b}$}
        \put(38,11){\line(3,-5){8}}\put(47,-2){${\scriptstyle c}$}
        \put(8,11){\circle*{3}}\put(38,11){\circle*{3}}
        \put(23,26){\circle*{3}}      
     \end{picture}
  \end{array}
 = ~\frac{\begin{array}{c}\setlength{\unitlength}{.8 pt}
     \begin{picture}(45,40)
        \put( 0,15){\line(1,-1){15}} \put(0,21){${\scriptstyle r}$}
        \put( 0,15){\line(1, 1){15}} \put(-1,4){${\scriptstyle b}$}
        \put( 0,15){\circle*{3}}
        \put(30,15){\line(-1, 1){15}} \put(26,21){${\scriptstyle t}$}
        \put(30,15){\line(-1,-1){15}} \put(25, 4){${\scriptstyle c}$}
        \put(30,15){\circle*{3}}
        \put( 0,15){\line(1,0){30}} \put(12,17){${\scriptstyle s}$}
        \put(15,30){\line(1,0){25}} \put(15,30){\circle*{3}}
        \put(15, 0){\line(1,0){25}} \put(15, 0){\circle*{3}}
        \put(40, 0){\line(0,1){30}} \put(43,13){${\scriptstyle a}$}
     \end{picture}\end{array}}
  {\begin{array}{c}\setlength{\unitlength}{.5 pt}
     \begin{picture}(40,40)
        \put(17,32.5){$\scriptstyle a$}
        \put(17,17){$\scriptstyle b$} 
        \put(17, 2){$\scriptstyle c$} 
        \put(20,15){\oval(40,30)} \put( 0,15){\line(1,0){40}} 
        \put( 0,15){\circle*{3}}  \put(40,15){\circle*{3}}
     \end{picture}
   \end{array}} 
 ~\cdot
  \begin{array}{c}\setlength{\unitlength}{1 pt}
    \begin{picture}(40,40)
       \put(15,18){\line(-1,-1){13}} \put(13,37){$\scriptstyle a$}
       \put(15,18){\line( 1,-1){13}} \put(29,0){$\scriptstyle c$}
       \put(15,18){\line(0,1){16}}   \put(-3,0){$\scriptstyle b$}
       \put(15,18){\circle*{3}}
    \end{picture}
   \end{array} ~.
\end{equation}
These formulae are straightforwardly proven by first observing that
uniqueness requires the left hand sides to be proportional 
to a 3-vertex. The constant factor is determined by closing the 
open networks with another 3-vertex that is multiplied from the left 
in (\ref{3vertex_reduction}), and from below in (\ref{3vertex_reduction2}).
It should also be noticed, that the two tetrahedral nets in
(\ref{3vertex_reduction}) and (\ref{3vertex_reduction2}) are of course  
chromatically evaluated to the same number.
Tetrahedral symmetry ensures their equality.

In addition, we also display some frequently used reductions of 
networks with two open edges, namely
\begin{equation}
  \begin{array}{c}\setlength{\unitlength}{1 pt}
     \begin{picture}(40,50)
        \put(5,21){$\scriptstyle b$}
        \put(22,11){\line(0,-1){15}} \put(24,44){$\scriptstyle a$}
        \put(22,35){\line(0,1){15}} \put(24,-2){$\scriptstyle a'$}
        \put(22,23){\circle{25}} \put(36,21){$\scriptstyle c$}
        \put(22,11){\circle*{3}}\put(22,35){\circle*{3}}
     \end{picture}
  \end{array}
 =~ \frac{\begin{array}{c}\setlength{\unitlength}{.5 pt}
      \begin{picture}(40,35)
        \put(20,32.5){$\scriptstyle a$}
        \put(20,17){$\scriptstyle b$} 
        \put(20, 2){$\scriptstyle c$} 
        \put(23,15){\oval(46,30)} \put( 0,15){\line(1,0){46}} 
        \put( 0,15){\circle*{3}}  \put(46,15){\circle*{3}}
      \end{picture}\end{array}} 
    {\begin{array}{c}\setlength{\unitlength}{.5 pt}
      \begin{picture}(40,35)
        \put(18,20){$\scriptstyle a$} 
        \put(20,15){\oval(40,30)[l]}    
        \put(26,15){\oval(40,30)[r]} 
        \put(20,-6){\line(0,1){12}}\put(26,-6){\line(0,1){12}}  
        \put(20,-6){\line(1,0){6}}\put(20,6){\line(1,0){6}}
        \put(20,30){\line(1,0){6}}
      \end{picture}\end{array}}
 \;\;
  \begin{array}{c}\setlength{\unitlength}{1 pt}
     \begin{picture}(15,40)
        \put(2,-2){\line(0,1){18}}\put(2,20){\line(0,1){18}}
        \put(-2,16){\line(0,1){4}}\put(6,16){\line(0,1){4}}
        \put(-2,20){\line(1,0){8}}\put(-2,16){\line(1,0){8}}
        \put(5,32){$\scriptstyle a$}
        \put(15,15){$\delta_{a,a'}$}
     \end{picture}
  \end{array}
\end{equation}
and
\begin{equation}
  \begin{array}{c}\setlength{\unitlength}{1 pt}
     \begin{picture}(50,40)
        \put(8,18){\line(1, 1){15}} \put(12,26){${\scriptstyle b}$}
        \put(8,18){\line(1,-1){15}} \put(12,6){${\scriptstyle c}$}
        \put(38,18){\line(-1,-1){15}} \put(31,6){${\scriptstyle d}$}
        \put(38,18){\line(-1,1){15}} \put(31,26){${\scriptstyle e}$}
        \put(8,18){\line(1,0){30}} \put(21,21){${\scriptstyle f}$}
        \put(23,33){\line(0,1){12}}\put(25,40){${\scriptstyle a}$}
        \put(23,3){\line(0,-1){12}}\put(25,-7){${\scriptstyle a'}$}
        \put(8,18){\circle*{3}}\put(38,18){\circle*{3}}
        \put(23,3){\circle*{3}}\put(23,33){\circle*{3}}      
     \end{picture}
  \end{array}
 \!\!\! = ~\frac{\begin{array}{c}\setlength{\unitlength}{.8 pt}
      \begin{picture}(45,40)
        \put( 0,15){\line(1,-1){15}} \put(-1,21){${\scriptstyle b}$}
        \put( 0,15){\line(1, 1){15}} \put(-1,4){${\scriptstyle c}$}
        \put( 0,15){\circle*{3}}
        \put(30,15){\line(-1, 1){15}} \put(26,21){${\scriptstyle e}$}
        \put(30,15){\line(-1,-1){15}} \put(26, 4){${\scriptstyle d}$}
        \put(30,15){\circle*{3}}
        \put( 0,15){\line(1,0){30}} \put(11,18.5){${\scriptstyle f}$}
        \put(15,30){\line(1,0){25}} \put(15,30){\circle*{3}}
        \put(15, 0){\line(1,0){25}} \put(15, 0){\circle*{3}}
        \put(40, 0){\line(0,1){30}} \put(43,13){${\scriptstyle a}$}
      \end{picture}\end{array}}
    {\begin{array}{c}\setlength{\unitlength}{.5 pt}
      \begin{picture}(40,35)
        \put(19,19){$\scriptstyle a$} 
        \put(20,15){\oval(40,30)[l]}    
        \put(26,15){\oval(40,30)[r]} 
        \put(20,-6){\line(0,1){12}}\put(26,-6){\line(0,1){12}}  
        \put(20,-6){\line(1,0){6}}\put(20,6){\line(1,0){6}}
        \put(20,30){\line(1,0){6}}
      \end{picture}\end{array}}
 \;\;
  \begin{array}{c}\setlength{\unitlength}{1 pt}
     \begin{picture}(15,40)
        \put(2,-2){\line(0,1){18}}\put(2,20){\line(0,1){18}}
        \put(-2,16){\line(0,1){4}}\put(6,16){\line(0,1){4}}
        \put(-2,20){\line(1,0){8}}\put(-2,16){\line(1,0){8}}
        \put(5,32){$\scriptstyle a$}
        \put(15,15){$\delta_{a,a'}$} 
     \end{picture}
  \end{array} ~~~~~.
\end{equation}

\paragraph{Twist Property.}
We also need quite often in the computations the \emph{twist property} 
of a three vertex, i.e. the fact that a change in the ordering of two 
lines yields a sign factor,
\begin{equation}
  \label{twist}
  \begin{array}{c}\setlength{\unitlength}{1 pt}
    \begin{picture}(40,50)
       \put(30,25){\line(-3, 2){30}} \put(-3,38){$\scriptstyle a$}
       \put(0,25){\line(3,2){13}} 
       \put(30,45){\line(-3,-2){13}} \put(31,38){$\scriptstyle b$}
       \put(15,25){\oval(30,20)[b]}  \put(15,15){\circle*{3}}
       \put(15,1){\line(0,1){14}}   \put(17,1){$\scriptstyle c$}
    \end{picture}
  \end{array} 
 = \; \lambda^{ab}_c \,
  \begin{array}{c}\setlength{\unitlength}{1 pt}
    \begin{picture}(40,40)
       \put(15,18){\line(-1, 1){13}} \put(1,33){$\scriptstyle a$}
       \put(15,18){\line( 1, 1){13}} \put(25,33){$\scriptstyle b$}
       \put(15,18){\line(0,-1){16}}   \put(17,2){$\scriptstyle c$}
       \put(15,18){\circle*{3}}
    \end{picture}
  \end{array} ~,
\end{equation}
where $\lambda^{ab}_c = (-1)^{(a'+b'-c')/2}$ and $x'=x(x+3)$.

\paragraph{Grasp Shifting Lemma.}
The last necessary relation is the \emph{grasp shifting lemma}
\begin{equation}
\label{graspshift}
  c ~\cdot \!
  \begin{array}{c}\setlength{\unitlength}{1.2 pt}
    \begin{picture}(40,40)
        \put(15,18){\line(-1, 1){13}} \put(1,33){$\scriptstyle a$}
        \put(15,18){\line( 1, 1){13}} \put(25,33){$\scriptstyle b$}
        \put(15,18){\line(0,-1){16}}   \put(17,2){$\scriptstyle c$}
        \put(15,18){\circle*{3}}
        \put(15,10){\circle*{3}}
        \put(15,10){\line(-1,0){10}}
        \put(3,11){${\scriptstyle 2}$}
    \end{picture}
  \end{array}
 = ~a~\cdot \!
  \begin{array}{c}\setlength{\unitlength}{1.2 pt}
    \begin{picture}(40,40)
        \put(15,18){\line(-1, 1){13}} \put(1,33){$\scriptstyle a$}
        \put(15,18){\line( 1, 1){13}} \put(25,33){$\scriptstyle b$}
        \put(15,18){\line(0,-1){16}}   \put(17,2){$\scriptstyle c$}
        \put(15,18){\circle*{3}}
        \put(8,25){\circle*{3}}
        \put(8,25){\line(-1,0){10}}
        \put(-4,26){${\scriptstyle 2}$}
    \end{picture}
  \end{array}  
 + ~b~\cdot \!\!\!
  \begin{array}{c}\setlength{\unitlength}{1.2 pt}
    \begin{picture}(40,40)
        \put(15,18){\line(-1, 1){13}} \put(1,33){$\scriptstyle a$}
        \put(15,18){\line( 1, 1){13}} \put(25,33){$\scriptstyle b$}
        \put(15,18){\line(0,-1){16}}   \put(17,2){$\scriptstyle c$}
        \put(15,18){\circle*{3}}
        \put(24,27){\circle*{3}}
        \put(24,27){\line(-1,0){10}}
        \put(12,28){${\scriptstyle 2}$}
    \end{picture} 
  \end{array} ~,
\end{equation}
which is proven as follows. Applying the recoupling theorem 
(\ref{recoupl_thm}) to the right hand side, using
definitions (\ref{6jsymbol}) and (\ref{tet_net}) of 
$6j$-symbols and the tetrahedral net, one finds 
straightforwardly the correct expression of the left hand side.

\paragraph{Final Remark.} In order to be consistent with the 
conventions used in 
the main part, we should have actually drawn the diagrammatic 
representations in this appendix surrounded by a dashed circle to 
indicate that manipulations take place at a point. The identities 
of recoupling theory are applicable only in the virtual part of the 
representation, not for example at real 
crossings. Naively, the reason for this is that we are not dealing with
flat connections like e.g. in BF theory. Consider for example
the evaluation of $\Delta_n$, which is the closed tangle of $n$ totally
antisymmetrized lines, or equivalently a closed color $n$-line.
Its chromatic evaluation gives (up to an $n$-dependent sign factor)
just the dimension of the $SU(2)$ representation in which the line 
lives. We could have considered $\Delta_n$ as a real spin network embedded 
in $\Sigma$, and not as being shrunk to a point (which is, in fact, what 
``virtual'' means). Expressing the corresponding state $\psi$ in 
terms of the holonomy of the connection around a closed (simply 
connected) curve $s$, 
$\psi_n(A) = (-1)^n\, \mbox{Tr}_n(\mathcal{P}\exp(\oint_s A))$, 
we would have obtained a non-trivial result for non-flat connections.
However, since the closed edge $s$ is virtual, the connection ``at a 
point'' is indeed flat, hence the integral is zero and the trace 
(up to a sign) nothing but the dimension of the representation. 
To sum up, we obtain the 
result (\ref{symmetrizer}) of this appendix, namely
$\psi_n^{\mbox{\tiny flat}}(A) \equiv \Delta_n = (-1)^n (n+1)$ 
(the sign factor appears due to the conventions of 
\cite{DePietriRovelli96}, which we use as well).

In expressions like the recoupling theorem (\ref{recoupl_thm}), 
only the internal part of the drawing is virtual. The 
open legs which are actually part of a larger spin network that
is not explicitly drawn, and the same on both sides of the equation,
would stick out of the virtual region.

\end{appendix}



\nocite{*}
\begin{thebibliography}{99}

\bibitem{Rovelli97a} Rovelli C 1998 Loop Quantum Gravity {\it Liv. Rev. 
    Rel.} 1998-1 {\it E-print} gr-qc/9710008

\bibitem{loops} Rovelli C and Smolin L 1990 Loop space representation
  of quantum general relativity {\it Nucl. Phys.} B {\bf 331} 80
  \par\item[] Rovelli C and Smolin L 1988 Knot theory and quantum gravity 
  {\it Phys. Rev. Lett.} {\bf 61} 1155

\bibitem{Early-hamiltonians} Rovelli C 1991 Ashtekar formulation of
  general relativity and loop space nonperturbative quantum gravity: A
  Report {\it Class.  Quant.  Grav.}  {\bf 8} 1613
  \par\item[] Husain V 1989 Intersecting Loop Solutions Of The Hamiltonian 
  Constraint {\it Nucl.  Phys.}  {\bf B313} 711 
  \par\item[] Br\"ugmann B 1991 Intersecting N loop solutions of the 
  Hamiltonian constraint {\it Nucl.  Phys.} B {\bf 363} 221  
  \par\item[] Gambini R 1989 Loop Space Representation of Quantum General 
  Relativity {\it Phys.  Lett.} B {\bf 255} 180
  \par\item[] Br\"ugmann B, Gambini R and Pullin J 1992 Jones Polynomials for
  Intersecting Knots as Physical States of Quantum Gravity  
  {\it Nucl. Phys.} B {\bf 385} 587 {\it E-print} hep-th/9202018
  
\bibitem{Late-hamiltonians} Di Bartolo C, Gambini R, Griego R and Pullin J 2000 
  Consistent canonical quantization of general relativity in the space of 
  Vassiliev knot invariants {\it Phys.  Rev.  Lett.} {\bf 84} 2314
  {\it E-prints} gr-qc/9909063, gr-qc/9911009 and gr-qc/9911010
  \par\item[] Gambini R and Pullin J 2000 Making classical and quantum canonical 
  general relativity computable through a power series expansion in the 
  inverse cosmological constant {\it Phys.  Rev.  Lett.} {\bf 85} 5272
  {\it E-print} gr-qc/0008031
  
\bibitem{Thiemann96a} Thiemann T 1998 Quantum Spin Dynamics (QSD) 
  {\it Class. Quant. Grav.} {\bf 15} 839 {\it E-print} gr-qc/9606089
  
\bibitem{Thiemann96b} Thiemann T 1998 Quantum Spin Dynamics (QSD) II 
  {\it Class. Quant. Grav.} {\bf 15} 875 {\it E-print} gr-qc/9606090

\bibitem{HamiltonianRS} Rovelli C and Smolin L 1994 The physical hamiltonian 
  in nonperturbative quantum gravity {\it Phys. Rev. Lett.}  {\bf 72} 446
  {\it E-print} gr-qc/9308002
          
\bibitem{HamiltonianR} Rovelli C 1995 Outline of a generally covariant 
  quantum field theory and a quantum theory of gravity {\it J. Math.  Phys.}
  {\bf 36} 6529 {\it E-print} gr-qc/9503067

\bibitem{Gambinietal97} Gambini R, Lewandowski J, Marolf D and 
  Pullin J 1998 On the consistency of the constraint algebra in spin network 
  quantum gravity {\it Int. J. Mod. Phys.} D {\bf 7} 97 {\it E-print} 
  gr-qc/9710018

\bibitem{LewandowskiMarolf97} Lewandowski J and Marolf D 1998 Loop constraints: 
  A habitat and their algebra {\it Int. J. Mod. Phys.} D {\bf 7} 299 
  {\it E-print} gr-qc/9710016

\bibitem{Smolin96} Smolin L 1996 The classical limit and the form of the 
  hamiltonian constraint in nonperturbative quantum gravity
  {\it E-print} gr-qc/9609034

\bibitem{Thiemann00a} Thiemann T 2000 Gauge Field Theory Coherent States (GCS): 
  I. General Properties {\it E-print} hep-th/0005233

\bibitem{ThiemannWinkler00a} Thiemann T and Winkler O 2000 Gauge Field Theory 
  Coherent States (GCS) : II. Peakedness Properties {\it E-print} hep-th/0005237 

\bibitem{ThiemannWinkler00b} Thiemann T and Winkler O 2000 Gauge Field Theory 
  Coherent States (GCS) : III. Ehrenfest Theorems {\it E-print} hep-th/0005234

\bibitem{DPF} De Pietri R and Freidel L Private communication
 
\bibitem{ReisenbergerRovelli97} Reisenberger M and Rovelli C 1997 
  \lq \lq Sum over Surfaces'' form of Loop Quantum Gravity {\it Phys. Rev.} 
  D {\bf 56} 3490 {\it E-print} gr-qc/9612035

\bibitem{Baez98} Baez J C 1998 Spin Foam Models {\it Class. Quant. Grav.} 
  {\bf 15} 1827 {\it E-print} gr-qc/9709052

\bibitem{Baez99} Baez J C 2000 An Introduction to Spin Foam Models of
  Quantum Gravity and BF Theory {\it Geometry and Quantum Physics}
  ed Gausterer H and Grosse H (Berlin: Springer) LNP 543 p~25 
  {\it E-print} gr-qc/9905087

\bibitem{BarrettCrane98} Barrett J D and Crane L 1998 Relativistic spin 
  networks and quantum gravity {\it J. Math. Phys.} {\bf 39} 3296 
  {\it E-print} gr-qc/9709028

\bibitem{Borissovetal97} Borissov R, De Pietri R and Rovelli C 1997
  Matrix elements of Thiemann's Hamiltonian constraint in loop quantum gravity 
  {\it Class. Quant. Grav.} {\bf 14} 2793 {\it E-print} gr-qc/9703090

\bibitem{GaulRovelli99} Gaul M and Rovelli C 2000 Loop Quantum 
  Gravity and the Meaning of Diffeomorphism Invariance
  {\it Towards Quantum Gravity: Proc. XXXV Karpacz Int. Winter School 
    on Theor. Phys.} ed Kowalski-Glikman J (Berlin: Springer) LNP 541 p~277
  {\it E-print} \mbox{gr-qc/9910079}

\bibitem{DePietriRovelli96} De Pietri R and Rovelli C 1996 Geometry 
  Eigenvalues and Scalar Product from Recoupling Theory in Loop Quantum 
  Gravity {\it Phys. Rev.} D {\bf 54} 2664 {\it E-print} gr-qc/9602023

\bibitem{DePietri97} De Pietri R 1997 On the relation between the connection 
  and the loop representation of quantum gravity {\it Class. Quant. Grav.} 
  {\bf 14} 53 {\it E-print} gr-qc/9605064

\bibitem{Barbero95} Barbero F 1995 Real Ashtekar Variables for Lorentzian 
  Signature Space-times {\it Phys. Rev.} D {\bf 51} 5507 
  {\it E-print} gr-qc/9410014

\bibitem{Lewandowski96} Lewandowski J 1997 Volume and Quantizations
  {\it Class. Quant. Grav.} {\bf 14} 71 {\it E-print} \mbox{gr-qc/9602035}

\bibitem{RovelliSmolin95} Rovelli C and Smolin L 1995 Discreteness of 
  Area and Volume in Quantum Gravity {\it Nucl. Phys.} B {\bf 442} 593
  {\it E-print} gr-qc/9411005 Erratum: {\it Nucl. Phys.} B {\bf 456} 
  734

\bibitem{AshtekarLewand98} Ashtekar A and Lewandowski J 1998 Quantum 
  Theory of Geometry II: Volume Operators {\it Adv.\ Theor.\ Math.\  Phys.}  
  {\bf 1} 388 {\it E-print} gr-qc/9711031

\bibitem{KauffmanLins94} Kauffman L H and Lins S L 1994 
  {\it Temperley-Lieb Recoupling Theory and Invariants of 3-Manifolds} 
  (Princeton: Princeton University Press) Annals of Mathematics Studies, 
  Number 134

\bibitem{AshtekarLewand97} Ashtekar A and Lewandowski J 1997 Quantum 
  Theory of Geometry I: Area Operators {\it Class. Quant. Grav.} {\bf 14} 
  A55 {\it E-print} gr-qc/9602046

\bibitem{GaulPhD} Gaul M 2001 \emph{Doctoral Thesis}

\bibitem{GaulRovelli00} Gaul M and Rovelli C On a crossing
  symmetric extension of Loop Quantum Gravity (in preparation)

\bibitem{BrinkSatchler68} Brink D M and Satchler G R 1968
  {\it Angular Momentum} 2nd ed (Oxford: Clarendon Press) 
               
\end{thebibliography}
\end{document}